\documentclass[preprint,11pt]{aastex}
\usepackage{graphics}
\usepackage{amssymb,amsmath}
\usepackage{subfig}

\begin{document}

\title{The {\em Chandra} View of Nearby $\mathsf{X}$-shaped Radio Galaxies}

\author{Edmund~J.~Hodges-Kluck$^{1}$, Christopher S. Reynolds$^{1}$,
Chi~C.~Cheung$^{2,3}$ \& M.~Coleman~Miller$^{1}$} 
\altaffiltext{1}{Department of Astronomy, University of Maryland, College
Park, MD 20742-2421}
\altaffiltext{2}{NASA Goddard Space Flight Center, Code 661, Greenbelt, MD
20771}
\altaffiltext{3}{Current address: National Research Council Research
Associate, Space Science Division, Naval Research Laboratory,
Washington, DC 20375}

\begin{abstract}
We present new and archival {\em Chandra X-ray Observatory}
observations of $\mathsf{X}$-shaped radio galaxies within $z\sim 0.1$
alongside a comparison sample of normal double-lobed FR~I and II radio
galaxies.  By fitting elliptical distributions to the observed diffuse
hot X-ray emitting atmospheres (either the interstellar or intragroup
medium), we find that the ellipticity and the position angle of the
hot gas follows that of the stellar light distribution for radio galaxy hosts
in general.  Moreover,
compared to the control sample, we find a strong tendency for
$\mathsf{X}$-shaped morphology to be associated with wings directed
along the minor axis of the hot gas distribution.  Taken at face
value, this result favors the hydrodynamic backflow models for the
formation of $\mathsf{X}$-shaped radio galaxies which 
naturally explain the geometry; the merger-induced rapid reorientation
models make no obvious prediction about orientation. 
\end{abstract}

\keywords{galaxies: active, galaxies: intergalactic medium}

\section{Introduction}

``Winged'' and $\mathsf{X}$-shaped radio galaxies (XRGs) are
centro-symmetric subclasses of Fanaroff-Riley~(FR) type I and II radio
galaxies \citep{fan74} which exhibit a second, fainter pair of wings
lacking terminal hot spots in addition to the symmetric double lobe
structure seen in ordinary FR~II galaxies \citep{leahy84}.  These
galaxies comprise up to 10\% of radio galaxies, although the galaxies
in which the length of the wings exceeds 80\% of that
of the primary lobes (``classical'' XRGs) are a very small subset of
radio galaxies \citep{leahy92,cheung07}.  The primary lobes of these
galaxies, as in other radio galaxies, are produced by a jet emanating
from an active galactic nucleus (AGN) which becomes decollimated as it
propagates into and interacts with the surrounding medium.  In FR~II
radio galaxies, the terminal shock where the jet rams into its
environment is characterized by a radio hot spot \citep[also often
seen in the X-ray band, e.g.][]{hardcastle04} which produces the
so-called ``edge-brightened'' morphology.  The absence of these hot
spots in the fainter wings of XRGs implies that they do not harbor an
active jet, although \citet{lal07} argue in favor of a dual-AGN origin
in which the pairs of lobes emanate from separate, unresolved AGN.

The $\mathsf{X}$-shaped morphology is of interest because two
remarkably disparate classes of models have been invoked to explain
it.  The first class is predicated on the reorientation of the jets
either by realignment of the supermassive black hole (SMBH) spin or
the accretion disk, whereas the second purports to explain the
distorted morphology as the result of hydrodynamic interaction between
the radio lobe and its surrounding gaseous environment on kiloparsec
scales.

In the first case, the most common explanation for the
$\mathsf{X}$-shaped morphology is that the SMBH has its spin axis
realigned, either via merger or precession.  The possibility that XRG
morphology is produced by a SMBH merger is of considerable interest as
a potential estimate of merger rates and hence source rates for the
{\em Laser Interferometer Space Antenna}; because fossil lobes will
quickly decay due to adiabatic expansion, an XRG would be a sign of a
recent merger.  In this scenario
\citep[e.g.][]{rottmann01,zier01,merritt02}, a SMBH binary formed by
galactic merger and dynamical friction hardens for an unknown length
of time via three-body interactions until gravitational radiation
becomes effective at radiating orbital energy.  At this point, the
binary quickly coalesces, and the radio jets realign along the
direction of the angular momentum of the final merged black hole.  The jet
begins propagating in the new direction, leaving a decaying pair of
radio lobes along the old spin axis.  A significant objection to the
merger model \citep{bogdanovic07} is that the linear momentum ``kick''
imparted to the coalesced SMBH can exceed several times $10^{3}$ km
s$^{-1}$ if the spins of the black holes are random at the time of
merger, sufficient to escape the galaxy.  Coaligned spins reduce the
magnitude of the kick, but would not result in an $\mathsf{X}$-shaped
source.  Although not a problem for rapid reorientation models which
rely on precession or other realigning mechanisms
\citep{ekers78,rees78,klein95,dennett02}, an additional objection to
the ``fossil'' lobe scenario is that the secondary lobe lengths in
some XRGs are inconsistent with the fast lobe decay expected when a
jet changes alignment \citep[][hereafter S09]{saripalli09}.

In contrast to the rapid-realignment models, hydrodynamic models
propose that the wings of XRGs were never directly inflated by a jet.
These models argue that XRGs form due to backflow (plasma flowing back
towards the AGN from the hot spot shocks) that interacts with the
surrounding gas.  As presently conceived, backflow models require
FR~II morphlogy to drive the strong backflows.  The existence of FR~I
XRGs challenges these hypotheses, but S09 argue that, since several of
the FR~I XRGs appear to be restarted AGN with inner FR~II morphology,
the FR~I XRGs could have had edge-brightened morphology when the wings
were inflated, implying an evolution of FR~II to FR~I sources
\citep{cheung09}.  In
this paper, we consider the backflow models as a unified class.  The
two most prominent backflow scenarios include the ``buoyant backflow''
model \citep{leahy84,worrall95} and the ``overpressured cocoon'' model
\citep[][hereafter C02]{capetti02}.  The buoyant backflow model
supposes that the buoyancy of the relativistic plasma cocoon in the
interstellar or intragroup/intracluster medium (ISM or IGM/ICM)
produces the additional wings.  Because of the collimation seen in the
more dramatic XRG wings, we refer to them hereafter as ``secondary
lobes'' even though, if the hydrodynamic premise is correct, they are
not of the same character as the primary lobes.

C02 propose a variant model in which backflowing plasma confined by an
envelope of hot gas continues to aggregate until the cocoon of radio
plasma is significantly overpressured, at which point it blows out of
the confining medium at its weakest point.  Supposing that the
confining medium is the ISM of an elliptical galaxy, if the jet is
aligned along the {\em major} axis of the galaxy, then the cocoon may
become overpressured before the jet bores through the ISM, and the
radio plasma will blow out along the {\em minor} axis of the galaxy,
forming the secondary lobes.  Conversely, if the jet is oriented along
the minor axis of the galaxy, then backflow will either escape along
the same axis or the jet will escape the ISM before the cocoon can
become overpressured.  C02 cite an intriguing correlation between the
orientation of the secondary lobes in XRGs and the minor axis of the
host galaxy as evidence for their model, which they further support
with two-dimensional hydrodynamic simulations.  In a follow-up study
including ``normal'' radio galaxies, S09 find that the primary lobes
of giant radio galaxies are preferentially aligned along the minor
axis of the host, whereas they extend the original C02 result to a
larger XRG sample.  Although the observed geometric correlation is
strong, much of the potentially relevant physics is absent in the C02
simulations, and \citet{kraft05} note that the buoyant backflow model
can explain the observed correlation by assuming an anisotropic medium
to divert the backflow.  Moreover, it is unclear whether real radio
lobes are actually overpressured.  \citet{reynolds01} argue that an
overpressured cocoon is inflated early on, but in the buoyant backflow
model, the secondary lobes are formed later.  In either case, the
geometric correlation favors a hydrodynamic origin for the secondary
lobes in the absence of an explanation for a relationship between the
angular momenta of two merging SMBHs and large-scale structure of the
galaxy.  For the remainder of this paper, we refer to the ``C02
geometry'' to describe the correlation noted in XRGs and the proposed
geometry of jet alignment that would produce them.

There are additional XRG formation models which are similar to the
ones presented above in that they rely solely on either the black
hole(s) involved or jet--gas interaction.  These include the
hypothesis that $\mathsf{X}$- and $\mathsf{Z}$-shaped distortions
arise via gravitational interaction with another galaxy
\citep{wirth82,vanbreugel83}, the idea that the jets are diverted by
the ISM of a smaller merging galaxy \citep{gopal03,zier05}, and the
aforementioned \citet{lal07} proposal that both lobes are powered by
active jets.  Because these models are not as easily probed by the hot
gas, we focus on the backflow models hereafter.

In this paper, we seek to characterize the properties of the hot gas
in these systems and determine whether the C02 optical--radio
geometric correlation also exists in the X-ray band.  One assumption
of the C02 proposal is that the stellar distribution
traces the hot gas which makes up the confining medium.  We will test this
assumption directly by determining the extent to which optical and X-ray
morphology are correlated. As a
parallel study, we present the results from new and archival {\em
Chandra X-ray Observatory} observations of XRGs and investigate
whether the hot gas in XRG systems differs from that in a comparison
sample of archival {\em Chandra} observations of ``normal'' FR~I and
II galaxies \citep[taken largely from the 3CRR catalog;][]{laing83}.

Any observational study of XRGs is necessarily limited by the small
number of known and candidate sources.  Our study is further limited
by two important factors: (1) the hot gas surrounding radio galaxies
becomes increasingly difficult to characterize at higher redshift, and
(2) useful X-ray data do not exist for most XRGs.  These
considerations strongly constrain our conclusions.  We therefore
discuss in detail our target selection criteria in \S2, as well as the
observational parameters and reduction techniques applied to the data.
In \S3 we discuss our primary analysis of the morphology of the hot
gas, and in \S4 we discuss the X-ray spectra of the AGN.  In \S5 we
summarize our results and interpretations.

Throughout this paper, we adopt the {\em Wilkinson Microwave
Anisotropy Probe} (WMAP) values of $H_0 = 71$ km s$^{-1}$ Mpc$^{-1}$,
$\Omega_M = 0.27$, and $\Omega_{vac} = 0.73$ with flat geometry.  We
calculate equivalent angular scale and distances at redshift using the
online calculator provided by \citet{wright06}, and for Galactic
absorption, we use the online HEASARC $N_H$ calculator with values
from the Leiden/Argentine/Bonn survey \citep{kalberla05}\footnote{See
http://heasarc.gsfc.nasa.gov/cgi-bin/Tools/w3nh/w3nh.pl}.

\section{{\em Chandra} Observations}

{\em Chandra} is well suited to a study of the hot environments of
radio galaxies thanks to its high sensitivity, 0.5$^{\prime\prime}$
spatial resolution, and 0.3-10 keV bandwidth.  {\em Chandra} has also
been able to resolve X-ray emission associated with the radio lobes
into hot spots and jets \citep{hardcastle04}; distinguishing this
emission from the gaseous halos is especially important for our
purposes.

Our analysis sample of {\em Chandra} data consists of eighteen
comparison sample galaxies and eight XRGs (Fig. 1 [online-only] \& 2 with
observational parameters in Table 1).  In this section, we describe
how we arrived at this sample, starting with the preliminary selection
criteria for the XRG and comparison samples from radio data and
availability in the {\em Chandra} archive (\S2.1).  After reducing
these data, we rejected a number of galaxies due to low quality or
pile-up (\S2.2), then used spectral fitting to find those galaxies
where the diffuse gas dominates the photon count in the relevant
regions (\S2.3).  These are the galaxies we include in our final
sample (Table 1); Figs. 1 (online-only) \& 2 are discussed along with the radio and
optical data we use in \S2.4.

We note that many of the archival data sets have been
published, and references are provided in Table \ref{obsparams} where
available.  Our goal is not to exhaustively study any individual
source.  Notes on individual sources are found in Appendix A.

\subsection{Preliminary Target Selection}

The most complete compilation of known and candidate $\mathsf{X}$-shaped
sources in the literature is that of \citet{cheung07}, who used the
NRAO\footnote{The National Radio Astronomy Observatory is a facility
of the National Science Foundation operated under cooperative
agreement by Associated Universities, Inc.} {\em Very Large Array}
\citep[VLA;][]{thompson80} Faint Images of the Radio Sky at Twenty cm
\citep[FIRST;][]{becker95} data to identify the candidate XRGs.  For
this paper, we define XRGs as comprising all radio galaxies in
\citet{cheung07} that have a wing length exceeding 80\% of the active
lobes \citep[][the ``classical'' XRGS]{leahy92} and those ``winged''
sources in \citet{cheung07} which have obvious $\mathsf{X}$-shaped
morphology in the angle between the active lobes and the wings.
Assuming the projected lengths on the sky are the real lengths of the
wings, in the backflow models these may be ``classical'' XRGs at an
earlier stage.  

We wish to study the hot gas surrounding the radio galaxy.  The rapid
decline in surface brightness and reduction in angular size with
increased redshift makes a cutoff redshift of $z \sim 0.1$ practical
for typical {\em Chandra} exposure times (even in this scheme, our
sample is biased towards high pressure systems).  Nineteen XRGs fall
within this cutoff; the highest redshift in this group is $z = 0.108$.
However, we have {\em Chandra} data for only thirteen of these
sources, and these comprise our preliminary sample.  These sources include 
3C~315, 3C~223.1, NGC~326, PKS~1422+26, 3C~433, 3C~403, 3C~192, B2~1040+31A,
Abell~1145, 4C~+00.58, 3C~136.1, 4C~+32.25, and 4C~+48.29.  

This sample is heterogenous in its radio
properties and morphology.  A few of the galaxies may be described as
$\mathsf{Z}$-shaped \citep{gopal03,zier05}.  NGC~326, for instance,
has long secondary lobes that do not seem to meet at a common center.
Others (e.g. B2~1040+31A) have obviously $\mathsf{X}$-shaped lobe
axes, but the lobes are not as well collimated as in NGC~326 or
3C~315.  Lastly, PKS~1422+26, B2~1040+31A and 3C~433 appear to be
hybrid FR~I/II radio galaxies with one FR~I lobe and one FR~II lobe
\citep[A ``HYMOR''; see][]{gopal00}.
It is unknown whether these distinctions are signatures of any
formation model, and we note that XRGs tend to lie close to the
FR~I/II break in other observables \citep{cheung09}.  In this paper we
use the definitions of \citet{cheung07} and so consider the XRG sample
as a unified whole, and we caution that our results are interpreted
within this framework.  In particular, there is no consensus in the 
literature on what constitutes an ``$\mathsf{X}$''-shaped galaxy.  In 
their optical study, for example, S09 classify 3C~76.1 ($z = 0.032$) as an 
XRG, whereas we do not.  They also classify several higher redshift galaxies
as XRGs which do not appear in the \citet{cheung07} list (e.g. 3C~401 or
3C~438); in this respect our sample is conservative.  At least one high
redshift XRG from the \citet{cheung07} list---3C~52 ($z = 0.285$)---has a
{\em Chandra} exposure, and 3C~197.1 ($z=0.128$) may be an XRG and is also
in the {\em Chandra} archive.  We choose not to use these exposures, although
we note them in Appendix A.3.

Within the same redshift cutoff, we identify normal FR I and II
galaxies in the {\em Chandra} archive as a comparison sample.  The aim
of this sample is to determine whether XRGs systematically differ in
X-ray properties from normal radio galaxies.  It is not obvious what
constitutes an appropriate comparison sample because although FR~II
lobes are deemed necessary to produce $\mathsf{X}$-shaped morphology
in the backflow scenario, FR~I XRGs exist (S09).  In the S09
interpretation of FR~I XRGs, the FR~I lobes must have had FR~II
morphology at the time the wings were generated.  Supposing that both
the C02 model and S09 interpretation are correct, the ``old'' FR~I
lobes must obey the same geometry as their ``active'' FR~II
counterparts.  A comparison of the XRG sample to both types is
therefore useful.  The combined sample also provides a larger
reservoir of sources for measuring the correlation between optical and
X-ray isophotes in the ISM.  We adopt a preliminary comparison sample
consisting of (1) all FR~II galaxies within $z \sim 0.1$ with data in
the {\em Chandra} archive and (2) FR~I galaxies from the DRAGN catalog
\footnote{Leahy, J.P., Bridle, A.H. \& Strom, R.G., editors ``An Atlas of
DRAGNs'' (http:/wwww/jb.man.ac.uk/atlas/)}
with the same conditions.  There are 41 sources
meeting these criteria.

\subsection{Data Reduction}

Thirteen XRGs and 41 normal radio galaxies comprise our preliminary
samples, but the data for many of these are not of sufficient quality
for our analysis.  Both the archival data and our new data use
Advanced CCD Imaging Spectrometer\footnote{See
http://cxc.harvard.edu/proposer/POG/pdf/ACIS.pdf} (ACIS) chips with no
transmission grating in place; we immediately rejected sources which
have only grating spectrometer observations due to severe pile-up in
the zeroth order image (only the brightest sources have grating 
data).  Most of the data sets use the nominal aim
point on the ACIS-S3 chip, but we consider several ACIS-I archival
observations. 

We reprocessed the {\em Chandra} data sets to generate 0.3--10~keV
level=2 files using the {\em Chandra} Interactive Analysis of
Observations (CIAO v4.0) data processing recipes
(``threads'')\footnote{ See
http://cxc.harvard.edu/ciao/threads/index.html} with the most recent
CALDB release (3.5.0).  Times with noticeable background flares were
excised by inspecting 0.3--10~keV lightcurves.  A few comparison
sample data sets contained the readout streak produced by bright
sources, which we replaced, following the CIAO thread, with a strip of
nearby background on either side of the central point source.
The bright sources NGC~6251 and 3C~264
were observed in a 1/8-size subarray mode of the ACIS-S3 chip, but
still produced a readout streak.  Exposure-corrected final images were 
then produced from the energy-filtered level=2 files following the
standard CIAO thread, using the aspect solution file associated with each
observation. 

The core AGN X-ray emission was identified by matching the position of
the X-ray source with the optical counterpart on the sky.  More
accurate positions were determined by binning the X-ray image and
calculating the centroid.

\subsection{Spectral Extraction}

Our analysis (\S3) is based on the properties and morphology of the ISM and
IGM/ICM.  Therefore, we reject galaxies from our preliminary sample where we
do not detect strong extended thermal plasma.  We determine the final sample in two
steps: (1) We use radial profile fitting to reject sources with no extended
emission and distinguish between multiple components of extended emission
in the remainder.  We extract spectra from regions corresponding to these
extended components, then (2) reject sources whose extended emission is not 
fit well
by strong thermal components.  We describe the first step presently and
the second in \S2.4.

Extended emission on the chip may be made up of multiple sources.  We use
surface brightness profile fitting to distinguish between these sources, and
based on these fits, define regions for spectral extraction to isolate each
source as much as possible.  We excised point sources from the level=2 file,
then extracted and fit radial profiles from annuli surrounding the AGN out
to many kpc (scale depends on $z$).  In many cases, no diffuse emission was
detected above background.  Where multiple components were found, we defined
spectral extraction apertures based on the characteristic radii ($r_0$ from
a $\beta$ model) for each region.  Compact emission centered on the AGN with
$r_0$ smaller than the host galaxy we identified as ``ISM'', whereas very
broad emission which declines slowly in surface brightness we identified with
the ``IGM/ICM''.  These are the labels used in Table \ref{thermalspecparams};
we note that the only sources included in Table \ref{thermalspecparams} are
those in which thermal emission has been verified.  At the present stage, the
purpose of these regions is solely to isolate different sources for spectral
fitting.

For the larger, dimmer extended sources, the fits (and hence $r_0$) were
less reliable.  However, the slow decline of the ``IGM'' surface brightness
means the spectra we extract are relatively insensitive to the size and shape
of the region.  Crucially, the spectral extraction apertures are {\em not}
the same as those we use for our analysis in \S3 and are defined only to
isolate different sources.  Spectra were extracted from these regions and 
were binned to at least 15 photons/bin to ensure the reliability of 
the $\chi^2$ statistic; in high-quality spectra, the binning was as high 
as 100.  In addition, we extract spectra for the unresolved AGN emission from
detection cells using the 90\% encircled energy radius as determined
by the CIAO {\em mkpsf} tool.  We return to the AGN spectra in \S4.

\subsection{Detection of Diffuse Gas}

We fit models to the spectra we extracted (\S2.3) to determine whether the
emission is dominated by thermal plasma.  Specifically, our sample
consists of those galaxies in which the thermal emission surrounding the
AGN appears dominant in number of photons, a point we return to below.
We first ascertain the presence of hot gas by fitting the spectra extracted from
each source using XSPEC v.12.5.0j \citep{arnaud96}.  In this paper we exclusively use
the {\tt apec} model.  This step is required since extended emission need not be thermal
in origin---plausible nonthermal sources include power law emission from the
boundary shock of the cocoon inflated by the radio jets or from X-ray binaries
(XRBs) in the host galaxy.  Our fits all use a frozen Galactic $N_H$ 
absorption component.  After thermal emission is established, we then require
the thermal component to be dominant at low energies.  This ensures that the
main contribution to the (energy-filtered) surface brightness is the thermal
plasma.  Of course, the hot atmospheres of many galaxies in our sample have 
been well studied (e.g. by {\em Chandra}; Table \ref{obsparams}), but we wish
to apply a uniform standard to all our sources including ones which have low
signal. 

For each spectrum, we first determine whether a single power law or thermal
model is a better baseline fit based on the $\chi^2$ statistic, then add
complexity as the degrees of freedom allow.  In the case of thermal models
we begin by freezing the abundances with an isothermal model (there exists an
abundance---normalization degeneracy in the absence of strong emission lines).
2-temperature (2-T) or multi-T models are invoked if an isothermal model is insufficient to
fit the spectrum and the additional thermal component produces a better fit
than a power law component.  It is worth asking whether a spectrum fit well
by a single thermal model is better fit by a blend of power law and thermal
components.  The answer is almost invariably ``no''.  In very low signal spectra
with few degrees of freedom we conclude hot gas is present if a thermal model 
with reasonable parameters ($kT < 5.0$ and $0.1 < Z < 1.0$) is a better fit 
than a power law model.  We justify this assumption by the slopes of soft
spectra: unabsorbed power law emission tends not to fall off at the lowest
energies, whereas thermal emission peaks near 1.0~keV.  A ``peaky'' spectrum,
even a low-signal one, is fit better by a thermal model.  
As described in \S2.3, we have attempted to isolate different sources, but
we see the ISM and IGM in projection.  We usually lack the signal
for a deprojection analysis, so we first extract and fit a spectrum from the
larger IGM region, then add the resultant models as frozen components in the
ISM spectrum (keeping the normalization thawed).  We find this produces better
results than using a ring of IGM as background for the ISM spectra.   It is
possible that the power law models are in fact describing thermal continua, but
the thermal models which fit these spectra have inordinately high temperatures
and often require suspiciously low abundances. 

For a galaxy to be included in our final sample we required the thermal
component to be dominant between $0.3 < kT < 2$~keV.  Due to {\em Chandra}'s
energy-dependent effective area, even components with higher {\em luminosity}
(e.g. an absorbed power law) may have many fewer photons in this range.  This
criterion allows us to characterize the morphology of only the hot plasma.

Up to this point we have treated our sources as uniformly as possible.  Our
analysis sample consists of the galaxies whose extended emission is dominated
by thermal plasma, but we also reject sources with very complex morphology
(3C~321, M84, M87, and 3C~305; Cyg~A was retained after excising the inner
cocoon).  We remind the reader that we began with an XRG sample comprising
those 13 galaxies within $z \sim 0.1$ with data in the {\em Chandra} archive.
Of these, we have 8 in our analysis sample.  The preliminary comparison sample
included 41 normal radio galaxies within the same cutoff, and of these we
retain 18 (ten of which are FR~II galaxies).  Unfortunately, in many of the
shorter XRG observations, we failed to detect significant diffuse emission.
Notes on individual galaxies are in Appendix A, including those XRGs not
included in our analysis sample (Appendix A.3).  Sources we considered but 
rejected are listed in Table \ref{rejectedobs}.

For those galaxies in our analysis sample we provide the model fit parameters
in Table \ref{thermalspecparams}.  We also present the emission-weighted
density $\bar{n}$ and average pressure $\bar{P} = \bar{n}kT_{\text{fit}}$.
$\bar{n}$ was computed by assuming the minor axis of the extraction region
on the sky is the true minor axis and that the ellipsoid is axisymmetric in
the minor axis (i.e. the axis along the line of sight is the minor axis).  We
report errors in Table \ref{thermalspecparams} for one parameter of interest
at $\Delta \chi^2 = 2.7$ (90\%), but we do not report errors on $\bar{n}$ or
$\bar{P}$ since there are unquantifiable sources of systematic error from
our assumptions on the volume.  The average densities and pressures are not
particularly useful for studying any one system because profile information
has been discarded; for most of our sources we cannot use
deprojection.  Our spectra with folded model fits are shown in their
entirety in the online edition in Fig.~3 (ISM), Fig.~4 (IGM), and
Fig.~5 (AGN). 

\subsection{Radio \& Optical Maps}

We require high signal and high resolution radio data in order to
study the interaction between the radio lobes and the X-ray emitting
gas.  As the secondary lobes of XRGs and ZRGs are typically much
fainter than the hot spots demarcating the terminal shocks of the
active lobes, deep radio observations are often required to accurately
determine their extent and angular offset from the active pair.  The
VLA is ideal for continuum band observations.  Reduced VLA data for many of
the sources are available through the NASA/IPAC Extragalactic Database
(NED); references are provided in Table \ref{obsparams} where
available.  In some cases, we have reprocessed archival VLA data
ourselves, and these are noted in Table \ref{obsparams}.

In contrast to the requirement for high quality radio data (needed to
constrain the orientation of the radio jets and lobes), images from
the (Palomar or UK Schmidt Telescope) Digitized Sky Survey (DSS) are
usually sufficient for comparing the extent and alignment of the
diffuse X-ray emitting gas to the distribution of optical light.
Except in the case of B2~1040+31A, a close triple system, the optical
emission on the relevant scales does not have significant
contamination from companion sources.  We note that the DSS spans our
entire sample, whereas the higher resolution Sloan Digital Sky Survey
\citep[SDSS;][]{adelman08} data are available for fewer than half of
our sources.  However, we compare the DSS images to the SDSS images
where available, and note that good agreement is found in most cases
for our model parameters (\S3.2; Table \ref{eccentricitytable}).  We
also use published {\em Hubble Space Telescope} (HST) images, in
particular those of \citet{martel99} for 3CR galaxies within $z <
0.1$.  However, this survey is not sufficiently complete to replace the
DSS data.

The images in Fig. 1 (available online) \& 2 show the data, using the higher quality SDSS
images for the optical band where possible.  Because of the
differences in scale between radio galaxies and the different media
which are important for lobe--gas interaction, the chip images we show
are all processed slightly differently.  In most cases, they have been
smoothed with a Gaussian kernel of fixed $\sigma$ (per galaxy) to
enhance diffuse emission.  Adaptive smoothing sometimes produces
artificial structure in our images and is not used.  Overlaid on the
images of the ACIS chips are radio contours, and accompanying each
X-ray image of the optical field on the same spatial scale.
Figures 1 \& 2 have been processed to represent the appearance of the
sources and distribution of surface brightness on the sky, but we did not use
these images directly in our analysis. 
In addition, the varying signal-to-noise ratios (S/N) in
X-ray images do not necessarily correspond to real variation in source
luminosity because the data sets come from observations of varying
depth and redshift.  Because some of the observations are shallow
($\le 10$~ks) or use the ACIS-I array, a non-detection of diffuse
emission is not conclusive.

\section{Thermal Atmosphere Properties}

The models which assert that XRGs result from the interaction of the
radio lobes with anisotropic gaseous environments
\citep{capetti02,kraft05} must ultimately be tested by direct
observations of these environments.  In particular, we wish to know
whether the correlation described in C02 exists in the X-ray band, i.e. 
whether XRGs are preferentially found in systems where the jets are directed
along the major axis of the gas distribution. 

This objective can be distilled into two distinct questions.  First, the
observed correlation between the XRG secondary lobes and the minor axis of
the host galaxy was taken by C02 to imply that the minor axis of the ISM in
these galaxies was similarly aligned.  But how well does the X-ray gas trace
the optical isophotes?  Second, is there a systematic difference between the
orientation of the radio lobes and the X-ray gas distribution in XRGs and
normal radio galaxies?  In other words, does the C02 correlation exist in
the media which interact with the radio lobes?  These questions are
distinct in part because the medium responsible for shaping the XRGs in 
the backflow models is not necessarily the inter{\em stellar} medium, but
may be the IGM/ICM.  Because these questions rely on our measurement of gas
morphology, we first describe our image fitting method, which consists of
fitting ellipses to measure ellipticity ($\epsilon \equiv 1 - b/a$) and 
position angle (PA) of the X-ray emission on the sky.

\subsection{Ellipse Fitting}

We determine the gross morphology of the hot gas by fitting ellipses to the
X-ray emission and finding the characteristic $\epsilon$ and PA---gross
elongation and orientation---of the surface brightness.  We treat the
surface brightness as a two-dimensional ``mass'' distribution and compute
the moment of inertia tensor within an aperture chosen by the scale of the
medium (a point we return to below).  From this we measure the principal
axes corresponding to the characteristic major and minor axes of the
distribution.  Of course, the gas distribution may well have complexity beyond
that captured by a simple elliptical model---however, most of our data do not
justify higher-order models.  We describe our method presently. 

Our generic ellipse-fitting routine takes as input a processed surface 
brightness distribution on the chip (see below).  To determine the principal 
axes, we use the QL method with implicit shifts \citep{press92} to find the 
eigenvalues of the tensor, whereas the position angle is determined by the 
orientation of one of the eigenvectors.  We use the {\em Interactive Data 
Language} {\tt eigenql} function\footnote{Source code available from
http://imac-252a.stanford.edu/programs/IDL/lib/eigenql.pro}
realization of this method.  The QL method with implicit shifts is a
method for finding the eigenvalues of a matrix by decomposing it into
a rotation matrix ``Q'' and a lower triangular matrix ``L''; by virtue
of the decomposition, the eigenvalues of the original matrix appear on
the diagonal of the lower triangular matrix at the end.  To determine the
2$\sigma$ error bars reported in Table \ref{eccentricitytable}, we use the
bootstrap resampling method over $10^4$ iterations \citep{efron82}.  We adopt
this method since it is better suited to low count rate images than the 
standard {\em Sherpa}\footnote{See http://cxc.harvard.edu/sherpa} 2D fitting 
routines.

Our treatment of the surface brightness prior to the fitting described above
differs slightly between the bright and compact ISM and the faint, extended 
IGM/ICM.  Within each regime we attempt to treat each data set uniformly; 
any additional processing is noted in Appendix A.
\begin{enumerate}
\item {\em Interstellar Medium} \ \\ 
For the ISM, we use no additional binning beyond the {\em Chandra} resolution
(and thus cannot subtract a constant background).  This choice is motivated
by the desire to minimize artificial smoothing due to the relatively small 
scales of the ISM.  However, we do apply the exposure correction across the 
medium as an adjustment to the brightness of each pixel, and we energy filter
the image based on the spectrum.  The radius of the 
(circular) aperture we use is straightforwardly determined from a 1D radial 
profile extracted from annuli centered on the AGN.  We excise the AGN emission 
through energy filtering and masking a point spread function (PSF) we created 
at the location of the AGN using the CIAO tool {\em mkpsf}.  We mask the region 
corresponding to the 95\% encircled energy ellipse.  One might worry that 
(especially with very bright AGN) this procedure would bias our measured 
$\epsilon$ or PA, but this appears not to be the case; the AGN 
is also usually centered on the nominal aim point, so $\epsilon_{\text{PSF}}$
is usually small.  The remaining counts 
within the aperture are fed into our ellipse-fitting routine.  Because 
background is not subtracted, the bootstrap method is likely more accurate, 
but our measured $\epsilon$ values are systematically low (though not 
{\em very} low thanks to the high contrast of the compact ISM).  We note that 
toggling pixel randomization appears to have no effect on our results.  

\item {\em Intragroup/Intracluster Medium} \ \\
Because the IGM is typically quite faint and extended, we must bin the chip
quite coarsely to see the enhanced surface brightness.  We mask the ISM and
all other point sources, then bin to $16 \times 16$ pixels and apply a 
similarly binned exposure correction.  Energy filtering is applied based on 
the spectrum.  Finally, we smooth the image with a 
Gaussian kernel ($\sigma = 2$ coarsely binned pixels).  We subtract background
levels using either empty regions on the chip (if the visible extent of the
IGM is small) or blank sky files if the medium fills the chip.  The background
files are similarly energy filtered and binned/smoothed.  We choose circular 
apertures motivated by the idea that all our radio galaxies have escaped the 
ISM, but only some have escaped the ``local'' IGM (\S3.3).  By ``local'' IGM we 
specifically mean we use an aperture guided by the characteristic radius $r_0$ 
determined in 1D radial profile fitting centered on the radio galaxy.  
Where $r_0$ is not well constrained, 
we take an aperture roughly bounding the ``2$\sigma$'' isophote (where 
``1$\sigma$'' is taken to be the
background level in the binned, smoothed image).  We then fit the IGM assuming
it is a smooth ellipsoid on the scales of interest (i.e. no internal structure)
so the range in plate scale is not important between galaxies.  A brief
inspection of the isophotes indicates
that this is a good first-order description on large scales, but wrong near
the center.  Inside our chosen aperture, the
bootstrap method essentially tracks the contrast of structure against noise,
so the uncertainty is also a function of the background level we subtract.
Nonetheless, the position angles are reasonably constrained.  Notably, there 
are a few cases where there is strong IGM/ICM not centered on the galaxy 
(NGC~326 and 3C~83.1B), but we ignore the emission that cannot be ``seen'' by
the radio galaxy.  Additionally, in B2~1040+31A, the most significant IGM is 
a smaller structure centered on the radio galaxy's host system enveloped in a 
larger, much dimmer atmosphere with a different PA.  Using the methods above,
we take the smaller structure to be the local IGM.
\end{enumerate}

We treat the optical DSS data similarly.  We do not further bin the images
and we measure the background near the host galaxy (but away from companions,
which we mask).  We use an aperture corresponding to the extent of the hot 
ISM emission.  We wish to use similar apertures in the X-ray and optical
images because $\epsilon$ may (physically) vary with $r$.  In the inner 
optical isophotes, both $\epsilon$ and PA may be (artificially) radially 
dependent due to convolution with the PSF and viewing a triaxial
object (isophotal ``twists'') respectively \citep{binney98}.  Inner isophotes
have dramatically lower values of $\epsilon$ than the outer ``real'' values
\citep[as in][]{tremblay07}.  The isophotal twists, on the other hand, are
a consequence of viewing a triaxial object at a viewing angle not aligned with
any of its axes.  Either is potentially a problem for our optical fitting when
the X-ray ISM aperture is small, but usually the X-ray emission is extended
sufficiently that most of the optical light in our aperture comes from regions
where $d\epsilon/dR$ is small; isophotal twists are only seen in the very 
central regions of a few nearby galaxies in our DSS data.  None of our sources 
are quasars or appear to be saturated. 

\subsection{Optical---X-ray Correlation}

Because long ($50-100$~ks) exposures are often required to see the diffuse 
X-ray gas even in nearby galaxies, it is plainly attractive to adopt the 
optical light as a proxy for the hot ISM.  To zeroth order the stars and
ISM should coincide, but the ISM may not be in hydrostatic equilibrium
with the host \citep{diehl07}, and may be disturbed by recent
mergers presumed to power the AGN.  In fact, \citet{diehl07} find (in an
analysis of 54 {\em Chandra} detections of hot ISM) that the gas ellipticity
and morphology differs significantly from the starlight, although their
sample only overlaps our own by a small amount.  Acknowledging that
a detailed view of the inner ISM shows significant differences from the
starlight, we ask whether the starlight is a sufficient proxy for the ISM in
gross morphology (i.e. we are interested in the outer X-ray ``isophotes'') or
whether it is of such different character that it weakens the \citet{capetti02}
and \citet{saripalli09} analyses. 

In fact, our work supports the identification of the optical light as an
appropriate proxy for the hot ISM.  
However, in contrast to C02, we do not find that
XRGs are preferentially in galaxies with higher $\epsilon$, although we
note that our X-ray $\epsilon$ values are probably underestimated.  
For the subset of our combined sample with strong ISM emission (19 of 26 
galaxies),
we find a correlation coefficient of $R = 0.60$ (Fig.~6) between the 
ellipticity of the
X-ray light ($\epsilon_{\text{X-ray}}$) and the ellipticity of the host
galaxy in the DSS images ($\epsilon_{\text{DSS}}$).  The correlation between
the position angles of the best-fit ellipses is even stronger ($R = 0.96$),
but this is not meaningful because we expect a positive correlation even if
the ellipses are misaligned.  This is because an ellipse may have a PA anywhere
between $0^{\circ} < \text{PA} < 180^{\circ}$, but when considering the
angular separation of the PAs of two superimposed ellipses, one of the
angles of separation must be acute.  Therefore, a better measurement of the
agreement between the optical and X-ray position angles is shown in Fig.~6
where the distribution of $\Delta \text{PA} = 
\text{PA}_{\text{ISM}}-\text{PA}_{\text{optical}}$ is shown.  The values in
Fig.~6 are presented in Table \ref{eccentricitytable}.  The
uncertainties are given at the 2$\sigma$ level from the bootstrap method.  For
$N = 21$ our results are significant at the 95\% level.  The values we use
are given in Table \ref{eccentricitytable} along with a comparison of DSS to
SDSS values (and our ISM values to the best-fit AGN PSF values).

We check our results for the optical values against the literature, where 
profile and isophotal fitting is standard.  Our distribution of optical 
$\epsilon$ peaks slightly below $\epsilon = 0.2$ and falls off quickly at 
higher values.  This is in agreement with the {\em HST}
study by \citet{martel99}, but is slightly rounder than the reported
distributions in the ground-based studies of radio galaxies 
by \citet{smith89} and \citet{lambas92}, whose distributions peak at 
$\epsilon = 0.2$ and 0.25 respectively.  The small excess of very round
hosts may be real (it persists even when $\epsilon$ is measured at larger
radii than reported) but our method may also underestimate $\epsilon$.  
The $\epsilon$ measurement has been made for far fewer sources in the X-ray
band so a systematic comparison is difficult.  In addition, we compare our
optical $\epsilon$ and PA values for the XRG sample specifically to any
matching sources in the C02 and S09 studies (Table \ref{opticalcomparison})
and find generally good agreement.  In other words, in the X-ray band we (to
some extent) reproduce the C02 and S09 results. 

Additionally, we note here a few caveats.  First, we do not take into
account the possible XRB contamination of the ISM.  We do not
consider this a significant problem based on our spectroscopic 
analysis---even though LMXB spectra are often stronger at lower energies, the 
thermal models are much better able to fit the peaks in the ISM spectra, and
are often sufficient for an adequate fit without requiring a power law.
This is consistent with \citet{sun05} who demonstrate that the XRB 
contribution to the emission is usually less than 10\% of the luminosity; in 
many of our spectra
we lack the signal to distinguish a $< 10$\% contribution.  Second, as noted
in \S3.1, our X-ray $\epsilon$ values tend to be systematically 
underestimated.  This makes a comparison with, for example, the \citet{diehl07}
work difficult.  Notably there
is one strong outlier in 3C~403, whose $\epsilon_{\text{X-ray}} = 0.47
\pm 0.03$ is much higher than its optical ellipticity.  \citet{kraft05} find
an even higher value of $\epsilon_{\text{X-ray}} = 0.57 \pm 0.04$ using
a profile fitting technique (they also compare $\epsilon_{\text{X-ray}}$ to 
{\em HST} $\epsilon_{\text{optical}}$), so it appears likely that the 3C~403
ISM is indeed out of equilibrium.  However, the case of 3C~403 is noteworthy
in that the small extent of the ISM relative to the AGN emission makes
energy filtering a more effective way to remove the AGN prior to fitting.
\citet{kraft05} use a $0.3-1.0$~keV window, but it is difficult to conclude
with certainty that this emission is ``dominated'' by the hot gas because it
is possible to obtain a good fit to the spectrum (shown online in
Fig.~5) using an unabsorbed
power law between $0.3-2.0$~keV.  If 3C~403 is indeed badly out of equilibrium,
it appears to be an isolated case in our sample. 

\subsection{X-ray---Radio Correlation}

We now measure the correlation between the position angles of the X-ray media
and the radio jets by asking whether the quantity $\Delta \text{PA} = 
\lvert\text{PA}_{\text{radio}} - \text{PA}_{\text{X-ray}}\rvert$ is
uniformly distributed for the XRGs and normal radio galaxies.  A uniform
distribution means that for a given sample there is no preferred alignment
between the radio jets and the PA of the surrounding hot gas.  The radio
jets are highly collimated and their orientation is determined ``by eye''; we
believe the values are accurate to within 10$^{\circ}$.  The ellipticity and
PA of the X-ray emission are determined as described in \S3.1, but to which
medium shall we compare the alignment of the jets?  Our goal is to determine
whether XRGs reside preferentially in media elongated along the jet axis.  In 
the overpressured cocoon model, this medium is assumed to be the ISM, but 
there are a few XRGS (NGC~326 and PKS~1422+26) for which the ISM cannot be the
``confining'' medium because the wings are produced outside the ISM.  
Therefore we must consider how to deal with the IGM/ICM.  We describe the
process of IGM/ICM aperture selection after our ISM comparison.  
In our discussion, the ``relevant'' medium is the one which could potentially 
confine the radio galaxy.  

\begin{enumerate}
\item {\em Interstellar Medium} \ \\ 
All of our radio galaxies have escaped the ISM, but it must have been the
relevant medium for all of them at early stages.  Assuming that the jet
orientation relative to the ISM has not changed over the life of the radio
galaxy, it is straightforward to compare $\Delta$PA for the normal radio
galaxies and the XRGs using the ISM.  In the C02 model, we expect that the
XRGs would have a small angular separation between the ISM major axis and
the radio jets, i.e. $\Delta$PA $\sim 0$.  Fig.~7 bears this expectation
out (values reported in Table \ref{eccentricitytable}).  
The aperture for the ISM was the same as described in \S3.2 and we
exclude the same five galaxies.  Fig.~7 includes 6 XRGs and shows the
distributions of both the active and secondary lobes.  As noted in \S3.2,
the agreement between the gas and the starlight is good.  Neither XRG
distribution is consistent with a uniform distribution ($P = 0.04$ for the
primary lobes and $0.01$ for the secondary lobes) despite the
small sample size, whereas the normal radio galaxies are consistent with 
uniform distributions ($P \sim 10$\%).  The normal FR~I $\Delta$PA 
distribution is only marginally consistent with uniform (although dramatically 
different from the XRGs) and may be influenced by the giant radio galaxies
which S09 noted tend to have jets aligned along the {\em minor} axis of the
host.  This result is, in our view, strong
evidence for the C02 model, although it was anticipated from \S3.2.  The IGM
may be the confining medium for several XRGs, but the strong agreement with
the C02 geometry in the ISM suggests that the ISM---jet orientation may be
intrinsically important.  The overpressured cocoon model put forth to explain
this geometry should not be confused with the observed result. 

\item {\em Intragroup/Intracluster Medium} \ \\
Because the XRGs exist whose wings could not have been produced by the ISM in
the C02 model, we want to see if the C02 geometry exists among IGM atmospheres
as well.  We thus have to consider how to choose apertures in the IGM/ICM for
the comparison sample as well.  There is no obvious way to compare these
media for various radio galaxies, especially since some of our control sample
radio sources extend far beyond the chip boundaries.  

However, using the same logic as above, we can measure the elongation and
orientation of the ``local'' IGM/ICM (\S3.1).  All sources larger
than the local IGM/ICM must have passed through it at one point, and it is
the relevant medium for all sources enclosed by it and outside the ISM.  As
mentioned in \S3.1, we assume that the IGM/ICM is described by a smooth
ellipsoid so the position angles are comparable between sources of vastly
different scale (the gradients are more important). 
We therefore expect the $\Delta$PA values to be uniformly distributed in the
control sample. 

We detect IGM/ICM emission in fewer galaxies, but we include the five galaxies 
with IGM/ICM but no clearly distinguishable ISM 
(3C~338, 3C~388, 3C~445, 3C~452, and Cyg~A).  We thus
have 5 normal FR~I galaxies, 7 normal FR~II galaxies, and 5 XRGs.  Of the 
XRGs, the IGM/ICM is the relevant medium for NGC~326 and PKS~1422+26,
and may be the relevant medium in 3C~315, 4C~+00.58 and B2~1040+31A.  Although 
the sample sizes are smaller, the $\Delta$PA distributions for the FR~I and II
galaxies are consistent with uniform ($P = 0.3$ and $0.4$ respectively; Fig.~8). 
The XRG sample, on the other hand, is not consistent with a uniform distribution
($P = 0.02$ for both the primary and secondary lobes).  Our values are found
in Table \ref{igmeccentricity}. 
 
In 3C~452, the X-ray emission traces the radio emission so
well that the geometry of the extended emission probably represents the radio
galaxy cocoon or a bounding shock 
and not the actual IGM (even when a purely ``thermal'' image
is reconstructed; Appendix A.1).  Similarly, the radio lobes of 3C~388 may be
responsible for the geometry of the surrounding medium and the inner
isophotes of the surrounding ICM are elongated perpendicular to the jet.  
If we exclude 3C~452 and use a smaller aperture in 3C~388 as the ``local'' 
ICM, our results for the IGM/ICM comparison outlined above are unchanged (the 
normal FR~II galaxies still look uniformly distributed). 
\end{enumerate}

We cannot easily compare a mix of ISM and IGM/ICM relevant media in the XRG
sample to the comparison sample.  However, we can do this for the XRG sample
because we have the additional spatial information of the secondary lobes. 
When we use the IGM values for NGC~326 and PKS~1422+26, the distribution of
$\Delta$PA obeys the C02 correlation even more strongly than for the ISM or
IGM alone.  The probability that the $\Delta$PA distribution for this 
``best guess'' sample is $P = 0.006$ for the primary lobes and $P = 0.001$ for
the secondary ones.  Regardless, the samples individually are distinguishable
from normal radio galaxies by their jet---medium geometry. 

Does this geometry necessarily implicate the overpressured cocoon model?  
NGC~326, 3C~433, and 4C~+00.58 obey the C02 geometry but would be difficult
to form in the overpressured cocoon model alone.  NGC~326,
for example, is a well studied XRG which has longer secondary lobes than
primary ones.  This poses a problem for the overpressured cocoon model, since
we expect the jets to expand supersonically and the wings to be (at most)
transonic.  The long secondary lobes may therefore imply buoyant evolution
of the backflow \citep{worrall95}, although subsonic expansion of the primary
lobes is possible.  3C~433 is an especially odd case since the southern
radio lobes are of qualitatively different character than the northern
counterparts.  \citet{miller09} argue that the difference between the northern
(FR~I) and southern (FR~II) lobes is due to propagating into a very asymmetric
medium.  The secondary lobes are close to the ISM, so it is ambiguous which 
medium is relevant for XRG formation.  Lastly, 4C~+00.58 appears to
violate the C02 geometry as the jet appears to come out of the minor axis of
the ISM.  4C~+00.58 almost certainly disagrees with the C02 geometry in the
optical image (due to the high $\epsilon$ the major axis is likely to be close
to the plane of the sky) and we detect an X-ray jet cospatial with the
northern radio jet.  However, it is unclear that the extended X-ray emission
comes entirely from the ISM.  Deeper followup observations are required to 
assess the role of the IGM in this source and establish whether it is 
actually a counterexample to the C02 geometry.  3C~192 is an ambiguous case
because although it obeys the C02 geometry, the eccentricity of the ISM is
small.  For 3C~192 to be produced in the overpressured cocoon model, small
differences in pressure gradients must be important.  

\section{Properties of the Central Engine}

In the backflow models, FR~II morphology is thought to be necessary to drive
strong backflows.  Although the backflows begin at the hot spots, 
edge-brightened FR~II morphology is strongly associated with absorbed 
power law AGN emission.  We note that several galaxies in our XRG sample
are not unambiguously FR type II (some of the galaxies with poor data are
FR~I), so we attempt to characterize the XRG sample in terms of absorbed or
unabsorbed AGN spectra.  Highly absorbed spectra tend to have higher $L_X$ for
a given flux 
than unabsorbed spectra owing to the large amount of ``absorption'' blocking
the low-energy photons, whereas unabsorbed spectra tend to be lower energy
with small X-ray luminosities above $\sim 2$~keV.  

Our model fitting of the AGN spectra started by fitting a single
power law (either absorbed or unabsorbed depending on the appearance of
the spectrum).  Additional complexity was added as the degrees of freedom
allowed until an acceptable fit was achieved.  We note that $N_H$ and 
$\Gamma$ are often degenerate and require one of the two to be frozen (to 
values ranging from $\Gamma = 1.0 - 2.0$; Table \ref{agnspecparams}).  All
the fits also incorporated Galactic $N_H$.  Our
model parameters and luminosities are given in Table \ref{agnspecparams}
and are largely in accordance with previous studies of the 3CRR 
{\em Chandra} sample \citep{evans06,balmaverde06}.  These papers have also
conducted a more exhaustive study of the AGN X-ray emission, whereas our
purpose is to determine whether this emission lends insight to the XRG
sample.  

In fact, the XRG sample cannot neatly be classified
in terms of asorbed or unabsorbed spectra (Fig.~9).  This is in agreement
with other evidence \citep[e.g. optical data from][]{cheung09}, although 
the highly absorbed XRGs do not tend to have a smaller $L_X$ than the
normal FR~II galaxies.  Indeed, the XRG sample, while small, is consistent
with the distribution of $L_X$ between $10^{40}$ and $10^{44}$ erg s$^{-1}$
for the full comparison sample, and not with either the normal FR~I or II
galaxies by themselves. 
We note that we detect X-ray jets in some of our systems, but we find
no evidence for misaligned jets in any of the XRGs.  The sample size is too
small to rule out the \citet{lal07} model on this basis, but we might expect
to see misaligned jets in their scheme. 

\section{Discussion \& Summary}

We present new and archival {\em Chandra} data of XRGs alongside
normal FR~I and II galaxies within $z \leq 0.1$.  We extend the
\citet{capetti02} and \citet{saripalli09} geometric correlation
between the orientation of the secondary lobes in XRGs and the 
minor axis of the interacting medium to the X-ray band, including
the intragroup/intracluster medium.  We find that this geometry
strongly distinguishes the XRGs from normal radio galaxies
(although we are unable to strongly distinguish between XRGs and normal 
radio galaxies using only the local IGM/ICM values), and that
XRGs may be produced in galaxies with both absorbed and unabsorbed
AGN.  Our results in the X-ray band are consistent with \citet{cheung09}
who find that XRGs lie close to the FR~I/II division in a comparison
of radio---optical luminosity.  We note that in properties of the hot
atmospheres (temperature, density, and pressure), the XRG sample appears
indistinguishable from the comparison sample (Table \ref{thermalspecparams})
but that these parameters are spatial averages of profiles and may not be
directly comparable.  

The remarkable geometric distinction between the normal and $\mathsf{X}$-shaped
radio galaxies may be an important clue to the genesis of the secondary
lobes.  We cannot rigorously examine the formation models in our data, but
identify potential problems for the backflow models in several galaxies.  The
backflow schemes naturally explain the observed geometry, but also have yet
to explain other features of XRGs in detailed modeling \citep[e.g. the flat
spectral indices of the secondary lobes][]{lal07}.  However, the models are
presently immature: the C02 overpressured cocoon model has only been modeled
in a highly elliptical 2D ISM, whereas the \citet{kraft05} suggestion that
the buoyant backflow mechanism can produce $\mathsf{X}$-shaped wings in the
presence of an anisotropic buoyant medium has not been modeled.  3D modeling
and deeper observations of the relevant media in XRGs are required to make a
convincing case for either of these models.  Notably, the
rapid reorientation models have yet to present a reason for the C02 
correlation.  Any successful model must account
for it.  Lastly, deeper observations of the XRGs in our sample for which no
diffuse emission was detected are required.  The total sample of nearby XRGs
is quite small, so every member is important. 

\acknowledgments

C.~S.~R. and E.~H.-K. thank the support of the {\em Chandra} Guest Observer
program under grant GO89109X.  E.~H.-K. thanks A.~I.~Harris for useful
discussion of the ellipse-fitting method.  C.~C.~C. was supported by an
appointment to the NASA Postdoctoral Program at Goddard Space Flight Center, 
administered by Oak Ridge Associated Universities through a contract with NASA.
This research has made use of the NASA/IPAC Extragalactic Database (NED)
which is operated by the Jet Propulsion Laboratory, California Institute of
Technology, under contract with the National Aeronautics and Space 
Administration.  
The Digitized Sky Surveys were produced at the Space Telescope Science 
Institute under U.S. Government grant NAG W-2166. The images of these 
surveys are based on photographic data obtained using the Oschin Schmidt 
Telescope on Palomar Mountain and the UK Schmidt Telescope. The plates were 
processed into the present compressed digital form with the permission of 
these institutions.
 
{\it Facilities:} \facility{Chandra}, \facility{VLA} 

\clearpage

\singlespace


\clearpage


\begin{figure}
\begin{center}
\includegraphics[scale=1.0,angle=0]{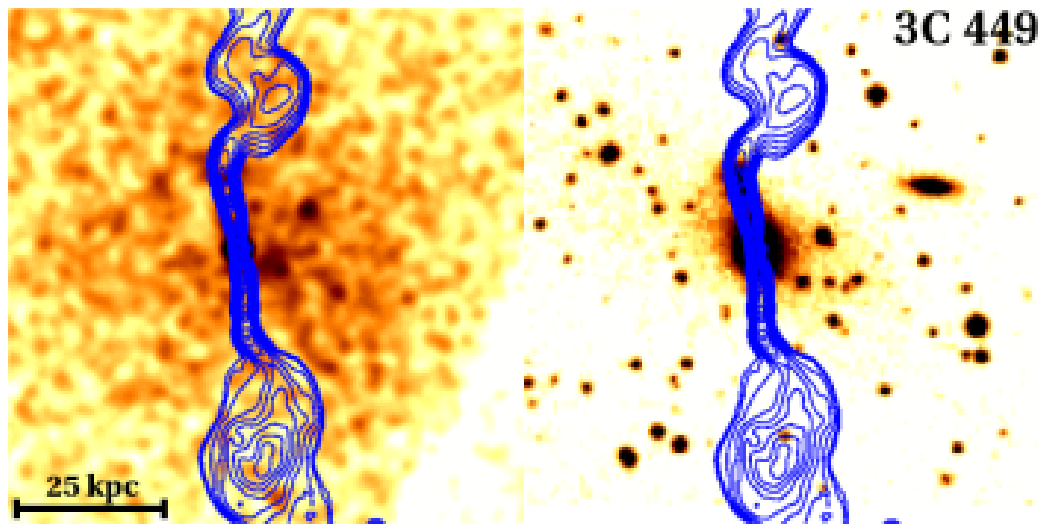}
\includegraphics[scale=1.0,angle=0]{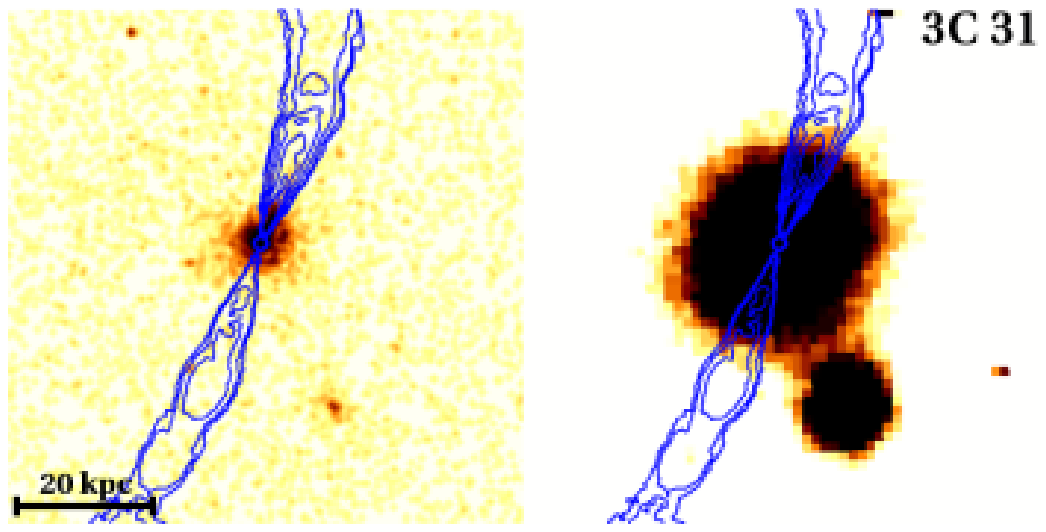}
\includegraphics[scale=1.0,angle=0]{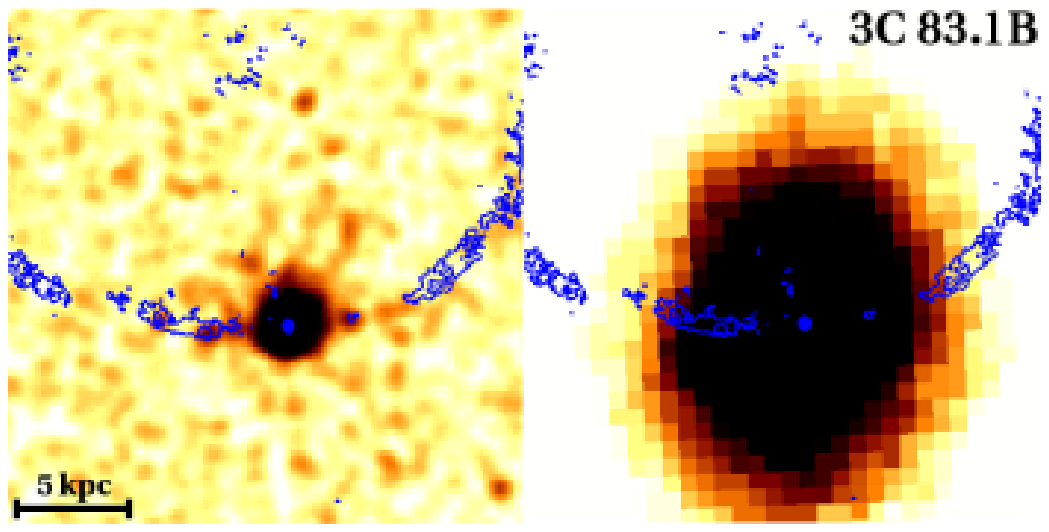}
\caption[]{}
\label{cs_images}
\end{center}
\end{figure}
\begin{figure}
\begin{center}
\ContinuedFloat
\includegraphics[scale=1.0,angle=0]{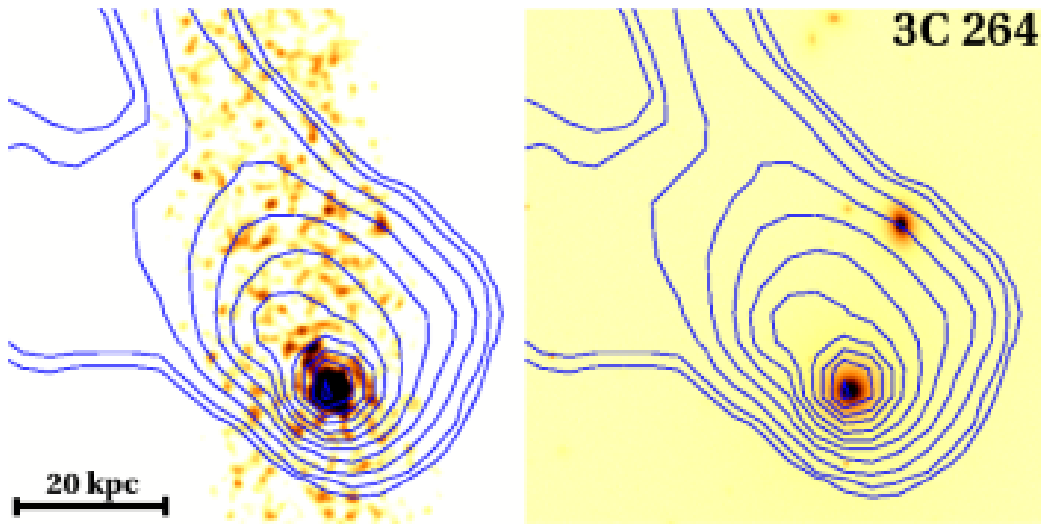}
\includegraphics[scale=1.0,angle=0]{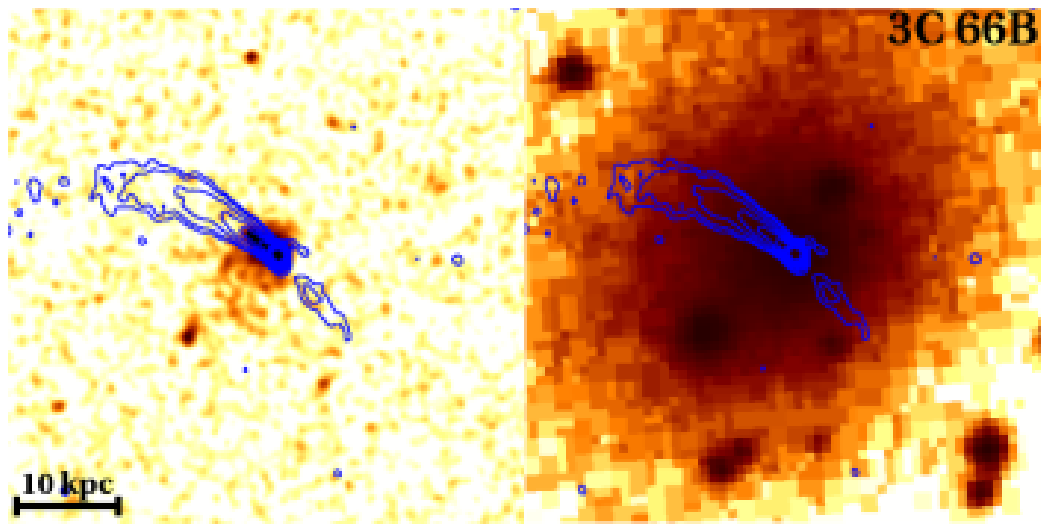}
\includegraphics[scale=1.0,angle=0]{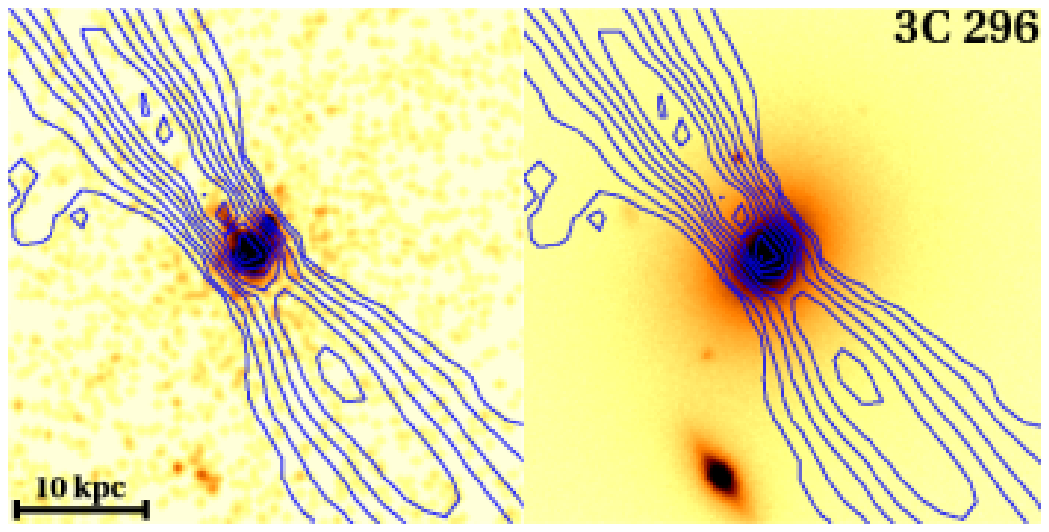}
\caption[]{}
\end{center}
\end{figure}
\begin{figure}
\begin{center}
\ContinuedFloat
\includegraphics[scale=1.0,angle=0]{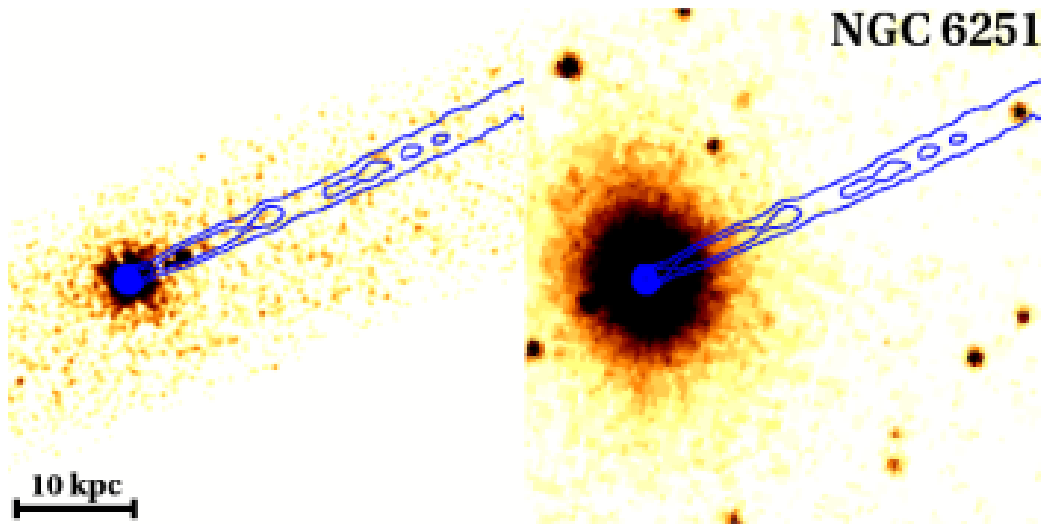}
\includegraphics[scale=1.0,angle=0]{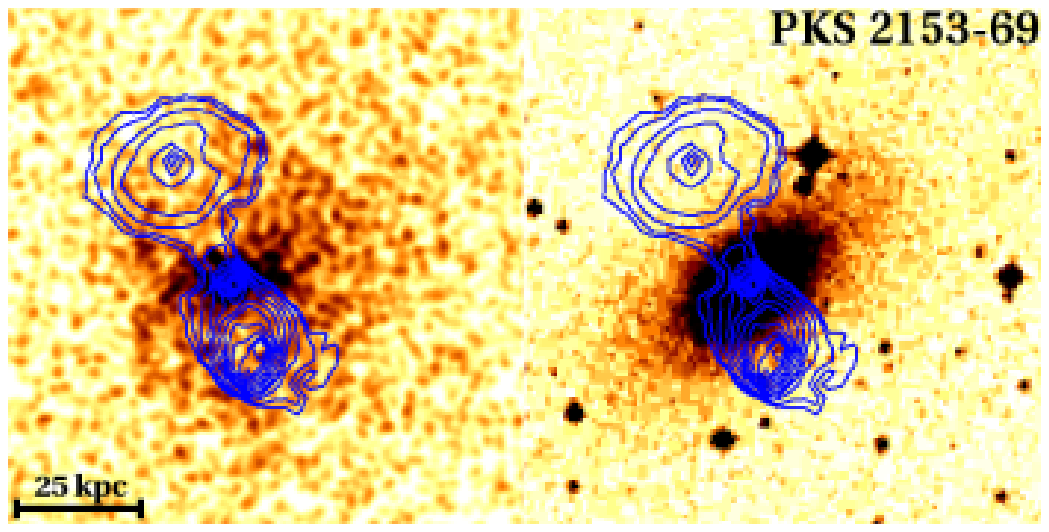}
\includegraphics[scale=1.0,angle=0]{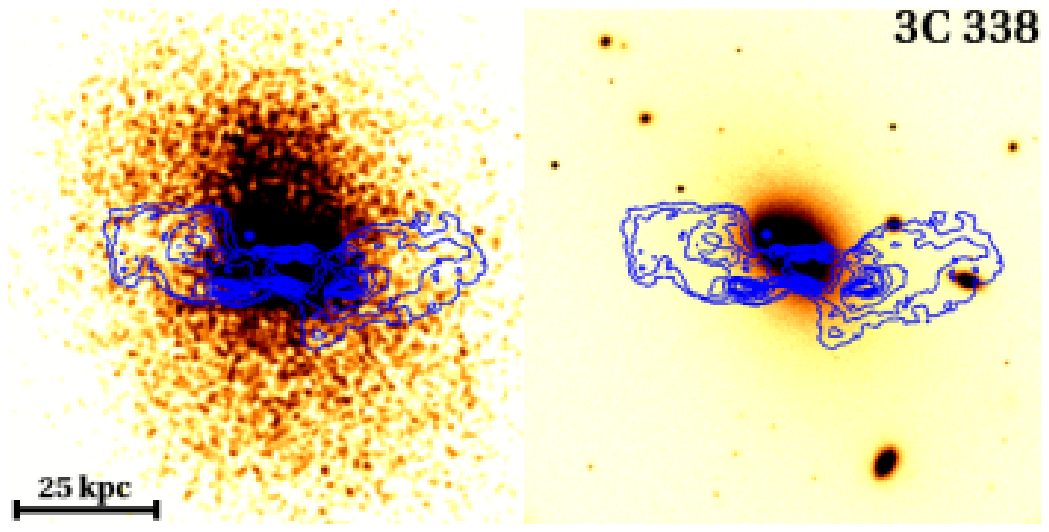}
\caption[]{}
\end{center}
\end{figure}
\begin{figure}
\begin{center}
\ContinuedFloat
\includegraphics[scale=1.0,angle=0]{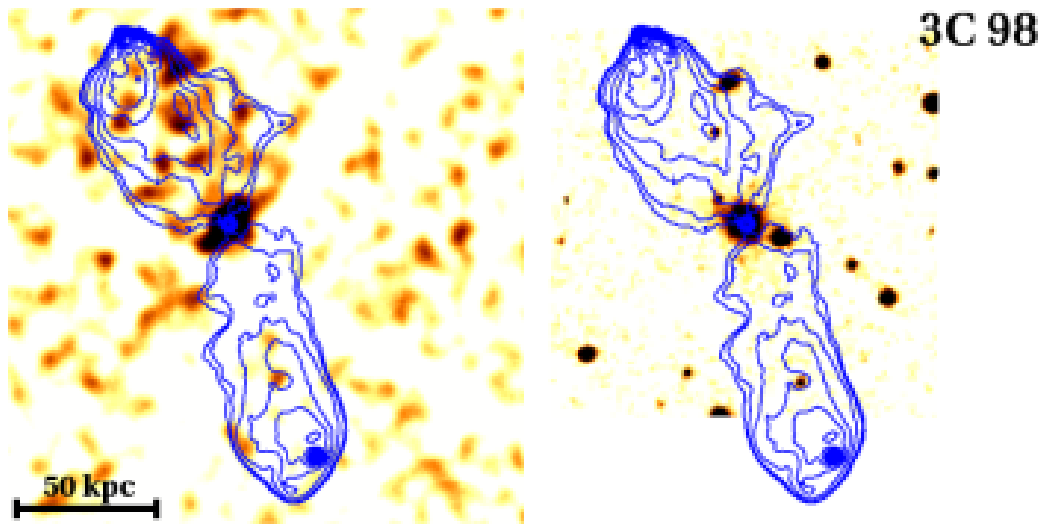}
\includegraphics[scale=1.0,angle=0]{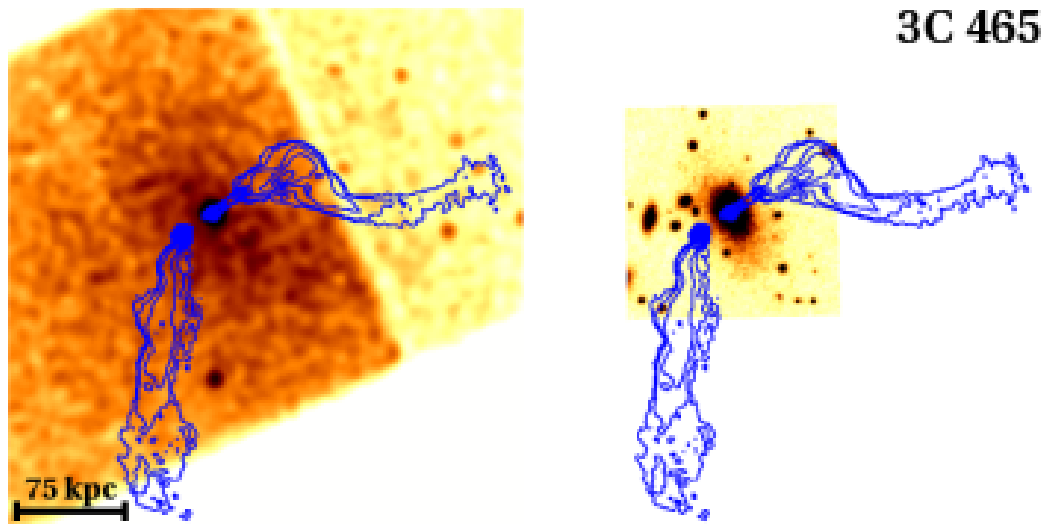}
\includegraphics[scale=1.0,angle=0]{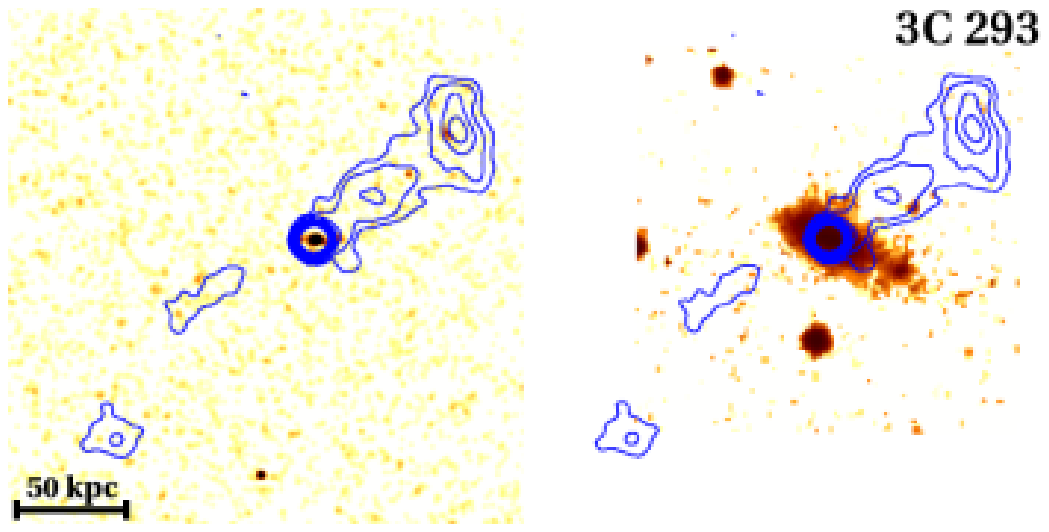}
\caption[]{}
\end{center}
\end{figure}
\begin{figure}
\begin{center}
\ContinuedFloat
\includegraphics[scale=1.0,angle=0]{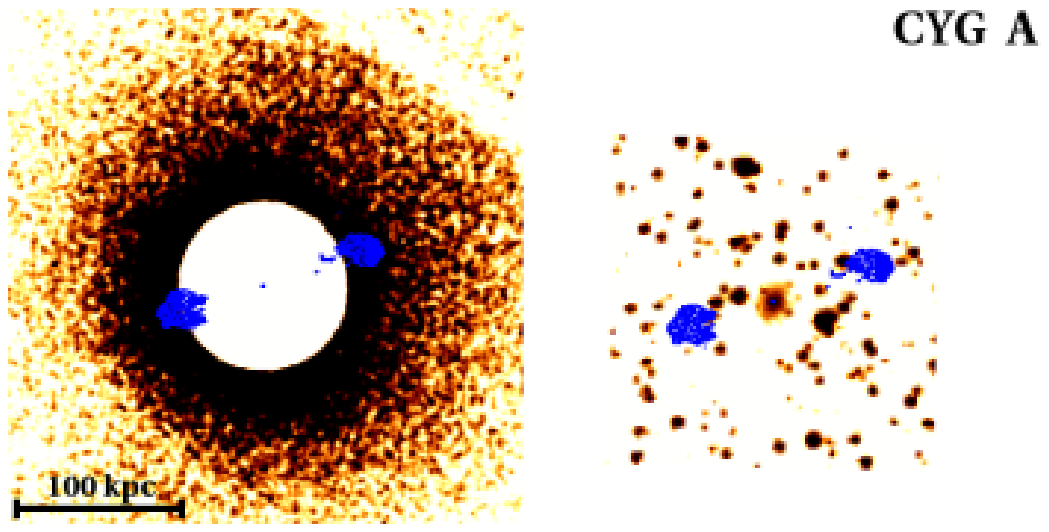}
\includegraphics[scale=1.0,angle=0]{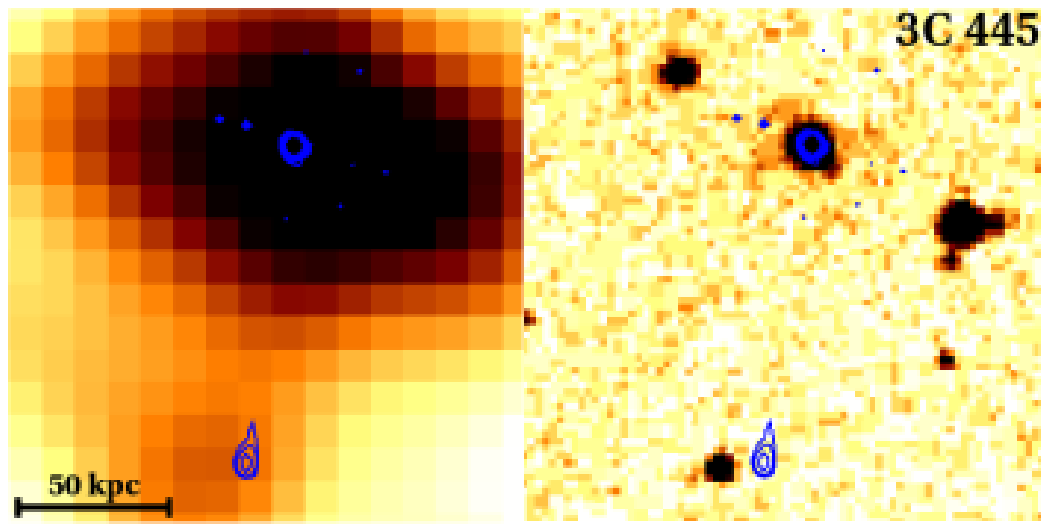}
\includegraphics[scale=1.0,angle=0]{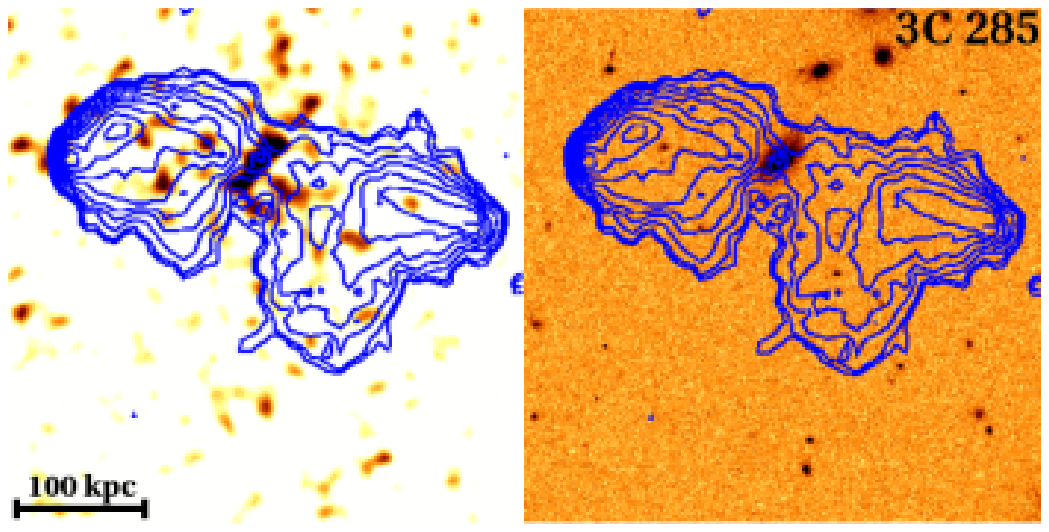}
\caption[]{}
\end{center}
\end{figure}
\begin{figure}
\begin{center}
\ContinuedFloat
\includegraphics[scale=1.0,angle=0]{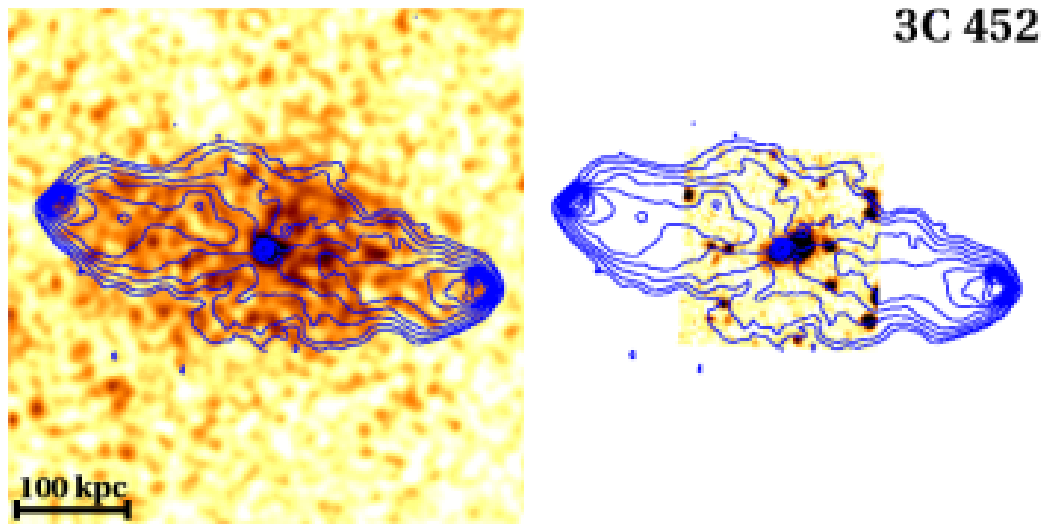}
\includegraphics[scale=1.0,angle=0]{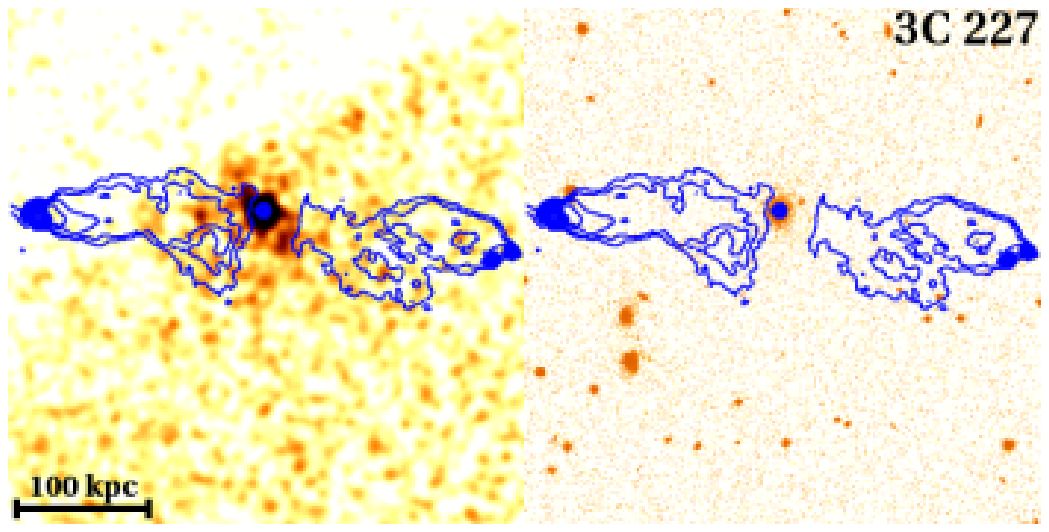}
\includegraphics[scale=1.0,angle=0]{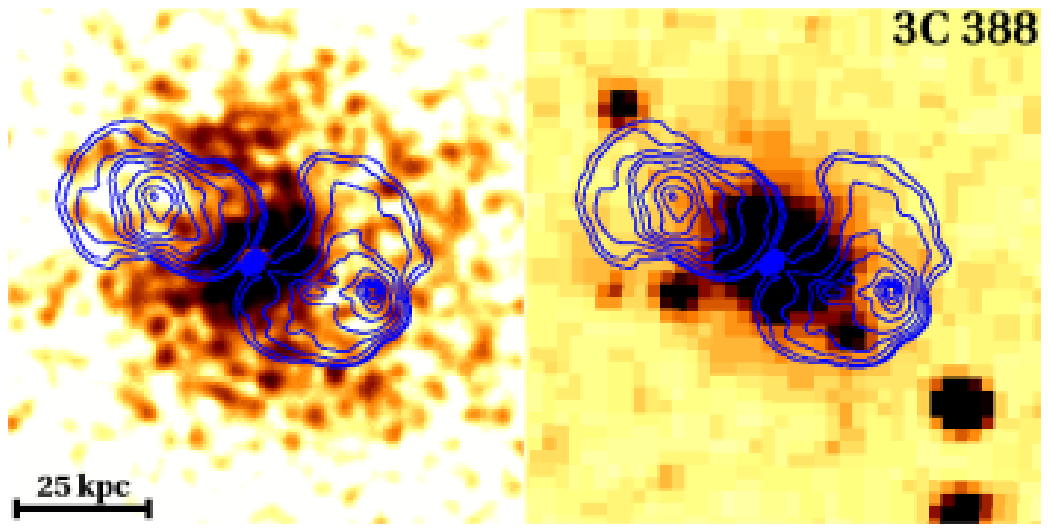}
\caption{{\em Chandra} and optical images of our control
  sample galaxies with significant thermal emission.  The left panel in each
  case is an image of the {\em Chandra} ACIS-S3 chip with uniform
  Gaussian smoothing and point sources removed.  The right image is an
  optical image from DSS or SDSS.  Overlaid in blue are VLA radio map contours.
  The images are sorted by ascending redshift.  Note that for 3C~98 and 
  3C~388 the chip image is of the ACIS-I chip(s) instead of ACIS-S3.
  This figure is available in the online edition of the journal.}
\end{center}
\end{figure}

\clearpage

\begin{figure}
\begin{center}
\label{xrg_images}
\includegraphics[scale=1.0,angle=0]{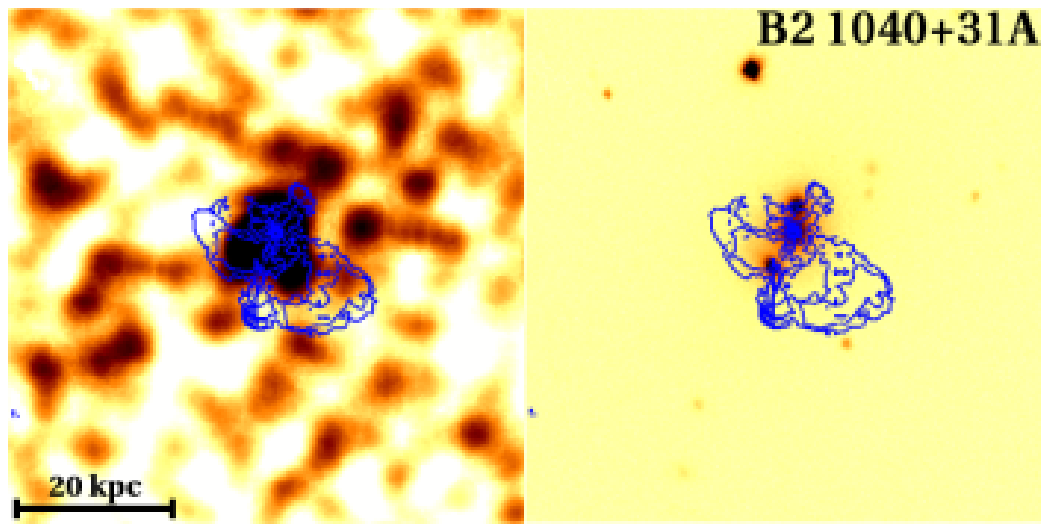}
\includegraphics[scale=1.0,angle=0]{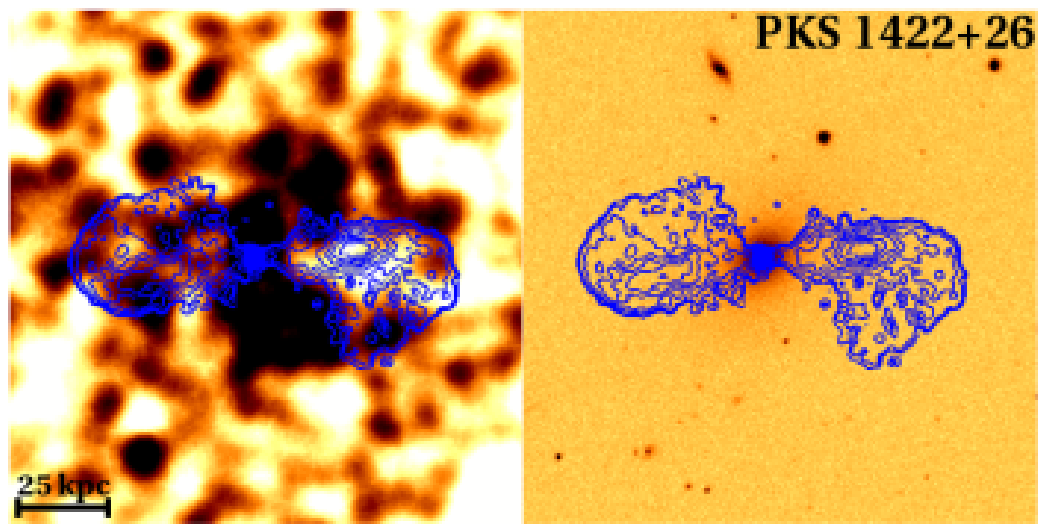}
\includegraphics[scale=1.0,angle=0]{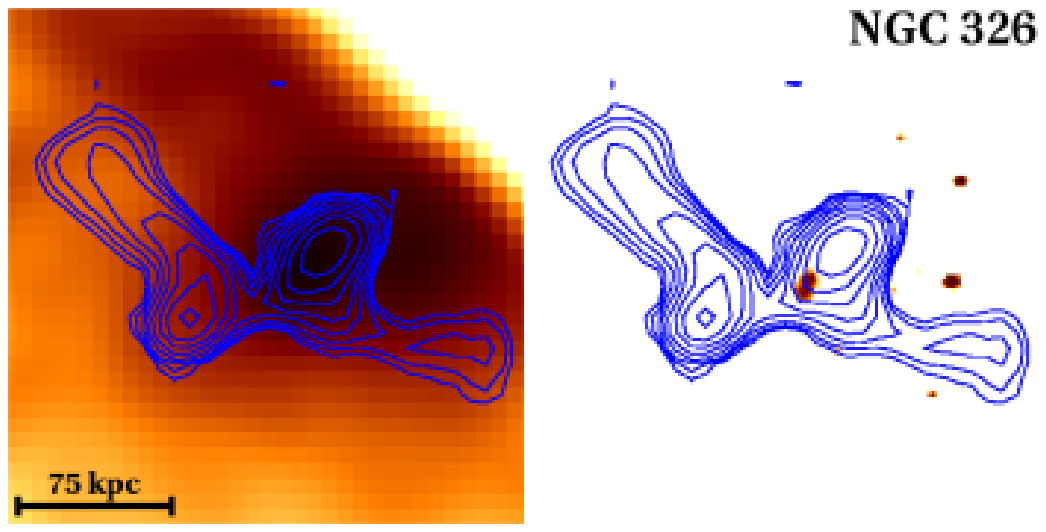}
\caption[]{}
\end{center}
\end{figure}
\begin{figure}
\begin{center}
\ContinuedFloat
\includegraphics[scale=1.0,angle=0]{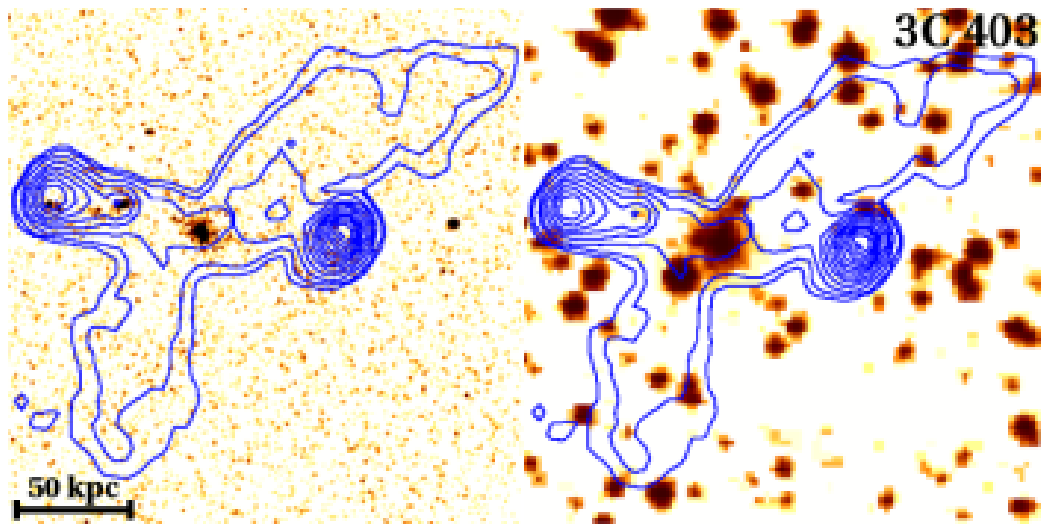}
\includegraphics[scale=1.0,angle=0]{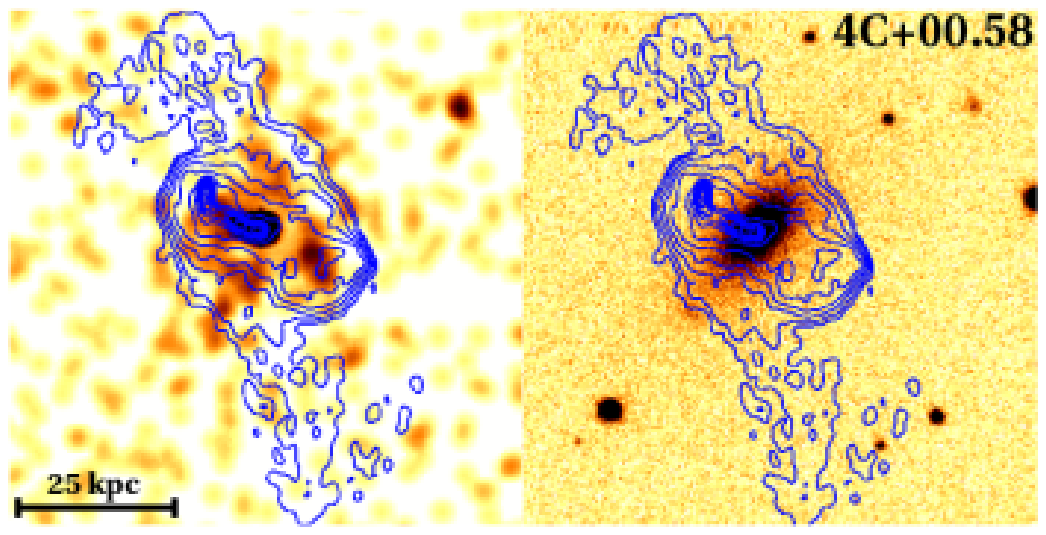}
\includegraphics[scale=1.0,angle=0]{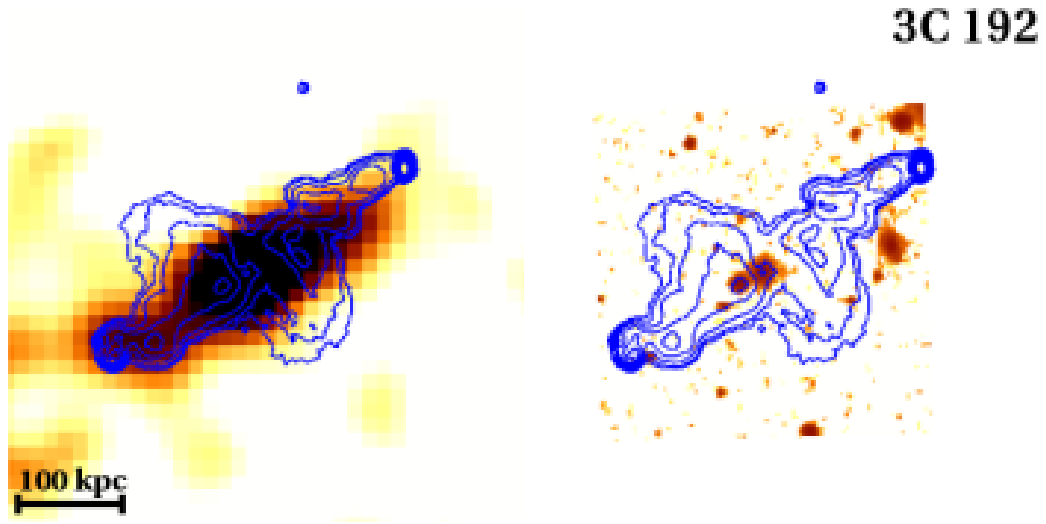}
\caption[]{}
\end{center}
\end{figure}
\begin{figure}
\begin{center}
\ContinuedFloat
\includegraphics[scale=1.0,angle=0]{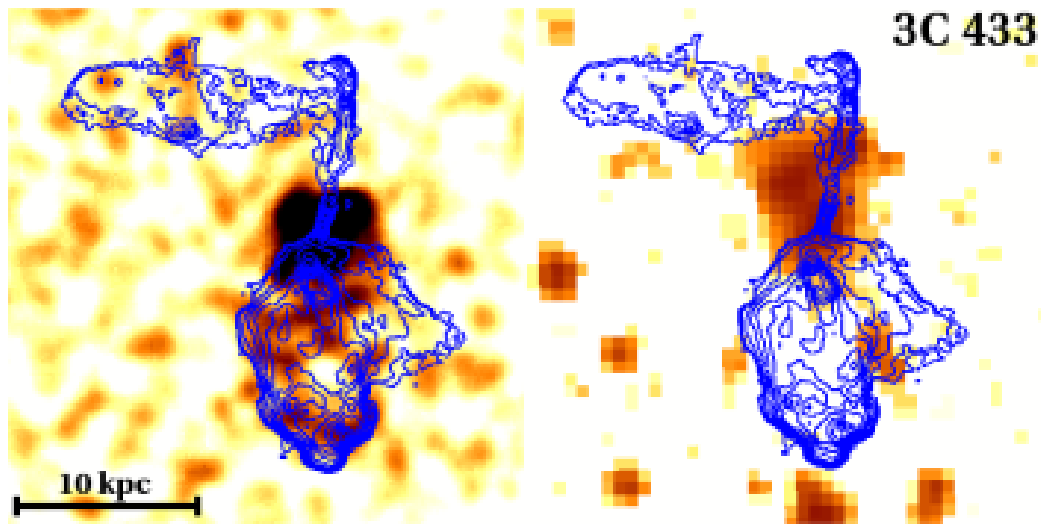}
\includegraphics[scale=1.0,angle=0]{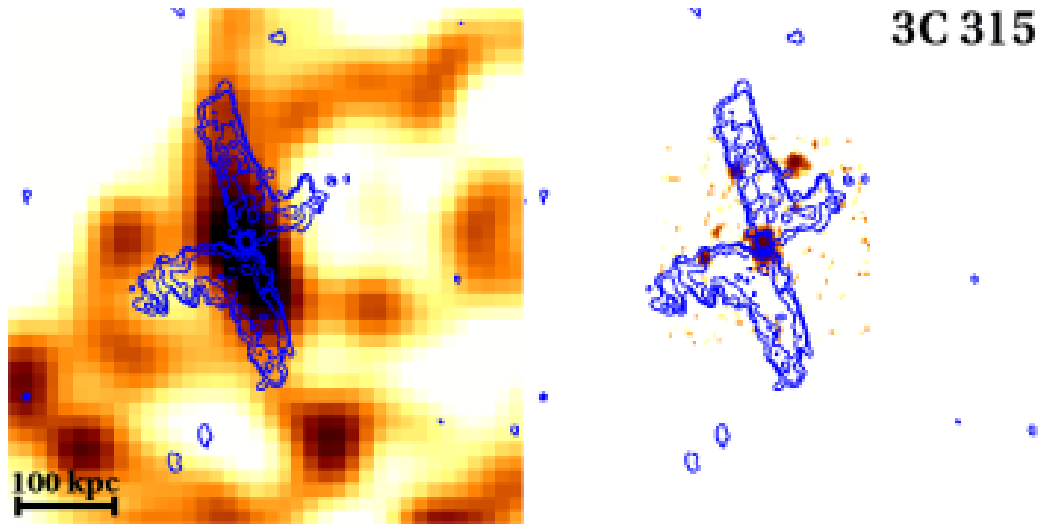}
\caption{{\em Chandra} and optical images of our XRG
  sample galaxies with significant thermal emission.  The left panel in each
  case is an image of the {\em Chandra} ACIS-S3 chip with uniform
  Gaussian smoothing and point sources removed.  The right image is an
  optical image from DSS or SDSS.  Overlaid in blue are VLA radio map contours.
  The images are sorted by ascending redshift.}
\end{center}
\end{figure}

\clearpage
\begin{figure}
\begin{center}
\includegraphics[scale=0.8,angle=0]{ism_spectra.ps_page_1}
\label{ism_spectra}
\caption[]{}
\end{center}
\end{figure}
\begin{figure}
\begin{center}
\ContinuedFloat
\includegraphics[scale=0.8,angle=0]{ism_spectra.ps_page_2}
\caption[]{}
\end{center}
\end{figure}
\begin{figure}
\begin{center}
\ContinuedFloat
\includegraphics[scale=0.8,angle=0]{ism_spectra.ps_page_3}
\caption[]{Spectra of the X-ray ISM we use in our morphology comparison
(Table \ref{eccentricitytable}).  The XRG sample is shown with galaxy
names in green and the total model fit has been plotted over the data in 
red.  The values can be found in Table \ref{thermalspecparams}.  Note that
3C~403 is shown in Fig.~5 since the ISM is small compared to the PSF and
is isolated via energy filtering \citep{kraft05}.  This image is
available in the online edition of the Journal}
\end{center}
\end{figure}

\clearpage
\begin{figure}
\begin{center}
\includegraphics[scale=0.8,angle=0]{igm_spectra.ps_page_1}
\label{igm_spectra}
\caption[]{}
\end{center}
\end{figure}
\begin{figure}
\begin{center}
\ContinuedFloat
\includegraphics[scale=0.8,angle=0]{igm_spectra.ps_page_2}
\caption[]{}
\end{center}
\end{figure}
\begin{figure}
\begin{center}
\ContinuedFloat
\includegraphics[scale=0.8,angle=0]{igm_spectra.ps_page_3}
\caption[]{Spectra of the X-ray IGM we use in our morphology comparison
(Table \ref{igmeccentricity}).  The XRG sample is shown with galaxy
names in green and the total model fit has been plotted over the data in 
red.  The values can be found in Table \ref{thermalspecparams}. This
figure is available in the online edition of the Journal. }
\end{center}
\end{figure}

\clearpage
\begin{figure}
\begin{center}
\includegraphics[scale=0.8,angle=0]{agn_spectra.ps_page_1}
\label{agn_spectra}
\caption[]{}
\end{center}
\end{figure}
\begin{figure}
\begin{center}
\ContinuedFloat
\includegraphics[scale=0.8,angle=0]{agn_spectra.ps_page_2}
\caption[]{}
\end{center}
\end{figure}
\begin{figure}
\begin{center}
\ContinuedFloat
\includegraphics[scale=0.8,angle=0]{agn_spectra.ps_page_3}
\caption[]{}
\end{center}
\end{figure}
\begin{figure}
\begin{center}
\ContinuedFloat
\includegraphics[scale=0.8,angle=0]{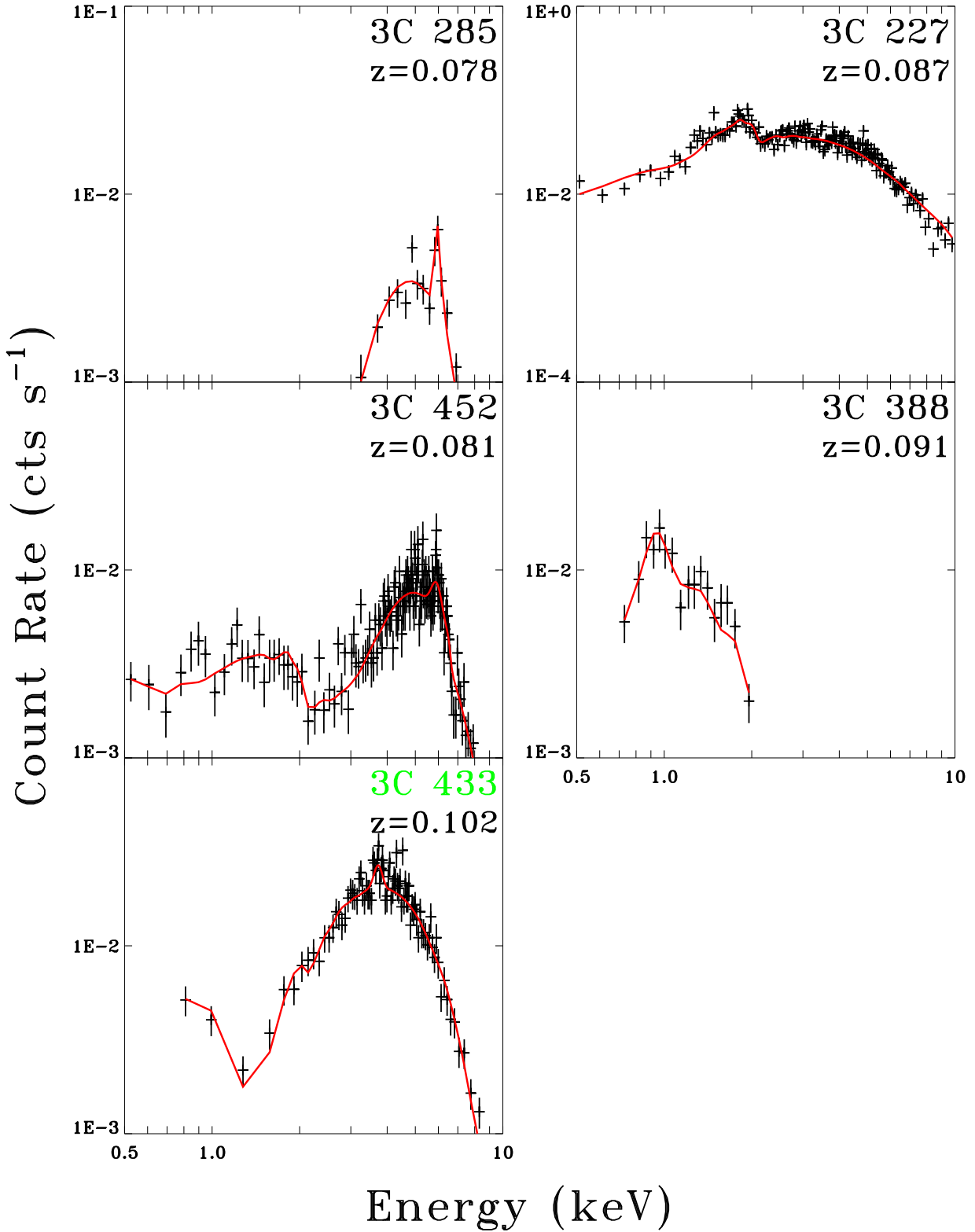}
\caption[]{Spectra extracted from the central PSF of the
host galaxy; emission may come from either the parsec-scale
jet or accretion disk region.  The total model fit has been
plotted over the data in red and XRG galaxies have their names
in green.  The spectra are essentially broken into absorbed
and unabsorbed spectra; some of the unabsorbed spectra require
thermal components.  Note that 3C~338 has no detected AGN 
emission and 3C~403 has a small ISM represented in the
spectrum of the area near the AGN.  This figure is available in the
online edition of the Journal.}
\end{center}
\end{figure}

\clearpage
\begin{figure}
\begin{center}
\includegraphics[scale=0.5,angle=90]{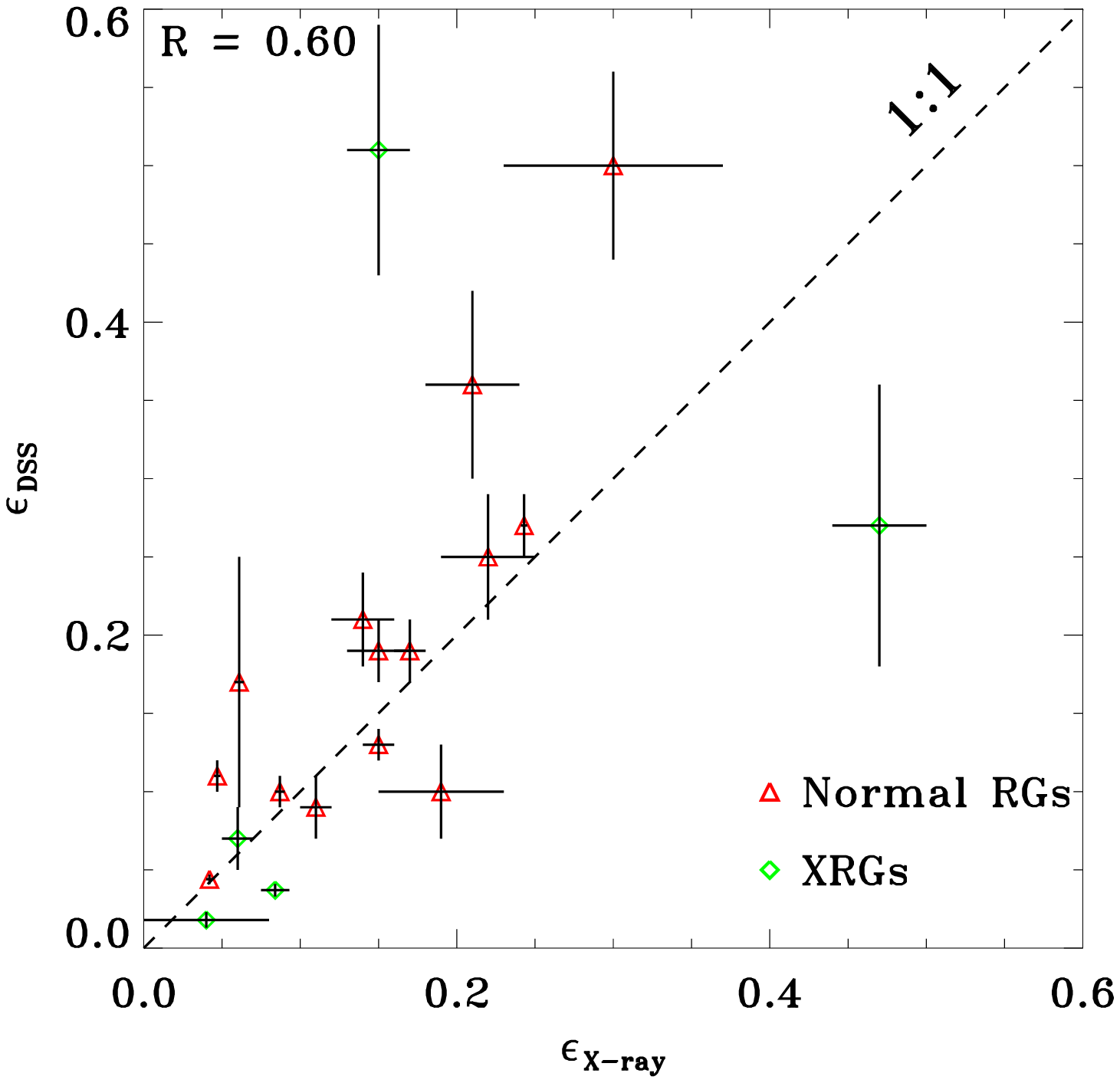}
\includegraphics[scale=0.5,angle=90]{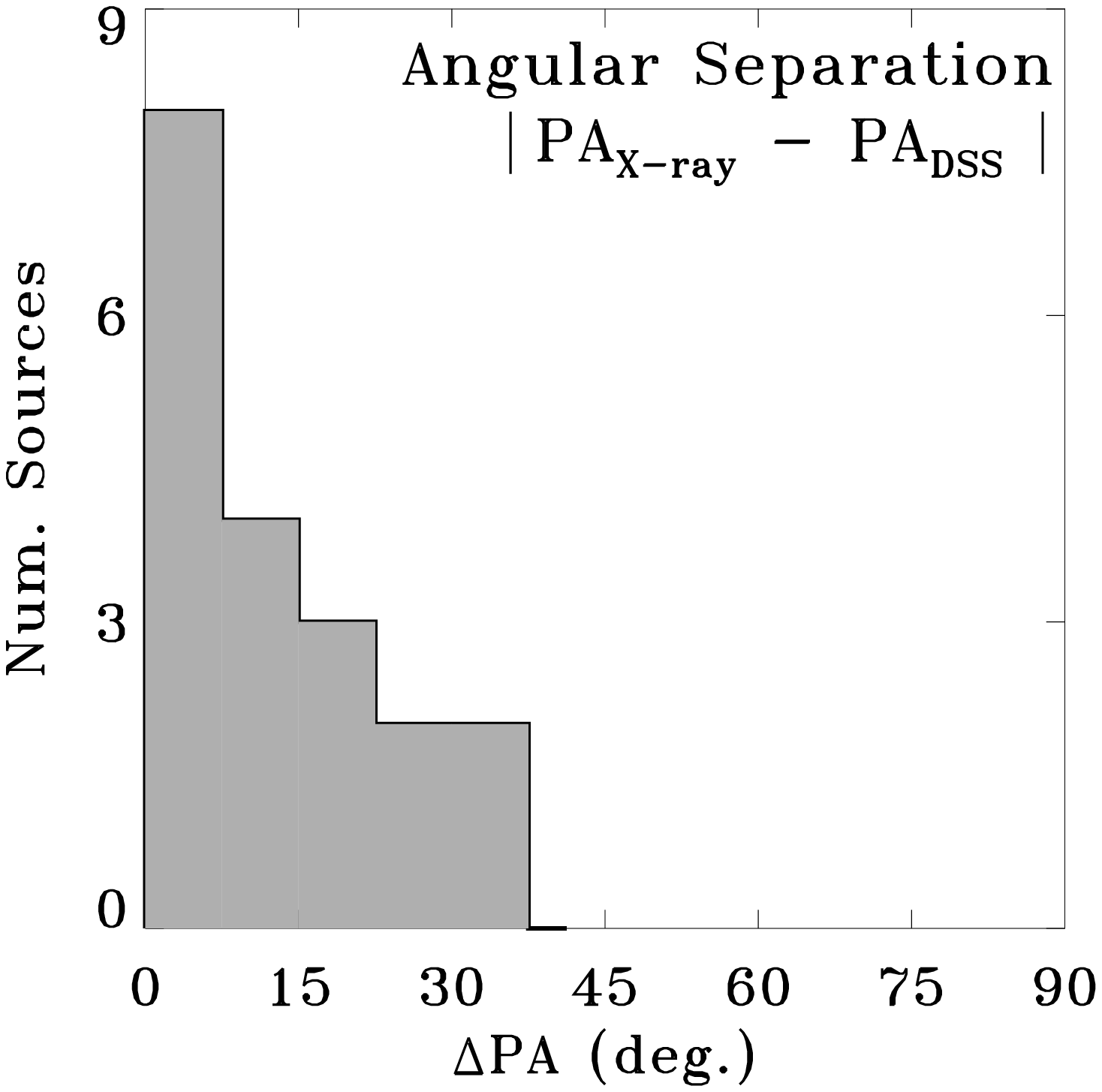}
\caption{{\sc Left:} A comparison of the eccentricities of the
diffuse ISM to the optical light where $\epsilon$ in each case
has been generated by our ellipse-fitting method (\S3.1).
The dashed line represents a 1:1 correlation rather than a best-fit
line. 
{\sc Right:} Comparison of the position angles of the best-fit
ellipses for the diffuse ISM to those of the optical light where 
the values are determined by our ellipse-fitting method (\S3.1).}
\label{eccentricity}
\end{center}
\end{figure}

\clearpage

\begin{figure}
\begin{center}
\includegraphics[scale=0.6,angle=0]{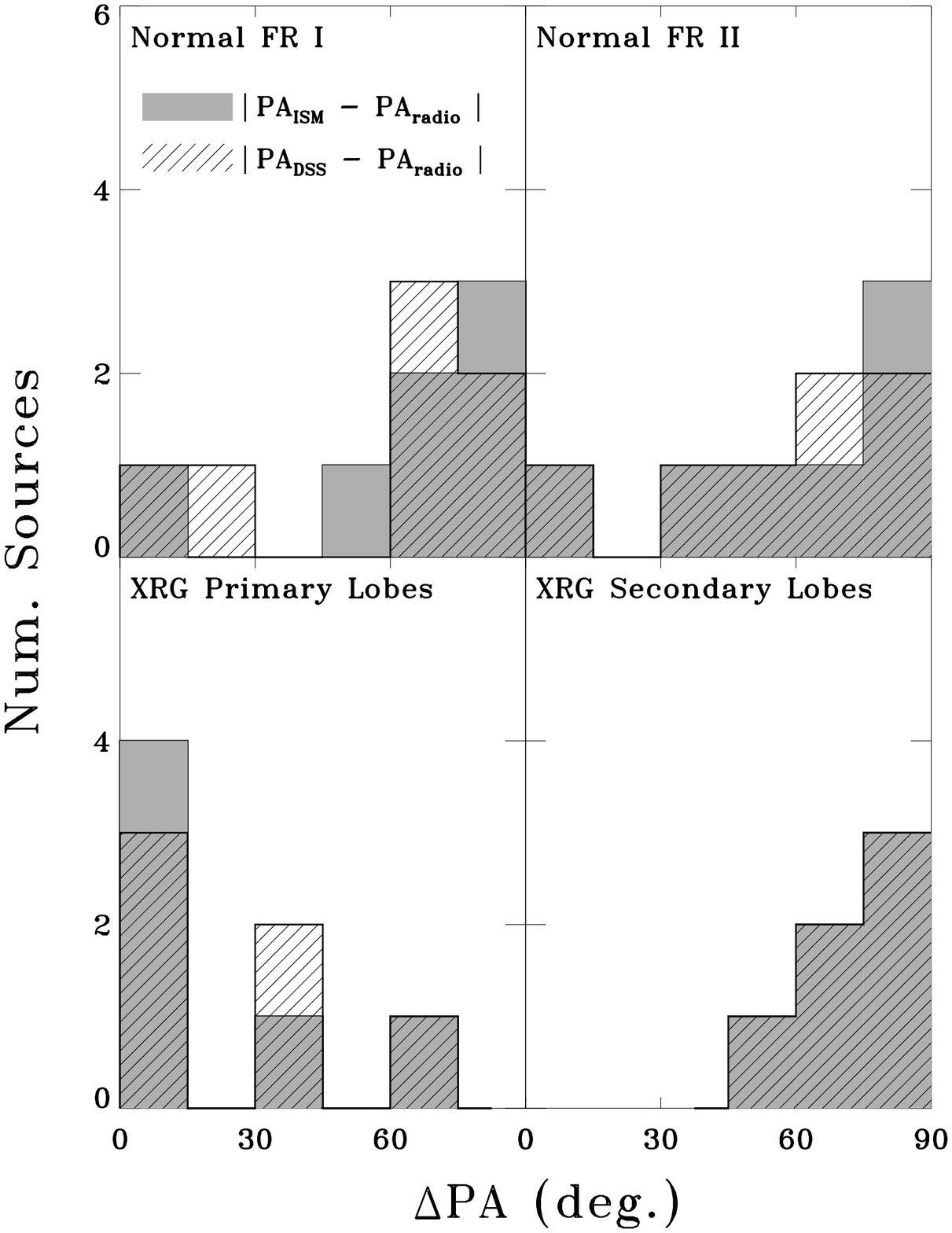}
\caption{Comparison of the alignment of the radio jets with 
the interstellar medium (filled) and
the optical light (hashed).  $\Delta$PA is the acute angle between
the major axes of the best-fit ellipses (\S3.1) and the orientation
of the radio jets.  $\Delta$PA $\sim 0^{\circ}$ indicates alignment of the {\em major}
axis of the medium with the radio lobes, whereas $\Delta$PA $\sim 90^{\circ}$ 
indicates alignment with the {\em minor} axis.  Note not all galaxies are
represented here.}
\label{radio_alignment_ism}
\end{center}
\end{figure}

\clearpage

\begin{figure}
\begin{center}
\includegraphics[scale=0.6,angle=0]{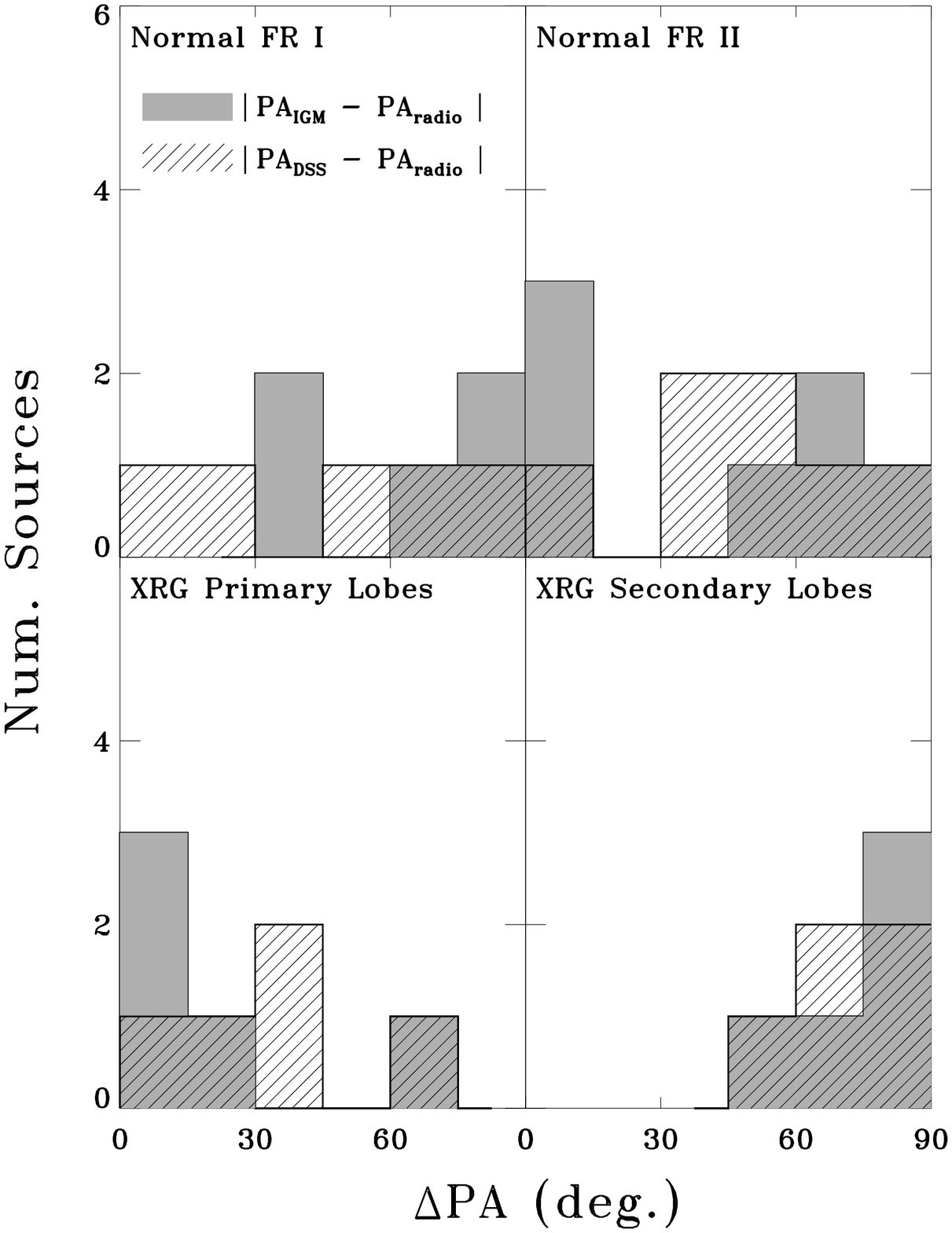}
\caption{Comparison of the alignment of the radio jets with 
the local IGM (filled, see text) and 
the optical light (hashed).  $\Delta$PA is the acute angle between
the major axes of the best-fit ellipses (\S3.1) and the orientation
of the radio jets.  $\Delta$PA $\sim 0^{\circ}$ indicates alignment of the {\em major}
axis of the medium with the radio lobes, whereas $\Delta$PA $\sim 90^{\circ}$ 
indicates alignment with the {\em minor} axis.  Not all galaxies are
represented here.}
\label{radio_alignment_igm}
\end{center}
\end{figure}

\clearpage

\begin{figure}
\begin{center}
\includegraphics[scale=0.75,angle=90]{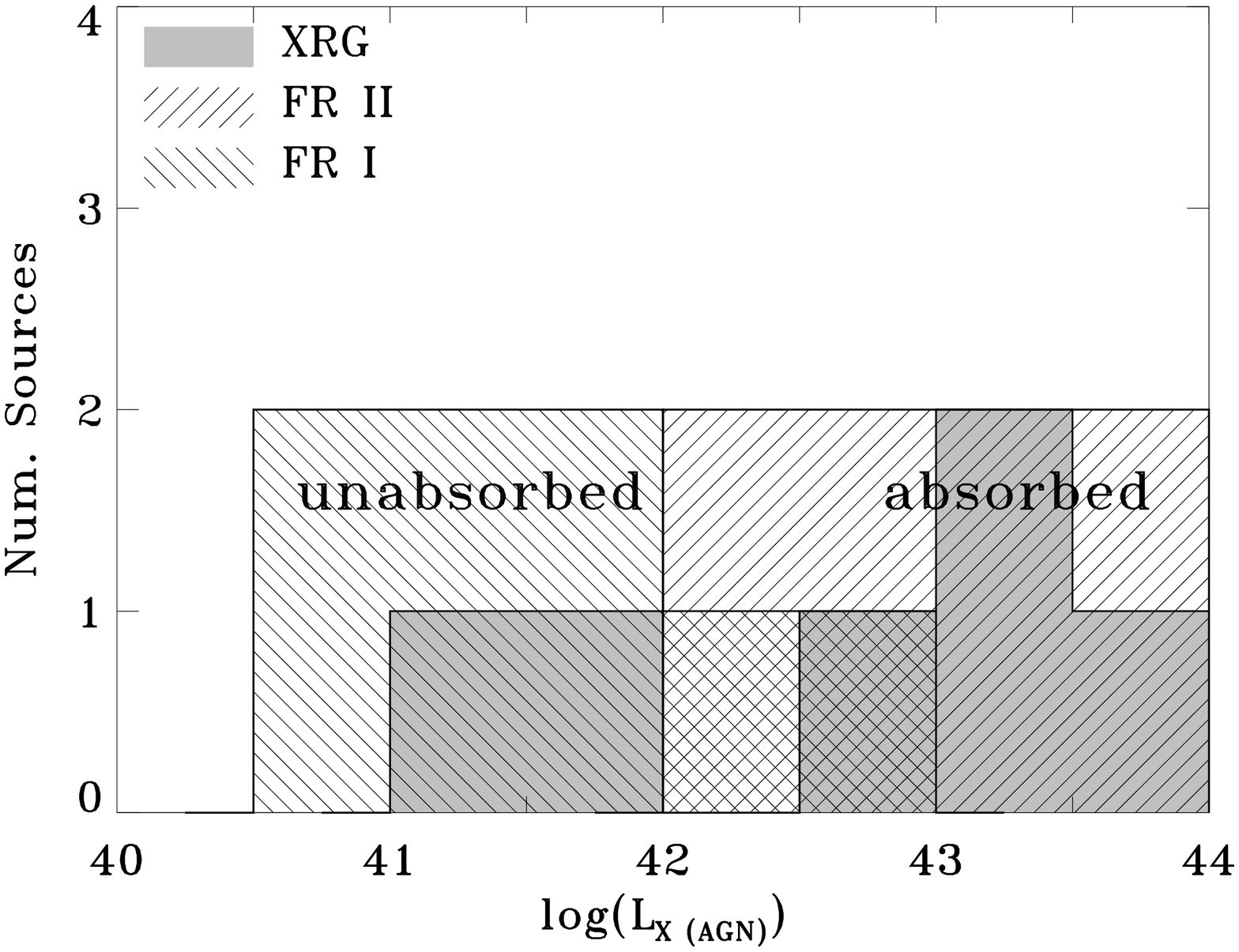}
\caption{Distribution of the XRG and comparison sample model luminosities for
spectra extracted from the central PSF.  The luminosities are calculated for
the full model fit.  FR~II galaxies tend to have highly absorbed spectra
with corresponding high luminosities, whereas FR~I galaxies tend to have
unabsorbed spectra.  XRGs have both absorbed and unabsorbed spectra.}
\label{agn_lumhist}
\end{center}
\end{figure}

\clearpage



\begin{deluxetable}{llccccclll}
\tablenum{1} 
\tabletypesize{\scriptsize} 
\tablecaption{{\em Chandra} Observational Parameters} 
\tablewidth{0pt} 
\rotate
\tablehead{ 
\colhead{Name} & \colhead{Secondary Identifier} & \colhead{$z$} & \colhead{Ang. Scale} & \colhead{obs. IDs}
& \colhead{Exp. Time} & \colhead{CCD Array} & \colhead{FR Type} & \colhead{Selected {\em Chandra} Refs.} & 
\colhead{Radio Refs.} \\ & & & (kpc/$^{\prime\prime}$) & & (ks) & & & 
}

\startdata 
\cutinhead{XRGs} 
B2 1040+31A    & J1043+3131 & 0.036 & 0.706 & 9272 & 10 & ACIS-S & I/II & this work & AM221+AM222 \\
PKS 1422+26    & J1424+2637 & 0.037 & 0.725 & 3983 & 10 & ACIS-S & I/II & this work & AM364 \\
NGC 326        & J0058+2651 & 0.048 & 0.928 & 6830 & 94 & ACIS-S & I/II & 1,2,9,10  & a \\
3C 403         & J1952+0230 & 0.059 & 1.127 & 2968 & 49 & ACIS-S & II   & 3,4,5 	& b \\
4C +00.58      & J1606+0000 & 0.059 & 1.127 & 9274 & 10 & ACIS-S & II   & this work & AC818 \\
3C 192         & J0805+2409 & 0.060 & 1.144 & 9270 & 10 & ACIS-S & II   & 6 		& c \\
3C 433         & J2123+2504 & 0.102 & 1.855 & 7881 & 38 & ACIS-S & I/II & 7 		& b \\
3C 315		   & J1513+2607 & 0.108 & 1.951 & 9313 & 10 & ACIS-S & I    & this work & AM364\\
\cutinhead{Comparison sample}
3C 449         & J2229+3921 & 0.017 & 0.341 & 4057 				  & 30 			& ACIS-S & I    & 4,5,8,9,10,26 	& AK319 \\
3C 31          & J0107+3224 & 0.017 & 0.341 & 2147 				  & 45 			& ACIS-S & I    & 4,5,9,11,26   	& a \\
3C 83.1B       & J0318+4151 & 0.018 & 0.361 & 3237 				  & 95 			& ACIS-S & I    & 4,5,12        	& AK403 \\
3C 264         & J1145+1936 & 0.021 & 0.419 & 4916 				  & 38 			& ACIS-S & I    & 4,8 		    	& d \\
3C 66B         & J0223+4259 & 0.022 & 0.439 & 828  				  & 45 			& ACIS-S & I    & 4,5,8,13      	& e \\
3C 296         & J1417+1048 & 0.024 & 0.477 & 3968 				  & 50 			& ACIS-S & I    & 4,5		    	& f \\
NGC 6251       & J1632+8232 & 0.024 & 0.477 & 4130 				  & 50 			& ACIS-S & I/II & 4,14 	 	    	& g \\
PKS 2153$-$69  & J2157$-$6941 & 0.028 & 0.554 & 1627 			  & 14 			& ACIS-S & II   & 15,16		 		& h \\
3C 338         & J1628+3933 & 0.030 & 0.593 & 497+498 			  & 20+20		& ACIS-S & I    & 4,5,8,9,14,17 	& i \\
3C 98          & J0358+1026 & 0.030 & 0.593 & 10234				  & 32 			& ACIS-I & II   & this work    	  	& j \\
3C 465         & J2338+2702 & 0.031 & 0.612 & 4816 				  & 50 			& ACIS-S & I    & 4,8,9,27         	& a \\ 
3C 293         & J1352+3126 & 0.045 & 0.873 & 9310 				  & 8 			& ACIS-S & II   & 25		 		& k \\
Cyg A		   & J1959+5044 & 0.056 & 1.073 & 360+5831 			  & 35+51 		& ACIS-S & II   & 18,19 	    	& j \\
3C 445		   & J2223$-$0206 & 0.056 & 1.073 & 7869 		      & 46			& ACIS-S & II   & 26		 		& l \\
3C 285         & J1321+4235 & 0.079 & 1.474 & 6911 				  & 40 			& ACIS-S & II   & 20  		    	& k \\
3C 452         & J2245+3941 & 0.081 & 1.508 & 2195 				  & 81 			& ACIS-S & II   & 4,21 		    	& b \\
3C 227         & J0947+0725 & 0.087 & 1.609 & 6842+7265 		  & 30+20 		& ACIS-S & II   & 22     	    	& b \\
3C 388         & J1844+4533 & 0.091 & 1.675 & 4756+5295 		  & 8+31		& ACIS-I & II   & 23 		    	& m \\
\enddata
\tablecomments{\label{obsparams}Observational parameters for archival
  and proprietary {\em Chandra} datasets with significant thermal emission
  (see Table \ref{thermalspecparams}).  Redshifts are obtained either
  from SIMBAD or compilation of \citet{cheung07}.  X-ray references with more detail
  on the particular data set are provided above; not all references are
  provided.  VLA program codes are provided where we processed data ourselves.}
\tablerefs{
{\bf X-ray:} (1) \citet{worrall95}; (2) \citet{murgia01};
(3) \citet{kraft05}; (4) \citet{evans06}; (5) \citet{balmaverde06};
(6) \citet{hardcastle06}; (7) \citet{miller09}; 
(8) \citet{donato04}; (9) \citet{canosa99}; 
(10) \citet{worrall00}; (11) \citet{jeltema08}; (12) \citet{sun05};
(13) \citet{hardcastle01}; (14) \citet{evans05}; 
(15) \citet{ly05}; (16) \citet{young05}; (17) \citet{johnstone02};
(18) \citet{young02}; (19) \citet{smith02}; (20) \citet{hardcastle07a};
(21) \citet{isobe02}; (22) \citet{hardcastle07b}; (23) \citet{kraft06}; 
(24) \citet{sun09}; (25) \citet{massaro08}; (26) \citet{perlman09};
(27) \citet{hardcastle05} \\
{\bf Radio:} (a) \citet{condon91}; (b) \citet{black92}; (c) \citet{baum88};
(d) NRAO VLA Archive Survey; 
(e) \citet{hardcastle96}; (f) \citet{leahy91}; (g) \citet{sambruna04};
(h) \citet{fosbury98}; (i) \citet{ge94}; (j) \citet{perley84}; 
(k) \citet{alexander87}; (l) \citet{leahy97}; (m) \citet{roettiger94}
}
\end{deluxetable}

\clearpage

\begin{deluxetable}{lcccll}
\tablenum{2} 
\tabletypesize{\scriptsize} 
\tablecaption{Rejected {\em Chandra} Observations} 
\tablewidth{0pt} 
\tablehead{ 
\colhead{Name} & \colhead{$z$} & \colhead{Obs. IDs} & \colhead{Exp. Time} &
\colhead{Obs. Type} & \colhead{Reason} \\
& & & (ks) & & 
}
\startdata
\cutinhead{XRGs}
4C +32.25	& 0.052 & 9271 & 10		& ACIS-S & a\\
4C +48.29	& 0.053 & 9327 & 10		& ACIS-S & a\\
3C 136.1	& 0.064	& 9326 & 10 	& ACIS-S & a\\
J1101+1640	& 0.068 & 9273 & 10		& ACIS-S & mispointing \\
3C 223.1	& 0.107	& 9308 & 8 		& ACIS-S & a\\
\cutinhead{Archival Normal Radio Galaxies}
M84			& 0.003 & 803+5908+6131 & 29+47+41  & ACIS-S & b\\
M87			& 0.004 & 352+1808		& 38+14		& ACIS-S & b\\
3C 84		& 0.018 & 333+428		& 27+25		& HETGS	 & b+pileup\\
3C 442A		& 0.026 & 5635+6353+6359+6392 & 29+14+20+33 & ACIS-I & b\\
3C 353		& 0.030 & 7886+8565		& 72+18		& ACIS-S & c\\
3C 120		& 0.033 & 3015+5693		& 58+67		& HETGS+ACIS-I & c\\
DA 240		& 0.036 & 10237			& 24		& ACIS-I & a\\
3C 305		& 0.042 & 9330			& 8 		& ACIS-S & b\\
3C 390.3	& 0.056 & 830			& 34		& ACIS-S & c\\
3C 382		& 0.058 & 4910+6151		& 55+65		& HETGS  & c+pileup\\
3C 33		& 0.059 & 6910+7200		& 20+20		& ACIS-S & c\\
3C 35		& 0.067 & 10240			& 26		& ACIS-I & a\\
0313-192	& 0.067 & 4874			& 19		& ACIS-S & c\\
3C 105		& 0.089 & 9299			& 8			& ACIS-S & a\\
3C 326		& 0.090 & 10242+10908	& 19+28		& ACIS-I & target not on chip\\
3C 321		& 0.096 & 3138			& 48		& ACIS-S & b\\
3C 236		& 0.099 & 10246+10249	& 30+41		& ACIS-I & c\\
3C 327		& 0.104 & 6841			& 40		& ACIS-S & c\\
4C +74.26	& 0.104 & 4000+5195		& 38+32		& HETGS	 & c\\
3C 184.1	& 0.118 & 9305			& 8			& ACIS-S & a\\
\enddata
\tablenotetext{a}{No diffuse emission detected}
\tablenotetext{b}{Complex morphology}
\tablenotetext{c}{No diffuse thermal emission detected}
\tablecomments{\label{rejectedobs}Galaxies were selected for the preliminary
sample as described in \S2.1 and rejected for lack of diffuse gas or 
prohibitively complex morphology for our analysis (e.g. obvious multiple
bubbles).}
\end{deluxetable}

\clearpage


{\small
\begin{deluxetable}{lccccccccclcc}
\tablenum{3}
\rotate
\tabletypesize{\scriptsize}
\tablecaption{Thermal X-ray Spectral Parameters}
\tablewidth{0pt}
\tablehead{

\colhead{Galaxy} & \colhead{Region} & \colhead{$a$} & \colhead{$b$} & \colhead{$z$} & 
\colhead{Gal. $N_H$} & \colhead{$kT$} & \colhead{$Z$} & 
\colhead{$\bar{n}$} & \colhead{$P$} & \colhead{Models} & \colhead{$L_{X,\text{th}}$} &
\colhead{$\chi^2$/d.o.f.} \\ 
& & (kpc) & (kpc) &   & ($10^{20}$ & (keV) & & ($10^{-3}$ & ($10^{-11}$     &  & ($10^{41}$ & \\
& &       &       &   & cm$^{-2}$) &       & & cm$^{-3}$) &
dyne cm$^{-2}$) &  & erg s$^{-1}$) & \\ 
}
\startdata 
\cutinhead{XRGs} 
B2 1040+31A  & ISM     & 5   & 5   & 0.036 & 1.67 & 1.6  $\pm$ 0.4         & 0.3 (f)             & 38.  & 9.7  & apec                     & 1.6  & 4.66/5  \\
$\cdots$     & IGM (s) & 25  & 25  & $\cdots$& $\cdots$& 1.8  $\pm$ 0.2         & 1.0 (f)             & 3.8  & 1.1  & apec                     & 3.2  & 10.9/11 \\
$\cdots$     & IGM (l) & 75  & 75  & $\cdots$& $\cdots$& 1.3$^{+0.3}_{-0.1}$    & 0.3$^{+0.4}_{-0.2}$ & 1.1  & 0.24 & apec 					   & 4.7  & 20.2/23 \\
PKS 1422+26  & IGM     & 85  & 40  & 0.037 & 1.54 & 0.89 $\pm$ 0.09        & 0.3$^{+0.2}_{-0.1}$ & 2.0  & 0.29 & apec+PL                  & 5.2  & 23.1/20 \\
NGC 326      & ISM     & 6   & 6   & 0.048 & 5.86 & 0.68 $\pm$ 0.05        & 1.0 (f)             & 28.  & 3.0  & apec(+IGM)               & 5.4  & 13.3/17 \\
$\cdots$     & IGM     & 166 & 94  & $\cdots$& $\cdots$& 3.6  $\pm$ 0.4         & 0.8 (f)             & 1.9  & 1.1  & apec                     & 10.  & 94.1/100\\
3C 403       & ISM     & 9   & 6   & 0.059 & 12.2 & 0.24 $\pm$ 0.03        & 1.0 (f)             & 23.  & 0.89 & apec+PL+$N_H$(gauss+PL)  & 1.0  & 100.7/94\\
$\cdots$     & IGM     & 97  & 40  & $\cdots$& $\cdots$& 0.6  $\pm$ 0.2         & 1.0 (f)             & 0.37 & 0.04 & apec                     & 0.9  & 34.3/36 \\
4C +00.58    & ISM     & 17  & 17  & 0.059 & 7.14 & 1.2  $\pm$ 0.2         & 1.0 (f)             & 7.3  & 2.0  & apec                     & 3.0  & 5.8/6 \\
3C 192       & ISM+IGM & 19  & 19  & 0.060 & 4.08 & 1.0$^{+2.2}_{-0.2}$    & 1.0 (f)             & 4.9  & 0.79 & apec+$N_H$(PL)           & 2.5  & 3.9/5  \\
3C 433       & ISM+IGM & 44  & 26  & 0.102 & 7.77 & 0.96 $\pm$ 0.1         & 1.0 (f)             & 2.8  & 1.1  & apec+PL+gauss 		      & 4.0  & 16.2/12 \\
3C 315		 & IGM	   & 200 & 110 & 0.108 & 4.28 & 0.6$^{+0.4}_{-0.1}$    & 1.0 (f)			 & 1.0  & 0.1  & apec+PL				  & 8.2  & 4.5/10 \\
\cutinhead{Comparison sample}
3C 449       & ISM     & 5   & 5   & 0.017 & 8.99 & 0.77 $\pm$ 0.09        & 1.0 (f)             & 15.  & 1.9  & apec(+IGM)               & 0.22 & 21.1/19 \\
$\cdots$     & IGM     & 32  & 32  & $\cdots$& $\cdots$& 1.58 $\pm$ 0.06        & 0.8 $\pm$ 0.2       & 2.9  & 0.73 & apec                     & 3.9  & 179.8/169\\
3C 31        & ISM     & 4   & 4   & 0.017 & 5.36 & 0.68$^{+0.04}_{-0.03}$ & 1.0 (f)             & 22.  & 2.4  & apec(+IGM)               & 0.34 & 47.0/58  \\
$\cdots$     & IGM     & 30  & 30  & $\cdots$& $\cdots$& 2.0$^{+0.5}_{-0.2}$    & 0.4$^{+0.2}_{-0.1}$ & 2.0  & 0.69 & apec                     & 1.3  & 143.1/138\\
3C 83.1B     & ISM     & 3   & 3   & 0.018 & 13.4 & 0.65 $\pm$ 0.04        & 1.0 (f)             & 34.  & 3.6  & apec(+IGM)               & 0.13 & 30.1/35\\
$\cdots$     & ICM     & 56  & 30  & $\cdots$& $\cdots$& 8.$^{+6}_{-3}$         & 1.0 (f)             & 1.2  & 1.6  & apec                     & 1.7  & 80.9/86 \\
3C 264       & ISM     & 5   & 5   & 0.021 & 1.83 & 0.34 $\pm$ 0.04        & 1.0 (f)             & 6.6  & 0.37 & apec(+IGM)               & 0.18 & 19.7/30\\
3C 66B       & ISM     & 9   & 9   & 0.022 & 7.67 & 0.61$^{+0.06}_{-0.04}$ & 0.3$^{+1.}_{-0.1}$  & 7.2  & 0.71 & apec(+IGM)               & 0.36 & 49.8/42 \\
3C 296       & ISM     & 9   & 7   & 0.024 & 1.92 & 0.76 $\pm$ 0.02        & 1.0 (f)             & 9.0  & 1.1  & apec(+IGM)               & 1.3  & 49.0/46 \\
$\cdots$	 & IGM	   & 75  & 75  & $\cdots$& $\cdots$& 4.$^{+4}_{-1}$	   & 1.0 (f)			 & 0.8	& 0.5  & apec					  & 0.8  & 18.2/19\\
NGC 6251     & ISM     & 6   & 6   & 0.024 & 5.59 & 0.66 $\pm$ 0.03        & 1.0 (f)             & 14.  & 1.5  & apec(+IGM)               & 1.0  & 110./111 \\
PKS 2153-69  & ISM     & 7   & 7   & 0.028 & 5.59 & 0.77 $\pm$ 0.07        & 1.0 (f)             & 10.  & 1.2  & apec(+hot IGM)           & 1.4  & 26.0/24 \\
$\cdots$     & IGM     & 32  & 32  & $\cdots$& $\cdots$& $T_1=3.0^{+2.}_{-1.}$  & 1.0 (f)             & 2.6  & 1.3  & apec+apec                & 6.8  & 41.8/42\\
$\cdots$     & $\cdots$& $\cdots$ & $\cdots$& $\cdots$& $\cdots$& $T_2=0.9\pm 0.06$      & 1.0 (f)             &      &      &                          &      &  \\
3C 338       & ICM     & 30  & 30  & 0.030 & 0.89 & $T_1=1.8^{+0.2}_{-0.1}$& 0.7$\pm$0.05        & 28.  & 17.  & apec+apec                & 434. & 432.5/390\\
$\cdots$     & $\cdots$& $\cdots$ & $\cdots$& $\cdots$& $\cdots$& $T_2=5.0^{+1.}_{-0.5}$ & 0.7$\pm$0.05        &      &      &                          &      &   \\
3C 98        & IGM     & 46  & 31  & 0.030 & 10.8 & 1.1$^{+0.3}_{-0.2}$    & 0.2$^{+0.4}_{-0.1}$ & 1.8  & 0.32 & apec                     & 0.8  & 8.5/11\\
3C 465       & ISM     & 6   & 6   & 0.031 & 4.82 & 1.12$\pm$0.08          & 1.0 (f)             & 20.  & 3.6  & apec(+IGM)               & 1.4  & 30.4/32\\
$\cdots$     & IGM     & 80  & 56  & $\cdots$& $\cdots$& 4.2$^{+0.5}_{-0.3}$    & 1.0 (f)             & 2.3  & 1.6  & apec+gauss               & 25.  & 176.8/164\\
3C 293       & ISM     & 8   & 5   & 0.045 & 1.27 & 1.0 $\pm$ 0.3          & 0.2 (f)             & 28.  & 4.4  & apec(+IGM)               & 1.6  & 0.7/4 \\ 
Cyg A        & ICM     & 140 & 140 & 0.056 & 30.2 & $T_{\text{low}}=3.2^{+1.3}_{-0.1}$ & 0.6 (f) & 4.8  & 6.0  & cflow                    & 1600 & 433.1/403\\
$\cdots$     & $\cdots$& $\cdots$    & $\cdots$& $\cdots$& $\cdots$& $T_{\text{high}}=18. \pm 7.$                 &      &      &                          &      &   \\
3C 445       & IGM     & 73  & 54  & 0.056 & 4.51 & 0.7$^{+0.3}_{-0.4}$    & 1.0 (f)             & 0.26 & 0.03 & apec+PL                  & 0.6  & 9.38/14\\
3C 285\tablenotemark{a}       & ISM+IGM & 30  & 30  & 0.079 & 1.27 & $T_1=1.2^{+0.4}_{-0.2}$& 1.0 (f)             & 2.3  & 0.13 & apec+apec                & 2.9  & 6.9/7\\
$\cdots$     & $\cdots$& $\cdots$    & $\cdots$& $\cdots$& $\cdots$& $T_2=0.33^{+0.3}_{-0.05}$ & 1.0 (f)          &      &      &                          &      &    \\
3C 452       & IGM     & 201 & 82  & 0.081 & 9.64 & 3.8  $\pm$ 0.1         & 1.0 (f)             & 1.3  & 0.79 & apec+PL                  & 5.0  & 148.7/155.\\
3C 227       & ISM+IGM & 28  & 28  & 0.087 & 2.11 & $T_1=0.2$ (f)          & 1.0 (f)             & 5.9  & 1.0  & apec+apec+gauss          & 7.4  & 12.7/29\\
$\cdots$     & $\cdots$& $\cdots$    & $\cdots$& $\cdots$& $\cdots$& $T_2=1.2^{+0.3}_{-0.2}$& 1.0 (f)             &      &      &                          &      &    \\ 
3C 388       & ICM(bar)& 16  & 16  & 0.091 & 5.58 & 2.4 $\pm$ 0.2          & 1.0 (f)             & 24.  & 9.2  & apec                     & 41.  & 38.9/40\\
$\cdots$     & ICM(big)& 130 & 130 & $\cdots$& $\cdots$& 3.3 $\pm$ 0.1          & 1.0 (f)             & 3.1  & 1.7  & apec                     & 425. & 174.9/200\\
\enddata
\tablenotetext{a}{3C 285 is difficult to disentangle, but the larger IGM values 
agree with those coincident with the ISM.}
\tablecomments{\label{thermalspecparams} Diffuse atmosphere spectral parameters
for targets in Table \ref{obsparams}.  (f) denotes a frozen parameter. 
In calculating $\bar{n}$, we assume axisymmetry and
use a volume $V = ab^2$.  
Unabsorbed luminosities are reported for the thermal component only; in the case
of the ISM, this is the cooler component.  The notation (+IGM)
indicates that a thermal model with all parameters except the normalization
frozen was used in the fit; these parameters are based on fits to the IGM.}
\end{deluxetable}
}

\clearpage


\begin{deluxetable}{lccccccccc}
\tablenum{4} 
\tabletypesize{\scriptsize} 
\tablecaption{Optical Hosts Comparison} 
\tablewidth{0pt} 
\tablehead{ 
\colhead{Name} & \colhead{} & \colhead{$\epsilon_{\text{DSS}}$} & \colhead{} & \colhead{} & \colhead{PA$_{\text{DSS}}$} & \colhead{} & \colhead{} & \colhead{PA$_{\text{radio}}$} & \colhead{}\\
               & This work  & C02                               & S09        & This work  & C02                         & S09		 & This work  & C02 						  & S09
}
\startdata
\cutinhead{Working Sample}
B2 1040+31A		& 0.15	& -	& - 	 & 153	& -		& -		& 159	& -		& - \\
PKS 1422+26		& 0.16	& - & -		 & 118	& -		& -		& 92	& -		& - \\
NGC 326			& 0.04	& -	& 0		 & 153	& -		& b		& 120	& - 	& 135\\
3C 403			& 0.27 	& a	& 0.25	 & 39	& 35	& 39 	& 72	& 85	& 79 \\
4C +00.58		& 0.51	& -	& -		 & 139	& -		& -		& 65	& -		& - \\
3C 192			& 0.02	& a	& 0		 & 119	& 95	& b 	& 123	& 125	& 123 \\
3C 433			& 0.30	& - & 0.47	 & 167	& -		& 145	& 169	& - 	& 164 \\
3C 315			& 0.37	& a	& 0.46	 & 44	& 35	& 33	& 12	& 10	& 8\\
\cutinhead{Unused XRGS}
4C +32.25		& 0.12	& a	& 0.15	 & 90	& 90	& 84	& 61	& 60	& 64\\
4C +48.29		& 0.03	& -	& 0		 & 127	& -		& b		& 179	& -		& 170\\
3C 136.1		& 0.39	& a	& 0.37	 & 101	& 100	& 117	& 106	& 110	& 139\\
J1101+1640		& 0.34	& -	& 0.29	 & 62	& -		& 64	& 113	& -		& 114\\
3C 223.1		& 0.42	& a	& 0.45	 & 43	& 40	& 40	& 7		& 15	& 15\\
\enddata
\tablenotetext{a}{\citet{capetti02} do not provide $\epsilon$ values in their paper.}
\tablenotetext{b}{\citet{saripalli09} do not report an optical PA if $\epsilon$ is ``0''.}
\tablecomments{\label{opticalcomparison}A ``-'' signifies no data exist
in the literature for comparison.  Note that while radio position angles are
generally in agreement, large discrepancies exist (attributable to where the
jet/lobe PA is measured).  Uncertainties are not quoted in C02 or S09.}
\end{deluxetable}

\clearpage


\begin{deluxetable}{llccccccccc}
\tablenum{5} 
\tabletypesize{\scriptsize} 
\tablecaption{Ellipse Parameters for the ISM} 
\tablewidth{0pt} 
\rotate
\tablehead{ 
\colhead{Name} & \colhead{$\epsilon_{\text{X-ray}}$} & \colhead{PA$_{\text{X-ray}}$} &
\colhead{$\epsilon_{\text{X-ray PSF}}$} & \colhead{PA$_{\text{X-ray PSF}}$} & 
\colhead{$\epsilon_{\text{DSS}}$} & \colhead{PA$_{\text{DSS}}$} & \colhead{$\epsilon_{\text{SDSS}}$}
& \colhead{PA$_{\text{SDSS}}$} & 
\colhead{PA$_{\text{jets}}$} & \colhead{PA$_{\text{wings}}$} \\ & & ($^{\circ}$) & & ($^{\circ}$) & & 
($^{\circ}$) &  & ($^{\circ}$) & ($^{\circ}$) & ($^{\circ}$)
}
\startdata 
\cutinhead{XRGs} 
B2 1040+31A		& 0.21$\pm$0.07 	& 158$\pm$11 & 0.07 & 58  & -               & -          & 0.15$\pm$0.02  & 153$\pm$22 & 159 & 58 \\
NGC 326			& 0.084$\pm$0.009 	& 130$\pm$37 & 0.09 & 134 & 0.037$\pm$0.004 & 153$\pm$33 & -              & -          & 120 & 42 \\
3C 403			& 0.47$\pm$0.03		& 35$\pm$5   & 0.11 & 108 & 0.27$\pm$0.09   & 39$\pm$35  & -              & -          & 72  & 133\\
4C +00.58		& 0.15$\pm$0.02		& 138$\pm$17 & 0.07 & 161 & 0.51$\pm$0.08   & 139$\pm$6  & 0.44$\pm$0.02  & 138$\pm$3  & 65  & 14 \\
3C 192			& 0.04$\pm$0.01		& 126$\pm$22 & 0.06 & 146 & 0.018$\pm$0.005 & 119$\pm$40 & 0.018$\pm$0.002& 132$\pm$38 & 123 & 54 \\
3C 433			& 0.09$\pm$0.02		& 156$\pm$40 & 0.04 & 172 & 0.3$\pm$0.1	    & 172$\pm$21 & -              & -          & 169 & 84 \\
\cutinhead{Comparison Sample}
3C 449			& 0.14$\pm$0.02		& 6$\pm$15 	 & 0.16 & 30  & 0.21$\pm$0.03   & 8$\pm$6    & -              & -          & 9   & \\
3C 31			& 0.087$\pm$0.003	& 112$\pm$16 & 0.18 & 159 & 0.10$\pm$0.01   & 142$\pm$12 & -              & -          & 160 & \\
3C 83.1B		& 0.17$\pm$0.01		& 166$\pm$16 & 0.12 & 1   & 0.19$\pm$0.02   & 161$\pm$17 & -              & -          & 96  & \\
3C 264			& 0.042$\pm$0.002	& 129$\pm$27 & 0.17 & 148 & 0.044$\pm$0.003 & 129$\pm$27 & 0.011$\pm$0.005& 160$\pm$32 & 30  & \\
3C 66B			& 0.15$\pm$0.01		& 128$\pm$9  & 0.15 & 13  & 0.13$\pm$0.01   & 131$\pm$15 & -              & -          & 60  & \\
3C 296			& 0.15$\pm$0.02		& 137$\pm$6  & 0.17 & 147 & 0.19$\pm$0.02   & 151$\pm$13 & 0.23$\pm$0.01  & 147$\pm$6  & 35  & \\
PKS 2153-69		& 0.22$\pm$0.03		& 106$\pm$8  & 0.17 & 34  & 0.25$\pm$0.04   & 126$\pm$12 & -              & -          & 136 & \\
NGC 6251		& 0.047$\pm$0.002	& 25$\pm$23  & 0.15 & 72  & 0.11$\pm$0.01   & 24$\pm$13  & -              & -          & 115 & \\
3C 98			& 0.19$\pm$0.04		& 92$\pm$24  & 0.13 & 90  & 0.10$\pm$0.03   & 62$\pm$20  & -              & -          & 10  & \\
3C 465			& 0.061$\pm$0.003	& 51$\pm$29  & 0.17 & 74  & 0.17$\pm$0.08   & 33$\pm$25  & -              & -          & 126 & \\
3C 293			& 0.30$\pm$0.07		& 84$\pm$18  & 0.05 & 110 & 0.50$\pm$0.06   & 64$\pm$7   & 0.29$\pm$0.05  & 60$\pm$10  & 126 & \\
3C 285			& 0.21$\pm$0.03		& 141$\pm$18 & 0.19 & 96  & 0.36$\pm$0.06   & 133$\pm$13 & 0.36$\pm$0.05  & 125$\pm$9  & 73  & \\
3C 227			& 0.11$\pm$0.01		& 19$\pm$20  & 0.33 & 98  & 0.09$\pm$0.02   & 27$\pm$17  & -              & -          & 65  & \\
\enddata
\tablecomments{\label{eccentricitytable}Position angles are reported 
counter-clockwise from North (0$^{\circ}$) and chosen to reflect the
acute angle between the major axes of the X-ray, optical, and radio
ellipses.  The error bars are reported at 95\% using our ellipse-fitting
method (\S3.1).  Radio
position angles are estimated visually and have an estimated
error of $\Delta$PA $\sim 10^{\circ}$.  The PSF values are
reported for the best-fit PSF used to mask the AGN emission.}
\end{deluxetable}

\clearpage


\begin{deluxetable}{lcccc}
\tablenum{6} 
\tabletypesize{\scriptsize} 
\tablecaption{Ellipse Parameters for the Local IGM/ICM} 
\tablewidth{0pt} 
\tablehead{ 
\colhead{Name} & \colhead{$\epsilon_{\text{IGM}}$} & \colhead{PA$_{\text{IGM}}$} & 
\colhead{PA$_{\text{jets}}$} & \colhead{PA$_{\text{wings}}$}\\
& & ($^{\circ}$) & ($^{\circ}$) & ($^{\circ}$) 
}
\startdata
\cutinhead{XRGs}
B2 1040+31A	& 0.18$\pm$0.08	  & 130$\pm$30 & 159 & 58 \\
PKS 1422+26	& 0.13$\pm$0.03   & 88$\pm$14  & 92	 & 179 \\
NGC 326		& 0.20$\pm$0.04	  & 128$\pm$20 & 120 & 42 \\
4C +00.58\tablenotemark{a}	& 0.15$\pm$0.02	  & 138$\pm$17 & 65  & 14 \\
3C 315 		& 0.38$\pm$0.06	  & 23$\pm$ 7  & 12  & 120 \\
\cutinhead{Comparison Sample}
3C 449		& 0.3$\pm$0.1	  & 41$\pm$12  & 9 	 & -\\
3C 31		& 0.18$\pm$0.01	  & 123$\pm$4  & 160 & -\\
3C 296		& 0.23$\pm$0.04	  & 121$\pm$8  & 35  & -\\
PKS 2153-69 & 0.13$\pm$0.03	  & 140$\pm$10 & 136 & -\\
3C 338		& 0.243$\pm$0.002 & 65$\pm$6   & 90  & -\\
3C 465		& 0.37$\pm$0.02	  & 53$\pm$3   & 126 & -\\
Cyg A		& 0.2$\pm$0.1	  & 22$\pm$4   & 115 & -\\
3C 445		& 0.3$\pm$0.1	  & 77$\pm$8   & 169 & -\\
3C 285		& 0.28$\pm$0.06	  & 163$\pm$9  & 73  & -\\
3C 452		& 0.43$\pm$0.03	  & 83$\pm$2   & 78  & -\\
3C 388 (l)	& 0.13$\pm$0.02	  & 65$\pm$6   & 55  & -\\
3C 388 (s)	& 0.088$\pm$0.005 & 133$\pm$18 & 55  & -\\
\enddata
\tablenotetext{a}{The same values for 4C~+00.58 are used in the ISM ellipse
table.  The identity of the medium we fit is ambiguous.}
\tablecomments{\label{igmeccentricity}The error bars are reported at 95\%
using our ellipse-fitting method but are likely underestimates.  Images
are processed before fits (\S3.1) by binning and smoothing due to the extended
faint emission.  For 3C~388 we use the large (l) ICM value, but note the 
smaller ICM has a distinct position angle (\S3.3).}
\end{deluxetable}

\clearpage


\begin{deluxetable}{lcclclccc}
\tablenum{7}
\rotate
\tabletypesize{\scriptsize}
\tablecaption{Radio Galaxy Nuclei X-ray Spectral Parameters}
\tablewidth{0pt}
\tablehead{

\colhead{Galaxy} & \colhead{$z$} & \colhead{Galactic $N_H$} & \colhead{Models} &
\colhead{Local $N_H$} & \colhead{$\Gamma$} & \colhead{$kT$} & \colhead{Luminosity} &
\colhead{$\chi^2$/d.o.f} \\ & & ($10^{20}$ cm$^{-2}$) & & ($10^{22}$ cm$^{-2}$) & &
(keV) & ($10^{42}$ erg s$^{-1}$) & 

}

\startdata 
\cutinhead{XRGs} 
PKS 1422+26 & 0.037 & 1.54 & $N_H$(PL+Gauss)         		 & 4. $\pm$ 1.   & 0.5 $\pm$ 0.4 			& -               & 3.8  & 8.2/16\\
NGC 326     & 0.048 & 5.86 & $N_H$(PL)+apec         	 	 & 0.5 (f)       & 1.3 $\pm$ 0.4 			& 0.68 $\pm$ 0.06 & 0.15 & 11.1/16\\
3C 403      & 0.059 & 12.2 & $N_H$(PL+Gauss)+PL+apec 		 & 46.$\pm$ 5.   & $\Gamma_1 = 1.9 \pm 0.1$ & 0.24 $\pm$ 0.03 & 16.6 & 100.7/94\\
            &       &      &                         		 &               & $\Gamma_2 = 2.0$ (f) 	& & & \\
4C +00.58   & 0.059 & 7.14 & PL                      		 & -             & 1.3 $\pm$ 0.6  			& -               & 1.1  & 1.6/4\\
3C 192      & 0.060 & 4.08 & $N_H$(PL)                       & 16 (f)        & 2.0 (f)                  & -               & 3.1  & 0.9/2\\
3C 433      & 0.102 & 7.77 & $N_H$(PL+Gauss)+apec    		 & 8. $\pm$ 1.   & 1.1 $\pm$ 0.1 			& 1.0 $\pm$ 0.2   & 59   & 72.5/77\\
\cutinhead{comparison sample}	
3C 449      & 0.017 & 8.99 & $N_H$(PL)+apec	    	     	 & 0.5 (f)       & 1.6 $\pm$ 0.2 			& 0.59 $\pm$ 0.08 & 0.05 & 7.6/14\\
3C 31       & 0.017 & 5.36 & PL+apec        	         	 & -             & 1.9 $\pm$ 0.1 			& 0.70 $\pm$ 0.03 & 0.13 & 81./83\\
3C 83.1B    & 0.018 & 13.4 & $N_H$(PL)+apec      	    	 & 2.2$\pm$ 0.9  & 2.0 $\pm$ 0.3 			& 0.48 $\pm$ 0.08 & 0.05 & 22.7/36\\ 
3C 264      & 0.021 & 1.83 & PL+apec             	    	 & -             & 2.13$\pm$ 0.03			& 0.34 $\pm$ 0.06 & 1.7  & 152.7/154\\
3C 66B      & 0.022 & 7.67 & PL+apec               		  	 & -             & 2.17$\pm$ 0.06			& 0.52 $\pm$ 0.08 & 0.33 & 90.6/114\\
3C 296      & 0.026 & 1.92 & $N_H$(PL)+apec        		  	 & 0.10 (f)      & 1.1 $\pm$ 0.1 			& 0.68 $\pm$ 0.02 & 0.38 & 91.8/81\\
NGC 6251    & 0.024 & 5.59 & $N_H$(PL)+apec         		 & 0.05$\pm$0.01 & 1.5 $\pm$ 0.03			& 0.35 $\pm$ 0.05 & 8.0  & 391/386\\
PKS 2153-69 & 0.028 & 2.67 & pileup(PL)+apec          		 & - 	  	     & 1.49$\pm$ 0.05			& 0.4  $\pm$ 0.2  & 12.2 & 73.1/86\\
3C 98       & 0.030 & 10.8 & $N_H$(PL)+gauss+PL				 & 9. $\pm$ 1.   & $\Gamma_1 = 1.2$ (f)     & -  			  & 3.1  & 47.4/35\\
			&		&	   & 								 & 				 & $\Gamma_2 = 1.5$ (f)     & & & \\
3C 465      & 0.031 & 4.82 & $N_H$(PL)+apec  	        	 & 0.3 (f)       & 2.4 $\pm$ 0.2 			& 0.83 $\pm$ 0.06 & 0.31 & 62.0/62\\
3C 293      & 0.045 & 1.27 & $N_H$(PL)+apec    	        	 & 11 $\pm$ 4    & 1.5 (f)       			& 1.55 (f)        & 4.5  & 9.3/10\\
Cyg A       & 0.056 & 30.2 & pileup($N_H$(PL))+gauss+cflow 	 & 25 $\pm$ 2    & 2.0 (f)                  & -               & 87.6 & 294.3/217\\
3C 445      & 0.056 & 4.51 & pileup($N_H$(PL))+gauss\tablenotemark{1}		 & 16 $\pm$ 1	 & 1.7 (f)					& -				  & 47.	 & 62.4/44\\
3C 285      & 0.079 & 1.27 & $N_H$(PL)+gauss			 	 & 28 $\pm$ 4    & 1.2 (f) 					& - 			  & 8.7  & 12.6/9\\
3C 452      & 0.081 & 9.64 & $N_H$(PL)+gauss+apec+PEXRAV\tablenotemark{2} 	 & 38 $\pm$ 9	 & 1.65 (f)   				& 0.8 $\pm$ 0.2   & 25   & 156.6/145\\
3C 227      & 0.087 & 2.11 & pileup($N_H$(PL))+PL            & 3.3 $\pm$ 0.1 & $\Gamma_1 = 1.75$ (f)    & -               & 155  & 248.9/167\\
            &       &      &                                 &               & $\Gamma_2 = 1.0 \pm 0.2$ &  & & \\
3C 388      & 0.091 & 5.58 & PL+apec						 & -			 & 2.3 $\pm$ 0.3			& 1.2 $\pm$ 0.2   & 1.9  & 14.1/16\\
\enddata
\tablenotetext{1}{Complex emission between $0.5-3.0$~keV is ignored; see Appendix A}
\tablenotetext{2}{PEXRAV is a reflection model from \citet{magdziarz95}}
\tablecomments{\label{agnspecparams}Reported errors for $\Gamma$ are the 
{\em smaller} of the values given by 90\% confidence intervals determined in 
XSPEC; if there is significant asymmetry about the best-fit value, the larger
error bar is ill defined.  (f) denotes a frozen parameter,
and complex model components are discussed in Appendix A with the relevant
galaxy.  The ``fits'' to low-count spectra (e.g. 3C~192 or 4C~+00.58) are
not reliable, but they do constrain well whether the spectrum is absorbed or
unabsorbed, so we report them only for an order-of-magnitude luminosity estimate.}
\end{deluxetable}

\clearpage


\appendix

\section{Notes on Individual Galaxies}

\subsection{XRGs}

\paragraph{{\bf B2~1040+31A ($z = 0.036$)}:} 
The radio galaxy emanates from the largest galaxy in a close triple system
\citep{fanti77} which itself is contained within a large region of intragroup
medium.  The larger IGM is in agreement with \citet{worrall00}, but for the
relevant medium we use a smaller region of IGM which is bright against the
large region and centered on the system.  In the DSS image the host galaxy 
cannot be isolated from its companions, so we use the SDSS image alone.  In 
the X-ray band, the host galaxy is bright; whether it is AGN emission is 
unclear based on the spectrum.  Because the character of the emission does not
change much outside the PSF, we assume we are seeing primarily thermal 
emission. 

\paragraph{{\bf PKS 1422+26 ($z=0.037$)}:} 
This galaxy has striking wings
and ``warm'' spots which occur about 50\% of the distance between the lobe 
edge and the radio core.  The radio galaxy defies easy Fanaroff-Riley
classification and is here considered a hybrid FR~I/II galaxy.  
\citet{canosa99} found radial asymmetry in the ROSAT image of the IGM
which the {\em Chandra} image confirms.  The radio hot spots coincide
well with decrements in the X-ray image which must be cavities blown in
the gas by the radio lobes.  The surface brightness of the ``ridge'' this
leaves in the middle is consistent with the hypothesis that the actual
orientation of the IGM is elongated in the direction of the radio lobes.
This is in contrast to 3C~285 or 3C~388, where cavities alone cannot
explain the ``excess'' emission near the core.  In addition, the spectrum
of the AGN has a highly significant emission line-like feature at 4.0~keV 
which is presently unexplained. 

\paragraph{{\bf NGC 326 ($z=0.048$)}:} 
NGC~326 is a double galaxy with the radio lobes emanating from the
northern component; both galaxies show diffuse X-ray ISM emission, 
and the X-ray emission from the southern galaxy is exclusively thermal.
The extent of the {\em Chandra} diffuse emission agrees well with the
ROSAT data where there is overlap \citep{worrall95}.  The long tails
in the radio map appear to follow empty channels in the {\em Chandra}
data, and the length of the tails relative to the active lobes provides
constraints on backflow formation mechanisms \citep{gopal03}.  The
ISM emission for the northern component is extracted excluding the
central point source; the temperature of $kT = 0.68$~keV agrees 
precisely with the temperature of the thermal model included in the
AGN emission, where a PL component is also required.  

\paragraph{{\bf 3C 403 ($z=0.059$)}:} 
This dataset was previously studied in detail by \citet{kraft05} who attempted 
to isolate the ISM emission from the nucleus by performing a spectral analysis 
in which the soft part of the spectrum was fit by an absorbed power law between
$1.0-2.0$~keV, with a line-dominated thermal fit between $0.3-1.0$~keV. 
They then fit an elliptical profile to these thermal events.  This line of
reasoning is supported by the absence of a bright core corresponding to the
PSF at $0.3-1.0$~keV.  However, we find an acceptable fit in which an 
unabsorbed power law dominates between $0.3-2.0$~keV.  This analysis is also
complicated by the presence of a feature at $0.85$~keV which is not fit by
either \citet{kraft05} or ourselves.  Since this feature accounts for
$\sim 25$\% of counts between $0.3-1.0$~keV (Fig.~5; available online, we cannot distinguish
between the thermal and power law models.  It should be noted that the PA
of the soft emission agrees well with the host galaxy; we are doubtless seeing
{\em some} ISM.  However, we do not know if it is the dominant component. 

\paragraph{{\bf 4C +00.58 ($z=0.059$)}:} 
4C~+00.58 was classified as a candidate XRG \citep{cheung07} due to faint
wings in the FIRST image.  A new, higher resolution radio map reveals that the
jet to the east of the core is bent toward the northwest and enclosed in a more
symmetric radio lobe.  We detect the inner part of this jet in the X-ray band
(confirmed by line-fitting to determine the PA using the bootstrap 
resampling method).  The PA of the jet was determined very well, so we masked
it from our fit.  It is unclear whether the diffuse emission represents the
ISM or the IGM.  $\epsilon_{\text{X-ray}}$ does not agree well with the host
galaxy, but the PA does.  There is likely some of both media represented in
the image, and deeper observations are required to separate the components.
Notably, the jet appears to pass through the {\em minor} axis of the host
galaxy.  We include this galaxy in both the ISM and IGM X-ray---radio
comparisons due to the ambiguity. 

\paragraph{{\bf 3C 192 ($z=0.060$)}:} 3C~192 is formally a ``winged''
galaxy \citep{cheung07}, but the morphology is otherwise very similar
to $\mathsf{X}$-shaped radio galaxies.  Enhanced emission is detected in
an elliptical region encompassing the primary radio lobes, but a smaller
enhancement we identify with the ISM exists near the AGN.  It is also possible
that this is the central region of the IGM, but it appears compact.  Moreover,
the ``IGM'' is consistent with being fit only by a power law ($\Gamma = 2.5$) 
and is in good spatial agreement with the primary lobes.  It may also be fit
by a thermal model with $kT \sim 0.9$ and $Z < 0.1$; if it is indeed thermal
emission, the IGM is in perfect agreement with \citet{capetti02}.  However, a
close inspection of the radio galaxy suggests that the X-ray emission may
come from a bounding shock or otherwise have been influenced by the radio
galaxy rather than a pre-existing highly eccentric IGM.  In particular, even
though the bright X-ray emission is not as large as the active lobes, the
southwestern lobe appears to track enhanced X-ray emission to its hot spot. 
Notably,
3C~192 is in a very round host \citep{smith89,cheung07b}.  \citet{capetti02}
found XRGs typically occur in galaxies with high projected $\epsilon$, so
the existence of similar morphology in round hosts bears investigation. 

\paragraph{{\bf 3C 433 ($z=0.102$)}:} 
The northern secondary lobe of 3C~433 is strikingly bent compared to the
southern secondary \citep{vanbreugel83}.  \citet{miller09} argue that the
hybrid FR~I/II lobe morphology is due to interaction with the surrounding
IGM.  In this scenario, the galaxy is a typical FR~II source propagating into
a very asymmetric environment.  \citet{miller09} also include in their paper
a {\em HST} image of the host galaxy whose PA is in good agreement with the
parameters we derive from the DSS image.  The X-ray emission near the southern
lobe is consistent with either thermal emission \citep{miller09} or power law
emission.  We assume the ISM is the relevant medium and attempt to isolate it,
but we may also be incorporating some local IGM emission in the spectrum. 

\paragraph{{\bf 3C 315 ($z=0.108$)}:}
The AGN is detected at a significance of $5.07\sigma$ (see Appendix A.3 for
methods), but there are too few counts to make a spectrum.  The diffuse 
IGM stands out against the background on large scales and is fit well by 
thermal models and not by power law models.  No ISM is detected, but we can
use the galaxy for comparison to the IGM independently.  It is possible that
this is not real IGM, as we suspect in 3C~192, but the alignment of the radio
galaxy and the hot atmosphere is not as good as in 3C~192, nor is it quite as
eccentric.  Thus, there is no strong evidence that the radio galaxy created
the observed X-ray morphology, but the exposure is very short.

\subsection{Comparison Sample}

\paragraph{{\bf 3C 449 ($z=0.017$)}:} Both the ISM and IGM are
bright in this 30 ks exposure, but the chip is somewhat smaller than
the extent of the radio lobes.  Notably, these lobes decollimate
and show the characteristic FR~I plumed structure near the edge 
of the intragroup medium; this structure and the hot atmospheres have
been studied with {\em XMM-Newton} in detail by \citet{croston03}. 
\citet{tremblay07} find that the jet is parallel to a warped
nuclear disk.
The AGN was fit with the XSPEC models $N_H$(PL)+{\tt apec}, 
but the an acceptable fit was also found for a PL+{\tt apec} fit with
no absorption and a smaller $\Gamma$.  

\paragraph{{\bf 3C 31 ($z=0.017$)}:} 
The X-ray jet \citep{hardcastle02} is quite bright in the 
ACIS-S3 image and is readily
distinguishable from other core X-ray emission, so we mask it in 
the analysis of the diffuse ISM.  A weaker, nearby source
to the southwest (visible in the optical light) may contaminate 
the ISM with a small amount of diffuse emission, but appears very 
weak in the X-ray. 

\paragraph{{\bf 3C 83.1B ($z=0.018$)}:} 
The {\em Chandra} data from this head-tail galaxy was published 
by \citet{sun05} who argue that there
is a distinct southern edge to the X-ray emission which otherwise
shares the ellipticity and position angle of the optical
isophotes.  They apply a deprojection analysis to the ISM data
and find that the central ISM has a temperature of $kT = 0.45$~keV,
in agreement with our non-deprojected fit to the central PSF emission
with $kT = 0.48$~keV.  \citet{sun05} also find that the LMXB contribution
to the diffuse gas is on the order of $\sim 5$\% and they show that the
southern edge in the X-ray data cannot be produced by ICM pressure,
but argue that the edge is a sign of ISM/ICM interaction and that 
if the galaxy is moving south very rapidly, the long twin tails
to the north are naturally explained.  The radio emission is likewise
curtailed to the south. 

\paragraph{{\bf 3C 264 ($z = 0.021$)}:}
3C 264 is a head-tail galaxy whose radio lobes both extend
to the northeast, but there is a large envelope of radio
emission around the AGN which is larger than the extent of
the $2\sigma_{\text{sky}}$ DSS optical isophotes.  
This data set is subarrayed to 1/8 of the ACIS-S3 area and
suffers from a readout streak.  We subtracted the streak and
this removed a relatively small number of photons from the diffuse 
ISM.  Both the host galaxy and the diffuse X-ray emission
appear almost circular on the sky such that the position angle
of the ellipses we fit (Table \ref{eccentricitytable}) are 
not well constrained.  

\paragraph{{\bf 3C 66B ($z=0.022$)}:}
We detect strong ISM emission and the X-ray jet in this {\em Chandra}
exposure.  
The IGM is detected and studied (with regards to its interaction with the
radio lobes) using {\em XMM-Newton} data by \citet{croston03}.  In the
{\em Chandra} exposure, however, the IGM is very weak and we cannot measure
the morphology.  The different character of the two lobes is attributed
to interaction with the hot gas by \citet{croston03}, with the western lobe
being more confined.  There is a small unresolved source to the southeast in
the X-ray image which has an optical counterpart, but it is far enough from
the ISM emission that our ellipse fitting is unaffected. 

\paragraph{{\bf 3C 296 ($z=0.024$)}:}
Before fitting an ellipse we subtract a small companion to the
northwest as well as the X-ray jet.  The ISM is clearly detected 
\citep[see also][NGC 5532 in their paper]{diehl07} and the lobes begin 
expanding outside the IGM.  The IGM has a similar shape and orientation to
the ISM, but it is clearly distinct in temperature and where it is
centered (in between 3C~296 and a companion to the southwest) even though 
it is relatively compact.  

\paragraph{{\bf NGC 6251 ($z=0.024$)}:}
The very large (Mpc-scale) radio lobes of this radio galaxy dwarf
the entire ACIS array, and the 1/8-size S3 subarray contains
emission only from the central core and along the jet
\citep{evans05}.  
We detect some emission associated with the jet to the
northwest of the galaxy along the chip.  A bright readout
streak has been removed, and we correct the spectrum of the
central PSF for pileup.  The spectrum of the AGN is fit well by either
a power law with $\Gamma < 0.5$ and a thermal model or a
slightly absorbed power law with $\Gamma = 1.5$ and a thermal
model. 

\paragraph{{\bf PKS 2153-69 ($z=0.028$)}:}
This data has been published before with regard to interaction of the radio
jet with a cloud of gas \citep{ly05,young05}.  The gross 
morphology agrees well with the C02 geometric relation, although
the surface brightness decrement to the north and south of the
galaxy is due to the cavities blown out by the radio lobes, as
is evident from the visible shocked bubbles of gas in the 
intragroup medium.  Despite these cavities, the morphology of the
larger IGM is easily measurable on larger scales, assuming the 
internal structure is entirely due to the radio galaxy activity.

\paragraph{{\bf 3C 338 ($z=0.030$)}:}
Unlike other extended sources for which deprojection is possible but which
can be fit by an isothermal model to find an average temperature 
(most of the other galaxies in our sample), 3C~338 has sufficient signal to 
require at least a 2-T model to find average temperatures (signifying the 
presence of hotter and cooler gas) and these temperatures are roughly
in line with the outer and inner temperatures of the \citet{johnstone02}
deprojection analysis which also agrees with our own (we report the 2-T fit
for consistency).  Although the cluster gas is interacting with the radio
galaxy, it is otherwise smooth and we take it to be a single large
ellipsoid for the purpose of measuring ellipticity and position angle of the
ICM.  Notably, we do not detect the AGN or the ISM amidst the very bright
ICM.  This observation is studied more 
thoroughly in \citet{johnstone02}, including the unusual
radio bridge parallel to the jets. 

\paragraph{{\bf 3C 98 ($z=0.030$)}:}
The ISM of the galaxy in this ACIS-I observation is very close to
the chip boundaries. 
However, it has clear ellipticity and is not contaminated by any
companions, so we are able to fit an ellipse.  There is diffuse
thermal emission associated with the northern radio lobe that is
similar to the ISM, but we see no evidence for such gas in the
southern lobe.  Notably, the orientation of the major axis of the
HST image in \citet{martel99} disagrees with our DSS ellipse by 
$\sim 20^{\circ}$.  The HST image is more reliable, meaning that
the radio jet is aligned close to the {\em major} axis of the
host.  

\paragraph{{\bf 3C 465 ($z=0.031$)}:}
3C 465 is a wide-angle tail FR~I radio galaxy located in
cluster gas that has been well studied \citep[both with {\em Chandra} and
{\em XMM-Newton} by][]{hardcastle05}.  The ICM covers most of the chip and
is centered on the host of 3C 465.  The ICM has clear ellipticity and the 
bright extended emission is comparable to the size of entire radio galaxy
We fit ellipses to both the ISM and ``local'' ICM on a scale slightly
smaller than half the size of the chip.  \citet{hardcastle05} also present
a radial profile which is in agreement with our average temperature for the
cluster gas.  The
spectrum of the AGN can be fit well either by a power law
with $\Gamma < 0.5$ and a thermal model or a slightly
absorbed power law with $\Gamma = 2.4$ and a thermal model.

\paragraph{{\bf 3C 293 ($z=0.045$)}:}
The X-ray emission associated with the radio galaxy is quite
weak, but the hot spots are distinct and there is a small amount
of extended ISM emission.  Unfortunately, the spatial extent of 
this diffuse emission is small compared to the optical isophotes, 
so the disagreement between the $\epsilon_{\text{X-ray}}$ and 
$\epsilon_{\text{DSS/SDSS}}$ (measured at a somewhat larger radius)
is not especially surprising.  Notably, the host galaxy has two 
nuclei and is in the process of merging \citep{martel99}. 

\paragraph{{\bf Cyg A ($z=0.056$)}:}
Cygnus A has been extensively studied in the X-rays thanks to its
exceptionally bright filamentary structure and evident shock cocoon
tracing the radio galaxy.  This structure makes it impossible to
measure the ellipse parameters of the ISM, but the larger ICM appears
to have little structure beyond the cocoon and filaments
\citep[][show that the host galaxy has complex morphology as well]{young02}.  
Therefore, we believe the directions of pressure and density gradients on large 
scales are likely to be similar to the ones the radio galaxy encounters.

\paragraph{{\bf 3C 445 ($z=0.056$)}:}
The bright nucleus of 3C~445 is far off-axis in this exposure; its 
brightness and high ellipticity make it impossible to measure the
ellipse parameters of the diffuse ISM.  Moreover, the \citet{martel99}
{\em HST} image is dominated by the unresolved nucleus, so our DSS
parameters may not be accurate.  However, we include the galaxy in
our sample because of the hot IGM surrounding the host.  

\paragraph{{\bf 3C 285 ($z=0.079$)}:}
3C~285 is currently experiencing a major merger, and instead of
a well-defined ISM, we fit an ellipse to the ``ridge'' structure
described in \citet{hardcastle07a}.  The ridge appears to be a
unified structure (its X-ray properties do not differ along its
length) and \citet{hardcastle07a} argue that it is {\em not}
produced by the interaction of the radio galaxy with its
environment.  They reason that the agreement with the starlight
(and, in 3C 442A, the flow of material from tidal tails into 
a similar ridge) could not be generally predicted by the interaction
of the radio galaxy with its environment.  This is in agreement
with our ISM--optical light correlation.  

\paragraph{{\bf 3C 452 ($z=0.081$)}:}
This data set was studied in detail by \citet{isobe02} who find that
the thermal and power law emission is well mixed throughout the region
covered by the radio lobes.  We attempt to isolate the two components
using an analysis similar to \citet{diehl07} in which the image is 
broken into hard and soft components.  The thermal and power law emission
each make up a certain percentage of the luminosity in each component,
so a synthesized image of the thermal emission can be 
constructed by adding the correct percentage of each component to the 
final image.  Although clearly the method will not be able to identify
individual counts as thermal or nonthermal in origin, if the spatial
distribution of nonthermal counts is significantly different from that
of thermal ones, we would expect to see a difference in the synthesized
images.  In fact, we find general agreement with \citet{isobe02}, although
the nonthermal emission is dominant near the hot spots of the radio galaxy.
If the thermal X-ray emission traces the physical boundaries of the IGM,
then we would expect to see either an $\mathsf{X}$- or $\mathsf{Z}$-shaped
galaxy, but the extremely good coincidence between X-ray and radio
emission argues that this is instead a cocoon inside a larger (unseen)
IGM.  

\paragraph{{\bf 3C 227 ($z=0.087$)}:}
3C~227 exhibits both core ISM and dimmer IGM emission near the
host galaxy.  The IGM morphology is difficult to study due to the chip
boundary.  The radio lobes appear to be ragged and bend
around the major axis of the host galaxy, suggesting that strong
backflows similar to those in the hydrodynamic XRG formation
models are at work, but the IGM is too weak near these mini-wings
to assess this hypothesis.  The galaxy was included in the 
\citet{hardcastle07b} study of particle acceleration in hot spots. 

\paragraph{{\bf 3C 388 ($z=0.091$)}:}
3C~388 is a radio galaxy oriented along the major axis of its host,
but we detect no ISM in the X-ray exposure due to the strength of
the surrounding ICM.  The radio galaxy is comparable to the size of
X-ray ICM isophotes elongated in its direction, which \citet{kraft06}
attribute to the influence of the radio galaxy on its surroundings.
The radio lobes have blown cavities in the X-ray emission, but near
the core of the radio galaxy, the X-ray isophotes are elongated in
a direction perpendicular to the radio lobes.  Although much of this
asymmetry is likely due to the cavities eating out the sides of a
spheroid where the radio galaxy overlaps the core region (i.e. it is
a physical structure but may not precede the radio galaxy), the
western radio lobe seems to bend arounnd the ridge.  Since the fainter
portions of the radio lobes may be older plasma evolving buoyantly
in the ICM (the jet in the western lobe is farther to the south), the
ridge may also be influencing the shape of the radio galaxy. 

\subsection{Unused XRGs}

Short observations of the more distant XRGs produced mixed results, with
a number of sources exhibiting little-to-no diffuse emission.  Where
we cannot determine whether any diffuse emission is dominated by thermal
emission or measure morphology, or where no diffuse emission corresponding
to the relevant medium is detected, we cannot use the galaxy in our analysis.
In the following short observations, the central point source is detected in
all cases (with the weakest detection having a significance barely exceeding
$3\sigma$), and upper limits are given for the thermal luminosity.  We
use the methods of \citet{ayres04} (Equations 3, 13, and 14 in his paper)
to measure the detection significance and flux confidence limits.  The
measured number of counts is given by
\begin{equation*}
S = S_0 + \biggl(\frac{s^2+2}{3}\biggr ) \pm \Delta S
\end{equation*}
where $S_0 = N-B$ is the detected number of counts in the cell, $B$ is the
expected background in the cell based on a much larger area elsewhere on the
chip, and $s$ is the significance; we choose $s = 1.645$ for 95\% confidence
intervals.  The quantity $\Delta S$ is given by
\begin{equation*}
\Delta S \equiv s\sqrt{\text{max}[(N-\tfrac{1}{4}B),0]+1}
\end{equation*}
The detect cells we used were uniformly chosen to be circles with 
3$^{\prime\prime}$ in radius to enclose the {\em Chandra} PSF at most
energies.  The significance of the detection $s$ was determined by solving the
quadratic equation \citep[Equation 3 in][]{ayres04}:
\begin{equation*}
\frac{\text{min}[B^{0.1},1]}{7} s^2 + \sqrt{B} s + 
\biggl[B - N - \frac{2\cdot\text{min}[B^{0.1}, 1]}{7}\biggr] = 0 
\end{equation*}
These values are reported for each galaxy.  The thermal luminosity upper
limits were computed by extracting a spectrum from a (large) region around
the point source and increasing the strength of a thermal model in XSPEC
until it was no longer a good fit.  The temperature cannot be constrained
and was thus fixed at $kT = 1.0$~keV.  The unabsorbed luminosity was then
computed at the upper bound of the model normalization and reported below.

\paragraph{{\bf 4C +32.25 ($z=0.053$)}:}
The point source is detected with a significance of $8.52\sigma$ with
$S = 20.7^{+9.4}_{-6.3}$ counts in the detect cell.  The diffuse
emission within a $r = 24^{\prime\prime}$ region (much larger than the
ISM) has an upper limit unabsorbed thermal luminosity of $4\times10^{40}$
erg s$^{-1}$ at $\sim 1.0$~keV, but it does not appear to be centrally
concentrated on the chip.  Our geometric analysis is not possible in 
this case, especially since the ISM is the medium of interest given the
scale of the radio emission. 

\paragraph{{\bf 4C +48.29 ($z=0.052$)}:}
We detect 19 counts in the detect cell (positioned around the core of radio
emission) and detect the AGN with a significance of $7.56\sigma$ and
$S = 17.6^{+8.9}_{-5.7}$ counts in a 95\% confidence interval.  The 
diffuse emission in a $r = 24^{\prime\prime}$ region is barely distinguishable
from the background and we measure an unabsorbed thermal luminosity 
upper limit of $L = 5 \times 10^{40}$ erg s$^{-1}$.  
We attribute this to the faint IGM as opposed to the ISM
due to the scale and lack of central concentration.

\paragraph{{\bf 3C 136.1 ($z=0.064$)}:}
This source has relatively high background, so we detect the AGN at a
significance of only $3.3\sigma$ despite finding 11 counts in the 
detect cell.  The number of counts is $S = 7.9^{+7.1}_{-3.9}$.  
Any diffuse emission is similarly buried in the background, and we
measure an unabsorbed thermal luminosity upper limit of $L = 2 \times 10^{40}$
erg s$^{-1}$ for a region with $r = 30^{\prime\prime}$.

\paragraph{{\bf J1101+1640 ($z=0.068$)}:}
Unfortunately, this XRG was not positioned on any ACIS chip during the
observation.  The galaxy lies in a cluster (Abell 1145) and the radio
lobes just extend onto part of the S3 chip, but no significant diffuse
gas from the cluster is detected on the chip.  The observation was 
mispointed due to an error in the primary literature and so cannot be
used.

\paragraph{{\bf 3C 223.1 ($z=0.107$)}:}
3C~223.1 actually has a bright AGN core where spectroscopy is possible,
revealing a highly absorbed power law.  The detection significance is
$34.1\sigma$ and the number of counts detected $S = 206^{+25}_{-22}$. 
However, any diffuse emission is extremely faint, with an upper limit 
to unabsorbed thermal luminosity of $L = 8\times10^{40}$ erg s$^{-1}$ 
in a region with $r = 30^{\prime\prime}$.  We are therefore unable
to use this galaxy in our XRG sample. 

We also choose not to use the two {\em Chandra} observations of XRGs at
higher redshift ($z = 0.128, 0.2854$) because of the requirement to 
simultaneously expand the comparison sample (and 3C~197.1 is not included in
the compilation of \citet{cheung07}).  We address these here:

\paragraph{{\bf 3C 197.1 ($z=0.128$)}:}
3C~197.1 has a long northern wing which is comparable in spatial extent to
the active lobe in a 1.5~GHz image \citep{neff95}, but there is no 
obvious southern wing which could be described as symmetric about the
central AGN.  In the 5~GHz image of \citet{neff95} extensions in the southern
lobe appear to be essentially symmetric about the jet axis.  
3C~197.1 is included in the \citet{saripalli09} list of XRGs, but
its inclusion in our sample is questionable.  We discuss it here because of
the other ``XRGs'' listed in \citet{saripalli09} and not in \citet{cheung07},
3C~197.1 bears the most similarity to the classical XRGs listed in 
\citet{cheung07}.  We do not include it in our sample due to its relatively
high redshift and ambiguity of classification.  The {\em Chandra} exposure
is a short 8~ks snapshot which clearly detects the AGN and some diffuse 
emission which may be the ISM or the IGM, but has insufficient counts to
claim a spectroscopic detection of hot gas.  

\paragraph{{\bf 3C 52 ($z=0.285$)}:}
3C~52 is a classical XRG with highly collimated secondary lobes in the 1.5~GHz
image \citep{alexander87} and excluded from our sample on the basis of much
higher redshift than our other sources.  The {\em Chandra} image is an 8~ks
snapshot which clearly detects the central point source and also a diffuse
atmosphere larger than the radio galaxy itself (with a radius of about 
50 arcsec centered on the galaxy).  This atmosphere is in good agreement
geometrically with the \citet{capetti02} relation (i.e. the major axis of the
ellipsoidal atmosphere is coaligned with the active lobes), but proving the
presence of a hot atmosphere spectroscopically is difficult.  An isothermal
{\tt apec} model requires $T > 7$~keV and is otherwise poorly constrained; 
a power law with $\Gamma = 1.5$ fits the spectrum well.  Since the emission 
region is large compared to the radio galaxy, it seems likely this is hot gas, 
but a much deeper observation is necessary to establish this. 


\begin{thebibliography}{}
\bibitem[Adelman-McCarthy et al.(2008)]{adelman08}
 Adelman-McCarthy, J., Agueros, M.A., Allam, S.S., et al. 2008, ApJS, 175, 297
\bibitem[Alexander \& Leahy(1987)]{alexander87} Alexander, P. \& Leahy, J.P.
 1987, \mnras, 225, 1
\bibitem[Allen et al.(2006)]{allen06} Allen, S.W., Dunn, R.J.H.,
  Fabian, A.C., Taylor, G.B., Reynolds, C.S. 2006, \mnras, 372, 21
\bibitem[Arnaud(1996)]{arnaud96} Arnaud, K.A. 1996 in ASP. Conf. Ser. 101, 
  Astronomical Data Analysis Software and Systems V, ed. G.H. Jacoby \& 
  J. Barnes (San Francisco: ASP), 17
\bibitem[Ayres(2004)]{ayres04} Ayres, T.R. 2004, \apj, 608, 957
\bibitem[Balmaverde et al.(2006)]{balmaverde06} Balmaverde, B., Capetti, A. \&
  Grandi, P. 2006, \aap, 451, 35
\bibitem[Baum et al.(1988)]{baum88} Baum, S.A., Heckman, T., Bridle, A., 
  van Breugel, W. \& Miley, G. 1988, \apjs, 68, 643
\bibitem[Becker et al.(1995)]{becker95} Becker, R.H., White, R.L. \& 
  Helfand, D.J. 1995, \apj, 450, 559
\bibitem[Binney \& Merrifield(1998)]{binney98} Binney, J. \& Merrifield, M.
  1998, {\em Galactic Astronomy}, Princeton Series in Astrophysics 
\bibitem[Black et al.(1992)]{black92} Black, A.R.S., Baum, S.A., Leahy, J.P.,
  Perley, R.A., Riley, J.M. \& Scheuer, P.A.G. 1992, \mnras, 256, 186
\bibitem[Bogdanovi\'{c} et al.(2007)]{bogdanovic07} Bogdanovi\'{c},
  T., Reynolds, C.S. \& Miller, M.C. 2007, \apj, 661, L147
\bibitem[Canosa et al.(1999)]{canosa99} Canosa, C.M., Worrall, D.M.,
  Hardcastle, M.J. \& Birkinshaw, M. 1999, \mnras, 310, 30
\bibitem[Capetti et al.(2002)]{capetti02} Capetti, A., Zamfir, S.,
Rossi, P., Bodo, G., Zanni, C. \& Massaglia, S. 2002, \aap, 394, 39
\bibitem[Cheung(2007)]{cheung07} Cheung, C.C. 2007, \aj, 133, 2097
\bibitem[Cheung \& Springmann(2007)]{cheung07b} Cheung, C.C. \& Springmann, A.
 2007 in ASP Conf. Ser. 373, The Central Engine of Active Galactic Nuclei, ed.
 L.C. Ho \& J.-M. Wang (San Francisco: ASP), 259
\bibitem[Cheung et al.(2009)]{cheung09} Cheung, C.C., Healey, S.E., Landt, H.,
 Kleijn, G.V. \& Jord\'{a}n, A. 2009, \apjs, 181, 548
\bibitem[Condon et al.(1991)]{condon91} Condon, J.J., Frayer, D.T \&
 Broderick, J.J. 1991, \aj, 101, 362
\bibitem[Croston et al.(2003)]{croston03} Croston, J.H., Hardcastle, M.J.,
 Birkinshaw, M. \& Worrall, D.M. 2003, \mnras, 346, 1041
\bibitem[Dennett-Thorpe et al.(2002)]{dennett02} Dennett-Thorpe, J.,
  Scheuer, P.A.G., Laing, R.A., Bridle, A.H., Pooley, G.G. \& Reich,
  W. 2002, \mnras, 330, 609
\bibitem[Diehl \& Statler(2007)]{diehl07} Diehl, S. \& Statler,
  T. 2007, \apj, 668, 150
\bibitem[Donato et al.(2004)]{donato04} Donato, D., Sambruna, R.M.
  \& Gliozzi, M. 2004, \apj, 617, 915
\bibitem[Efron(1982)]{efron82} Efron, B. 1982, CBMS-NSF Regional Conf. Ser.
  in Applied Mathematics, The Jackknife, the Bootstrap and Other Resampling
  Plans (Philadelphia: Soc. for Industrial \& Applied Mathematics)
\bibitem[Ekers et al.(1978)]{ekers78} Ekers, R.D., Fanti, R., Lari,
  C. \& Parma, P. 1978, Nature, 276, 588
\bibitem[Evans et al.(2005)]{evans05} Evans, D.A., Hardcastle, M.J., Croston, J.H.,
  Worrall, D.M. \& Birkinshaw, M. 2005, \mnras, 359, 363
\bibitem[Evans et al.(2006)]{evans06} Evans, D.A., Worrall, D.M., Hardcastle, M.J.,
  Kraft, R.P. \& Birkinshaw, M. 2006, \apj, 642, 96
\bibitem[Evans et al.(2008)]{evans08} Evans, D.A., Fong, W.F., Hardcastle, M.J.,
  Kraft, R.P., Lee, J.C., Worrall, D.M., Birkinshaw, M., Croston, J.H. \& 
  Muxlow, T.W.B. 2008, \apj, 675, 1057
\bibitem[Fanaroff \& Riley(1974)]{fan74} Fanaroff, B.L. \& Riley,
  J.M. 1974, \mnras, 167, 31P
\bibitem[Fanti et al.(1977)]{fanti77} Fanti, C., Fanti, R., Gioia,
  I.M., Lari, C., Parma, P. \& Ulrich, M.-H. 1977, A\&AS, 29, 279
\bibitem[Fosbury et al.(1998)]{fosbury98} Fosbury, R.A.E., Morganti, R., 
  Wilson, W, Ekers, R.D., di Serego Alighieri, S. \& Tadhunter, C.N. 1998,
  \mnras, 296, 701
\bibitem[Ge \& Owen(1994)]{ge94} Ge, J. \& Owen, F.N. 1994, \aj, 108, 1523
\bibitem[Gopal-Krishna \& Wiita(2000)]{gopal00} Gopal-Krishna \& Wiita, P.J.
  2000, \aap, 363, 507
\bibitem[Gopal-Krishna et al.(2003)]{gopal03} Gopal-Krishna, Biermann,
  P.L. \& Wiita, P.J. 2003, \apj, 594, L103
\bibitem[Hardcastle et al.(1996)]{hardcastle96} Hardcastle, M.J.,
  Alexander, P., Pooley, G.G. \& Riley, J.M. 1996, \mnras, 278, 273
\bibitem[Hardcastle et al.(2001)]{hardcastle01} Hardcastle, M.J.,
  Birkinshaw, M. \& Worrall, D.M. 2001, \mnras, 326, 1499
\bibitem[Hardcastle et al.(2002)]{hardcastle02} Hardcastle, M.J.,
  Worrall, D. M., Birkinshaw, M., Laing, R. A. \& Bridle, A. H. 2002,
  \mnras, 334, 182
\bibitem[Hardcastle et al.(2004)]{hardcastle04} Hardcastle, M.J.,
  Harris, D.E., Worrall, D.M. \& Birkinshaw, M. 2004, \apj, 612, 729
\bibitem[Hardcastle et al.(2005)]{hardcastle05} Hardcastle, M.J.,
  Sakelliou, I. \& Worrall, D.M. 2005, \mnras, 359, 1007
\bibitem[Hardcastle et al.(2006)]{hardcastle06} Hardcastle, M.J.,
  Evans, D.A. \& Croston, J.H. 2006, \mnras, 370, 1893
\bibitem[Hardcastle et al.(2007a)]{hardcastle07a} Hardcastle, M.J.,
  Kraft, R.P., Worrall, D.M., Croston, J.H., Evans, D.A., 
  Birkinshaw, M. \& Murray, S.S. 2007, \apj, 662, 166
\bibitem[Hardcastle et al.(2007b)]{hardcastle07b} Hardcastle, M.J., 
  Croston, J.H. \& Kraft, R.P. 2007, \apj, 669, 893
\bibitem[Heckman et al.(1985)]{heckman85} Heckman, T., Illingworth, G.,
  Miley, G., van Breugel, W. 1985, \apj, 299, 41
\bibitem[Isobe et al.(2002)]{isobe02} Isobe, N., Tashiro, M.,
  Makishima, K., Iyomoto, N., Suzuki, M., Murakami, M.M., Mori, M. \&
  Abe, K. 2002, \apj, 580, L111
\bibitem[Jackson et al.(2003)]{jackson03} Jackson, N., Beswick, R., 
  Pedlar, A., Cole, G.H., Sparks, W.B., Leahy, J.P., Axon, D.J. \&
  Holloway, A.J. 2003, \mnras, 338, 643
\bibitem[Jeltema et al.(2008)]{jeltema08} Jeltema, T.E., Binder, B. \&
  Mulchaey, J.S. 2008, \apj, 679, 1162
\bibitem[Johnstone et al.(2002)]{johnstone02} Johnstone, R.M., Allen,
  S.W., Fabian, A.C. \& Sanders, J.S. 2002, \mnras, 336, 299
\bibitem[Kalberla et al.(2005)]{kalberla05} Kalberla, P.M.W., Burton,
  W.B., Hartmann, D., Arnal, E.M., Bajaja, E., Morras, R., P\"{o}ppel,
  W.G.L. 2005, \aap, 440, 775
\bibitem[Klein et al.(1995)]{klein95} Klein, U., Mack, K.-H.,
  Gregorini, L. \& Parma, P. 1995, \aap, 303, 427
\bibitem[Kraft et al.(2005)]{kraft05} Kraft, R.P., Hardcastle, M.J.,
  Worrall, D.M. \& Murray, S.S. 2005, \apj, 622, 149
\bibitem[Kraft et al.(2006)]{kraft06} Kraft, R.P., Azcona, J., Forman, W.R.,
  Hardcastle, M.J., Jones, C. \& Murray, S.S. 2006, \apj, 639, 753
\bibitem[Laing et al.(1983)]{laing83} Laing, R.A., Riley, J.M. \&
  Longair, M.S. 1983, \mnras, 204, 151
\bibitem[Lal \& Rao(2004)]{lal04} Lal, D.V. \& Rao, A.P. 2004,
  Bull. Astr. Soc. India, 32, 247
\bibitem[Lal \& Rao(2005)]{lal05} Lal, D.V. \& Rao, A.P. 2005, \mnras,
  356, 232
\bibitem[Lal \& Rao(2007)]{lal07} Lal, D.V. \& Rao, A.P. 2007, \mnras,
  374, 1085
\bibitem[Lal et al.(2008)]{lal08} Lal, D.V., Hardcastle, M.J. \& Kraft,
  R.P. 2008, \mnras, 390, 1105
\bibitem[Lambas et al.(1992)]{lambas92} Lambas, D.G., Maddox, S.J. \&
  Loveday, J. 1992, \mnras, 258 404
\bibitem[Leahy \& Williams(1984)]{leahy84} Leahy, J.P. \& Williams,
  A.G. 1984, \mnras, 210, 929
\bibitem[Leahy \& Perley(1991)]{leahy91} Leahy, J.P. \& Perley, R.A.
  1991, \aj, 102, 537
\bibitem[Leahy \& Parma(1992)]{leahy92} Leahy, J.P. \& Parma, P. 1992,
  in Extragalactic Radio Sources: From Beams to Jets, ed. J. Roland, H. Sol,
  \& G. Pelletier (Cambridge: Cambridge Univ. Press), 307
\bibitem[Leahy et al.(1997)]{leahy97} Leahy, J.P., Black, A.R.S., 
  Dennett-Thorpe, J., Hardcastle, M.J., Komissarov, S., Perley, R.A., 
  Riley, J.M. \& Scheuer, P.A.G. 1997, \mnras, 291, 20
\bibitem[Ly et al.(2005)]{ly05} Ly, C., de Young, D.S. \& Bechtold, J.
 2005, \apj, 618, 609
\bibitem[Magdziarz \& Zdziarski(1995)]{magdziarz95} Magdziarz, P. \& 
 Zdziarski, A.A. 1995, \mnras, 273, 837
\bibitem[Martel et al.(1999)]{martel99} Martel, A.R. et al. 1999,
 \apjs, 122, 81
\bibitem[Massaro et al.(2008)]{massaro08} Massaro, F. et al. 2008 in
  American Astronomical Society, HEAD meeting \#10, 26.19
\bibitem[Massaro et al.(2009)]{massaro09} Massaro, F., Harris, D.E., 
 Chiaberge, M., Grandi, P., Macchetto, F.D., Baum, S.A., O'Dea, C.P. \&
 Capetti, A. 2009, \apj, 696, 980
\bibitem[Merritt \& Ekers(2002)]{merritt02} Merritt, D. \& Ekers,
  R.D. 2002, Science, 297, 1310
\bibitem[Miller \& Brandt(2009)]{miller09} Miller, B.P. \& Brandt, W.N.
  2009, \apj, 695, 755
\bibitem[Murgia et al.(2001)]{murgia01} Murgia, M., Parma, P., de
  Ruiter, H.R., Bondi, M., Ekers, R.D., Fanti, R. \& Fomalont,
  E.B. 2001, \aap, 380, 102
\bibitem[Neff et al.(1995)]{neff95} Neff, S.G., Roberts, L. \& Hutchings, J.B.,
  1995, \apjs, 99, 349
\bibitem[Parma et al.(1986)]{parma86} Parma, P., de Ruiter, H.R., 
  Fanti, C. \& Fanti, R. 1986, \aaps, 64, 135
\bibitem[Perley et al.(1984)]{perley84} Perley, R.A., Dreher, J.W. \&
  Cowan, J.J. 1984, \apj, 285, L35
\bibitem[Perlman et al.(2009)]{perlman09} Perlman, E.~S., Georganopolous, M., 
  May, E.~M. \& Kazanas, D. 2009, \apj, in press (arXiv:0910.3021v1)
\bibitem[Press et al.(1992)]{press92} Press, W.H., Teukolsky, S.A., 
  Vetterling, W.T. \& Flannery, B.P. 1992 in {\it Numerical Recipes in C: The
  Art of Scientific Computing} (Cambridge: Cambridge Univ. Press)
\bibitem[Rees(1978)]{rees78} Rees, M. 1978, Nature, 275, 516
\bibitem[Reynolds et al.(2001)]{reynolds01} Reynolds, C.S., Heinz, S. \&
 Begelman, M.C. 2001, \apj, 549, L179
\bibitem[Roettiger et al.(1994)]{roettiger94} Roettiger, K., Burns, J.,
  Clarke, D.A. \& Christiansen, W.A. 1994, \apj, 421, L23
\bibitem[Rottmann(2001)]{rottmann01} Rottmann, H. 2001, PhD thesis,
  Univ. Bonn
\bibitem[Sambruna et al.(2004)]{sambruna04} Sambruna, R.~M., 
  Gliozzi, M., Donato, D., Tavecchio, F., Cheung, C.~C. \& Mushotzky, R.~F.
  2004, \aap, 414, 885
\bibitem[Sanderson et al.(2006)]{sanderson06} Sanderson, A.J.R., Ponman,
  T.J. \& O'Sullivan, E. 2006, \mnras, 372, 1496
\bibitem[Saripalli \& Subrahmanyan(2009)]{saripalli09} Saripalli,
  L. \& Subrahmanyan, R. 2009, \apj, 695, 156
\bibitem[Scheuer(1974)]{scheuer74} Scheuer, P.A.G. 1974, \mnras, 166,
  513
\bibitem[Silk \& Rees(1998)]{silk98} Silk, J. \& Rees, M. 1998, \aap,
  331, L1
\bibitem[Smith \& Heckman(1989)]{smith89} Smith, E.P. \& Heckman, T.M. 1989,
  \apj, 341, 685
\bibitem[Smith et al.(2002)]{smith02} Smith, D.A., Wilson, A.S., Arnaud, K.A.,
  Terashima, Y. \& Young, A.J. 2002, \apj, 565, 195
\bibitem[Sun et al.(2005)]{sun05} Sun, M., Jerius, D. \& Jones, C. 2005,
  \apj, 633, 165
\bibitem[Sun et al.(2009)]{sun09} Sun, M., Voit, G.M., Donahue, M., Jones, C.,
 Forman, W. \& Vikhlinin, A. 2009, \apj, 693, 1142
\bibitem[Thompson et al.(1980)]{thompson80} Thompson, A.R.,  Clark, B.G., 
 Wade, C.M. \& Napier, P.J. 1980, \apjs, 44, 151
\bibitem[Tremblay et al.(2007)]{tremblay07} Tremblay, G.R., Chiaberge, M.,
 Donzelli, C.J., Quillen, A.C., Capetti, A. Sparks, W.B. \& Maccheto, F.D.
 2007, \apj, 666, 109
\bibitem[van Breugel \& Jagers(1982)]{vanbreugel82} van Breugel, W. \&
  Jagers, W. 1982, A\&AS, 49, 529
\bibitem[van Breugel et al.(1983)]{vanbreugel83} van Breugel,
  W. Balick, B., Heckman, T., Miley, G. \& Helfand, D. 1983, \aj, 88, 1
\bibitem[Wan \& Daly(1996)]{wan96} Wan, L. \& Daly, R. 1996, \apj,
  467, 145
\bibitem[Werner et al.(1999)]{werner99} Werner, P.N., Worrall, D.M. \&
  Birkinshaw, M. 1999, \mnras, 207, 722
\bibitem[Wirth et al.(1982)]{wirth82} Wirth, A., Smarr, L. \&
  Gallagher, J.S. 1982, \aj, 87, 602
\bibitem[Worrall et al.(1995)]{worrall95} Worrall, D.M., Birkinshaw,
  M. \& Cameron, R.A. 1995, \apj, 449, 93
\bibitem[Worrall \& Birkinshaw(2000)]{worrall00} Worrall, D.M.,
  Birkinshaw, M. 2000, \apj, 530, 719
\bibitem[Wright(2006)]{wright06} Wright, E.L. 2006, PASP, 118, 1711
\bibitem[Young et al.(2002)]{young02} Young, A.J., Wilson, A.S., 
 Terashima, Y., Arnaud,  K.A. \& Smith, D.A. 2002, \apj, 564, 176
\bibitem[Young et al.(2005)]{young05} Young, A.J., Wilson, A.S., Tingay, S.J.
 \& Heinz, S. 2005, \apj, 622, 830
\bibitem[Zier \& Biermann(2001)]{zier01} Zier, C. \& Biermann,
  P.L. 2001, \aap, 377, 23
\bibitem[Zier(2005)]{zier05} Zier, C. 2005, \mnras, 364, 583
\end{thebibliography}
\end{document}